\documentclass[aps,prb,superscriptaddress,prb,floatfix,
twocolumn,amsmath,amssymb,showpacs]{revtex4}
\usepackage{amsfonts}
\usepackage{lipsum}
\usepackage{multirow} 
\usepackage{graphicx}% Include figure files
\usepackage{dcolumn}% Align table columns on decimal point
\usepackage{bm}% bold math
\usepackage{multirow}
\usepackage{booktabs}
\usepackage{mathtools}   % loads »amsmath«
\makeatletter
\def\user@resume{resume}
\def\user@intermezzo{intermezzo}
\newcounter{previousequation}
\newcounter{lastsubequation}
\newcounter{savedparentequation}
\renewenvironment{subequations}[1][]{%
      \def\user@decides{#1}%
      \setcounter{previousequation}{\value{equation}}%
      \ifx\user@decides\user@resume 
           \setcounter{equation}{\value{savedparentequation}}%
      \else  
      \ifx\user@decides\user@intermezzo
           \refstepcounter{equation}%
      \else
           \setcounter{lastsubequation}{0}%
           \refstepcounter{equation}%
      \fi\fi
      \protected@edef\theHparentequation{%
          \@ifundefined {theHequation}\theequation \theHequation}%
      \protected@edef\theparentequation{\theequation}%
      \setcounter{parentequation}{\value{equation}}%
      \ifx\user@decides\user@resume 
           \setcounter{equation}{\value{lastsubequation}}%
         \else
           \setcounter{equation}{0}%
      \fi
      \def\theequation  {\theparentequation  \alph{equation}}%
      \def\theHequation {\theHparentequation \alph{equation}}%
      \ignorespaces
}{%
  \ifx\user@decides\user@resume
       \setcounter{lastsubequation}{\value{equation}}%
       \setcounter{equation}{\value{previousequation}}%
  \else
  \ifx\user@decides\user@intermezzo
       \setcounter{equation}{\value{parentequation}}%
  \else
       \setcounter{lastsubequation}{\value{equation}}%
       \setcounter{savedparentequation}{\value{parentequation}}%
       \setcounter{equation}{\value{parentequation}}%
  \fi\fi
  \ignorespacesafterend
}
\makeatother
\begin{document}
\title{\bf 
Theory of anharmonic phonons in 2D crystals}

\author{K. H. Michel}
\email{karl.michel@uantwerpen.be}
\affiliation{Universiteit Antwerpen, Department of Physics, 
Groenenborgerlaan 171, BE-2020 Antwerpen, Belgium.
}

\author{S. Costamagna}
\email{costamagna@ifir-conicet.gov.ar}
\affiliation{
Instituto de
F\'{\i}sica Rosario, Bv. 27 de Febrero 210 bis, 2000 Rosario,
Argentina.}
\affiliation{Universiteit Antwerpen, Department of Physics, 
Groenenborgerlaan 171, BE-2020 Antwerpen, Belgium.
}

\author{F. M. Peeters}
\email{francois.peeters@uantwerpen.be}
\affiliation{Universiteit Antwerpen, Department of Physics, 
Groenenborgerlaan 171, BE-2020 Antwerpen, Belgium.
}
\date{\today}

\begin{abstract}
Anharmonic effects in an atomic monolayer thin crystal with honeycomb lattice structure 
are investigated by analytical and numerical lattice dynamical methods. Starting from a semi-empirical 
model for anharmonic couplings of third and fourth order, we study the in-plane and out-of-plane
(flexural) mode components of the generalized wave vector dependent Gr\"uneisen parameters,
the thermal tension and the thermal expansion coefficients as function of temperature and 
crystal size. From the resonances of the displacement-displacement correlation functions
we obtain the renormalization and decay rate of in-plane and flexural phonons as function of
temperature, wave vector and crystal size in the classical and in the quantum regime. 
Quantitative results are presented for graphene. % as a specific model, 
There we find that the transition temperature $T_{\alpha}$ from negative to positive thermal expansion
is lowered with smaller system size.
Renormalization of the flexural mode has the opposite effect and leads 
to values of $T_{\alpha} \approx$300~K for systems of macroscopic size.
% and by renormalization of the flexural mode frequency.
Extensive numerical analysis throughout
the Brillouin zone explores various decay and scattering channels.
The relative importance of Normal and Umklapp processes is investigated. 
The work is complementary to crystalline membrane theory and computational studies 
of anharmonic effects in two-dimensional crystals.

\end{abstract}

%\pacs{73.23.-b, 73.21.La, 72.15.Qm, 71.27.+a}

%%%%%%%%%%%%%%%%%%%%%%%%%%%%%%%%%%%%%%%%%%%%%%%%%%%%%%%%%

\maketitle
%%%%%%%%%%%%%%%%%%%%%%%%%%%%%%%%%%%%%%%%%%%%%%%%%%%%%%%%%%%
\section{Introduction}

Phonon-phonon interactions due to the anharmonicity of lattice forces are essential for the 
understanding of thermoelastic properties and heat transport in solids~\cite{book}.
While for three-dimensional (3D) crystals the subject is well established~\cite{maradudin},
the discovery of graphene and of other two-dimensional(2D) crystals~\cite{geim,geim2,castro} 
has given new impetus to experimental and theoretical studies of anharmonicity 
related phenomena in ultra-thin crystals with a countable number of layers.

Electron microscopy and diffraction studies have proven the existence of ripples 
in single- and bi-layer graphene membranes~\cite{meyer}.
The thermal expansion of graphene has been found to be negative~\cite{yoon}
in the measured temperature range between 200-400~K while earlier experiments~\cite{bao}
had led to the estimate that a transition to 
positive values occurs near 350~K. Measurements of thermal conductivity $\varkappa$ on 
suspended single-layer graphene revealed an anomalous large value above the in-plane bulk 
graphite value~\cite{Balandin-exp}. 
Since then the determination of $\varkappa$ %\mathfrak{} 
as function of temperature $T$ in suspended and in supported~\cite{science} 
single- and few-layer graphene~\cite{wang} 
is an important topic of experimental and theoretical research~\cite{bala}.

Theoretical explanations of all these phenomena are related to the anharmonic 
coupling between in-plane stretching and out-of-plane bending or flexural phonon modes.
Such a coupling was originally suggested~\cite{Lifshitz}
as a membrane effect and explains the negative coefficient of thermal 
expansion in layered structures. 
Ab-initio density functional theory (DFT) calculations~\cite{mounet}
show that the thermal contraction in graphene subsists up to $T>$~2000~K. 
Atomistic Monte Carlo simulations~\cite{faso3}
exhibit a crossover from contraction to expansion near 900~K. 
Most recently the thermal expansion in monolayer graphene has been calculated by the
unsymmetrized self consistent field method~\cite{franco}.
Monte Carlo simulations
also suggest that the formation of ripples~\cite{meyer}
due to anharmonic coupling leads to the stabilization of graphene as a 2D crystal~\cite{fasolino}.
Acoustic phonon lifetimes in free standing and in strained graphene have been 
calculated by DFT methods~\cite{nano-marzari,lorenzo} 
and the results have been used to estimate the $T$ dependence of the intrinsic anharmonic
thermal conductivity~\cite{lorenzo}.
Recent analytic studies~\cite{amorin} of thermodynamic properties by continuum field theory
methods exploit the equivalence~\cite{chaikin}
%methods exploit the equivalence~\cite{chaikin,mariani}
between graphene treated in the continuum approximation and a crystalline 
(polymerized) membrane~\cite{membrane}.
%Beyond the classical theory of membranes~\cite{membrane}
%quantum effects have been analyzed in the low $T$ regime. 
Within the continuum theory of thin sheets~\cite{chaikin,Landau}
the in-plane strains comprise terms which are quadratic in the out-of-plane fluctuations.
These terms then lead to anharmonic couplings in the elastic free energy. 
The resulting phonon mediated interactions between Gaussian curvatures increase
the bending rigidity and stabilize the membrane~\cite{Nelson-Peliti}.

In the present paper we start from a somewhat different 
approach based on a discrete atomistic model of a monolayer crystal. 
We consider a Hamiltonian where the potential energy has been 
expanded up to fourth order in the atomic displacements. 
The harmonic and anharmonic coupling coefficients are determined by means of empirical data of 
phonon dispersions~\cite{Mohr} and Gr\"uneisen parameters~\cite{nikas}.
As an advantage of such a concept we consider the fact that 
we can use to a large extent well established analytical and numerical methods from 
lattice dynamics and at the same time take into account the specific structural 
properties of a 2D hexagonal crystal.

We will restrict ourselves to the study of thermal tension 
(equivalently thermal expansion) and of phonon resonances (shifts and linewidths).
Although the thermal conductivity is from the technological point of view
the more important quantity, a comprehensive study is beyond the scope 
of the present work. 
For an outline of different theoretical approaches developed so far for 
phonon transport in graphene, see Refs.~\onlinecite{nika-1, hindawi,broido1,fisher1}.

The content of the paper is as follows. In Section~II we recall some basic concepts and definitions
of the theory of lattice dynamics and anharmonic phonons. Next, in Sect.~III, we first express
the thermal tension in terms of phonon related quantities such as 
vibrational energy and generalized Gr\"uneisen coefficients. We distinguish in-plane and out-of-plane
acoustic modes. Secondly we study the resonances of the corresponding displacement-displacement
Green's functions, thereby paving 
the way for a later discussion of phonon linewidths and bandshifts.  
In Sect.~IV we describe a central force constants 
model which will be used for quantitative calculations.
In Sect.~V, we present detailed analytical calculations of the generalized 
Gr\"uneisen coefficients and of the thermal expansion. 
%The latter is discussed as 
%function of temperature and crystal size. 
The competing interplay of out-of-plane modes 
which favor thermal contraction and of in-plane modes which favor 
thermal expansion is investigated as function of temperature and crystal size.
In Sect. VI we study lineshifts and decay rates of in-plane and out-of-plane modes
and the effect of flexural mode renormalization on thermal expansion.
In Sect. VII we present extensive
numerical calculations of Gr\"uneisen coefficients, 
thermal expansion, phonon lineshift and decay rates.  
Concluding remarks (Sect. VIII) close the paper.

%
%%%%%%%%%%%%%%%%%%%%%%%%%%%%%%%%%%%%%%%%%%%%%%%%%%%%%%%%%%%%%%%%%%%
\section{Basic Concepts}

We recall some elements  of lattice dynamics of a non-primitive 
non-ionic crystal~\cite{Born,maradudin} and apply these concepts to 
2D graphene~\cite{Saito,Reich}.

The crystal consists of $N$ unit cells, each unit cell contains two C atoms which we label
by an index $\kappa$=$1,2$. The positions of the unit cells are fixed 
by the lattice vectors 
\begin{equation}
\vec{X}(\vec{n})=n_1 \vec{a}_1+n_1 \vec{a}_2.
\end{equation}
Here $\vec{a}_{\alpha}$ $\alpha$=$1,2$ are two noncolinear basis vectors while 
$\vec{n}$=$(n_1,n_2)$, where $n_i$ are integers, labels the unit cells. 
The equilibrium positions of the atoms in the lattice plane are given by 
\begin{equation}
\vec{X}(\vec{n}\kappa)=\vec{X}(\vec{n})+\vec{r}(\kappa),
\label{eq2}
\end{equation}
where $\vec{r}(\kappa)$ specifies the positions of the $\kappa$-th
atom in the $\vec{n}$-th unit cell. We use $u_i(\vec{n}\kappa)$
for the $i$-th Cartesian components ($i$=$x,y,z$) of the instantaneous displacement
vector of atom ($\vec{n}\kappa$) away from its equilibrium position,
the $z$-component refers to the out-of-plane displacements. 

The crystal potential energy $\Phi$ is a function of the instantaneous positions 
$\vec{R}(\vec{n}\kappa)$=$\vec{X}(\vec{n}\kappa)+\vec{u}(\vec{n}\kappa)$
of the atoms. 
%While $\vec{R}$ and $\vec{u}$ are 3D vectors, 
%$\vec{X}(\vec{n}\kappa)$ has the components ($X_x,X_y,0$). 
Expansion of the 
potential energy in terms of displacements away from the equilibrium position gives
\begin{equation}
\Phi=\Phi^{(0)}+\Phi^{(2)}+\Phi^{(3)}+\Phi^{(4)}+...\ \ \ .
\label{eq3}
\end{equation}
Here $\Phi^{(0)}$ is the rigid lattice potential,
$\Phi^{(2)}$ the harmonic potential and 
$\Phi^{(3)}$, $\Phi^{(4)}$ are the third order and fourth order 
anharmonic potential contributions.
Explicitly we write
\begin{subequations}
\begin{align}
%\begin{equation}
%\begin{eqnarray}
\Phi^{(2)}&=\frac{1}{2}\sum_{\vec{n}\kappa}
\sum_{\vec{n}'\kappa'}
\sum_{ij}\Phi^{(2)}_{ij}(\vec{n}\kappa;\vec{n}'\kappa')
u_i(\vec{n}\kappa)u_j(\vec{n}'\kappa')\label{xxa}\\
%\end{equation}
%\begin{equation}
\Phi^{(3)}&=\frac{1}{3!}\sum_{\vec{n}\kappa}
\sum_{\vec{n}'\kappa'}\sum_{\vec{n}''\kappa''}
\sum_{ijk}\Phi^{(3)}_{ijk}(\vec{n}\kappa;\vec{n}'\kappa';\vec{n}''\kappa'')\nonumber \\
&\times u_i(\vec{n}\kappa)u_j(\vec{n}'\kappa')u_k(\vec{n}''\kappa'')\\
\Phi^{(4)}&=\frac{1}{4!}\sum_{\vec{n}\kappa}
\sum_{\vec{n}'\kappa'}\sum_{\vec{n}''\kappa''}\sum_{\vec{n}'''\kappa'''}\nonumber \\
&\sum_{ijkl}\Phi^{(4)}_{ijkl}(\vec{n}\kappa;\vec{n}'\kappa';\vec{n}''\kappa'';
\vec{n}'''\kappa''')\nonumber \\
&\times u_i(\vec{n}\kappa)u_j(\vec{n}'\kappa')u_k(\vec{n}''\kappa'')
u_l(\vec{n}'''\kappa''')
\end{align}
\end{subequations}
%\end{eqnarray}
%\end{equation}
The coupling parameters $\Phi^{(2)}_{ij}(\vec{n}\kappa;\vec{n}'\kappa')$,
$\Phi^{(3)}_{ijk}(\vec{n}\kappa;\vec{n}'\kappa';\vec{n}''\kappa'')$ and
$\Phi^{(4)}_{ijkl}(\vec{n}\kappa;\vec{n}'\kappa';\vec{n}''\kappa'';
\vec{n}'''\kappa''')$ are the second, third  and fourth 
order derivatives of the potential energy 
with respect to the displacements, taken at the equilibrium positions. 
%Since the rest positions of the atoms are taken as equilibrium positions,
%the first order term $\Phi^{(1)}$ is identically zero.

\begin{figure}[t]
\vspace{0.1cm}
\includegraphics[width=0.49\textwidth]{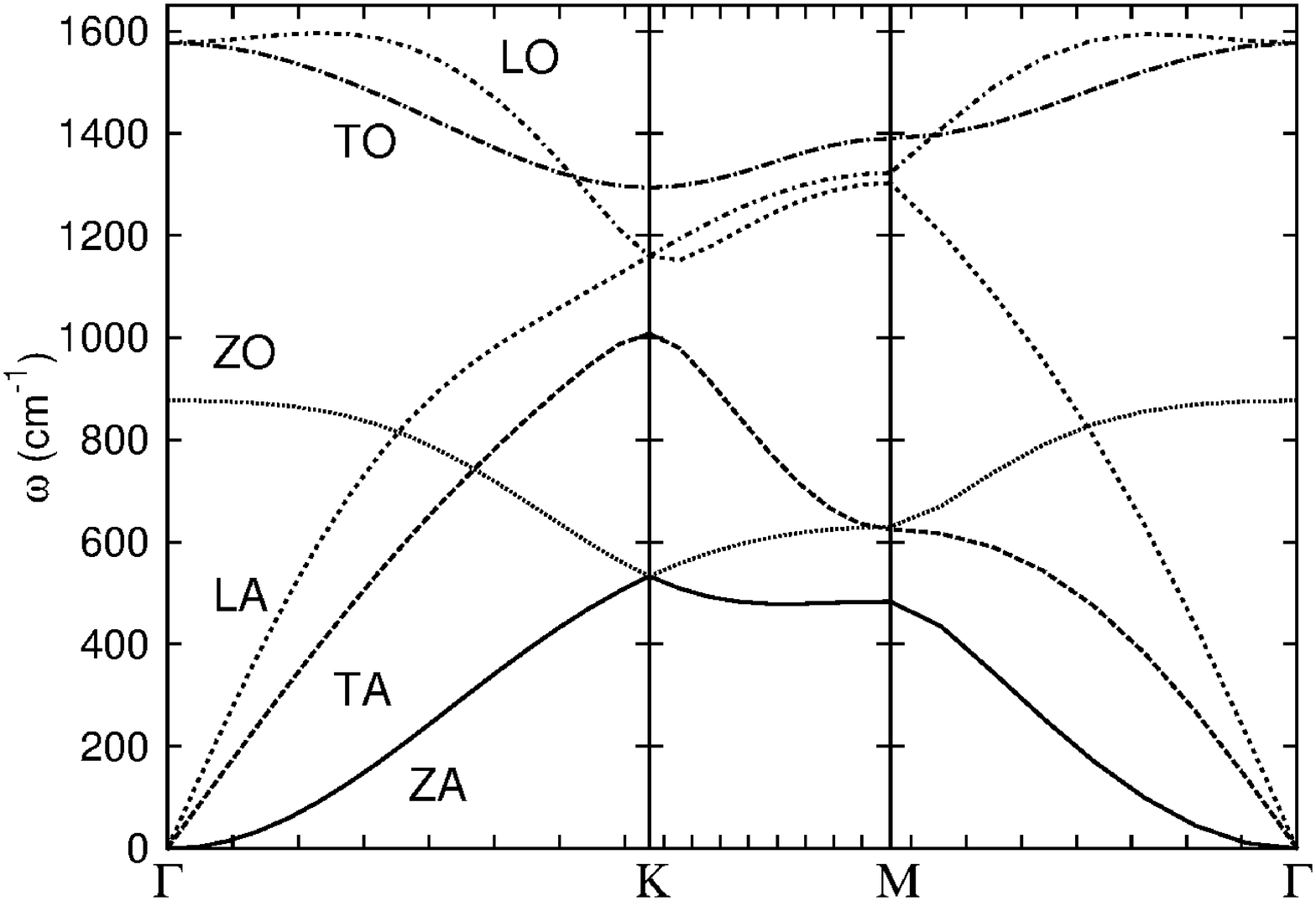}
\caption{Phonon modes of graphene along 
the high symmetry crystallographic direction 
$\Gamma$−K-M−$\Gamma$ (Ref.~\onlinecite{Karl}). 
}
\label{fig1}
\end{figure}

The kinetic energy of the crystal is given by 
\begin{equation}
T=\sum_{\vec{n}\kappa}\sum_i \frac{p^2_i(\vec{n}\kappa)}{2M_{\kappa}}, 
\end{equation}
where $p_i(\vec{n}\kappa)$ are the components of the momentum conjugate to 
$u_i(\vec{n}\kappa)$ and where $M_{\kappa}$ is the mass of the 
$\kappa$-th atom. In case of graphene 
with two C atoms per unit cell one has
$M_1$=$M_2$=$M_C$ where $M_C$=$12\ au$
is the mass of the carbon atom.
The area of the unit cell is given by $v_{2D}=a^2\sqrt{3}/2$, 
with $|\vec{a}_1|=|\vec{a}_2|=a=2.46\AA$.

In the following we restrict ourselves to lowest order anharmonicities and consider 
$\Phi^{(3)}$ and $\Phi^{(4)}$ as a perturbations to the harmonic Hamiltonian 
$H_h$=$(T+\Phi^{(2)})$. 
%
%
%The crystal dynamics in harmonic approximation is 
%described in case of graphene by $6N$ coupled equations of motion
%\begin{equation}
%M_{\kappa}\Ddot{u}_i(\vec{n}\kappa)=-
%\sum_{\vec{n}'\kappa'j}\Phi^{(2)}_{ij}(\vec{n}\kappa;\vec{n}'\kappa')
%u_j (\vec{n}'\kappa')
%\end{equation}
%for the (time dependent) atomic displacements.
We introduce Fourier transforms in space and time by writing 
\begin{equation}
u_i (\vec{n}\kappa) =
\frac{1}{\sqrt{M_C N}}
\sum_{\vec{q}}u_i^{\kappa} (\vec{q})
e^{i\vec{q}\cdot\vec{X}(\vec{n}\kappa)-i\omega t}
\end{equation}
Here $\vec{q}$ is the wave vector in the 2D Brillouin zone (BZ)
while $\omega$ is the frequency.
%The equation of motion reads 
%\begin{equation}
%\omega^2u_i^{\kappa}(\vec{q})=
%\sum_{\kappa' j} D_{ij}^{\kappa \kappa'}(\vec{q})
%u_j^{\kappa'}(\vec{q})
%\end{equation}
%where $D_{ij}^{\kappa \kappa'}$ are the elements of 
The $6\times6$ dynamical matrix $D(\vec{q})$ has the elements:
\begin{equation}
D_{ij}^{\kappa \kappa'}(\vec{q})=\frac{1}{M_C}\sum_{\vec{n}'}
\Phi_{ij}^{(2)}(\vec{n}\kappa;\vec{n}'\kappa')
e^{i\vec{q}\cdot[\vec{X}(\vec{n}'\kappa')-
\vec{X}(\vec{n}\kappa)]}
\label{eq9}
\end{equation}
and is Hermitian. 
Solving the secular equation one obtains the eigenfrequencies 
$\omega(\vec{q},\lambda)$, $\lambda=1,...,6$, 
%and the eigenvectors $e_i^{\kappa}(\vec{q},\lambda)$ 
%The eigenvalues and eigenvectors 
%re obtained  
%y solving the secular equation 
%begin{equation}
%1\omega^2  -D(\vec{q})|=0.
%label{eq10}
%end{equation}
%e obtain 
%begin{equation}
%omega^2(\vec{q},\lambda)=\sum_{\kappa' j}\sum_{\kappa i}
%^{\kappa*}_{i}(\vec{q},\lambda)
%^{\kappa \kappa'}_{i j}(\vec{q})e^{\kappa'}_j (\vec{q},\lambda),
%label{eq11}
%end{equation}
%here $\omega^2(\vec{q},\lambda)$,
%\lambda$=$1,2, ... ,6$, are the eigenvalues 
and the eigenvectors $\vec{e}(\vec{q},\lambda)$ %the orthonormalized eigenvectors 
with components $e_i^{\kappa}(\vec{q},\lambda)$,
$\kappa$=$1,2$ and $i$=$x,y,z$.
As is well known~\cite{Saito} there are three acoustical branches %$(\lambda=1,2,3)$
which we label by ZA, TA and LA and three optical branches %$(\lambda=4,5,6)$
which we label by ZO, TO and LO (See Fig.~\ref{fig1}). 
%See also Sect.III, Fig.\ref{fig1}.

In terms of phonon annihilation and creation operators 
$b_{\vec{q}}^{\lambda}$ and $b_{-\vec{q}}^{\lambda \dagger}$,
that satisfy the usual 
commutation relations for Bose operators, 
the harmonic part of the Hamiltonian reads
\begin{equation}
H_h=\sum_{\vec{q}}
\sum_{\lambda}
\hbar \omega(\vec{q},\lambda) 
\Bigg(b_{\vec{q}}^{\lambda\dagger}b_{\vec{q}}^{\lambda}+\frac{1}{2}\Bigg).
%\left(n\binom{\lambda}{\vec{q}}+
%\frac{1}{2}\right),
\end{equation}
%where $b_{\vec{q}}^{\lambda \dagger}b_{\vec{q}}^{\lambda}$
%is the density operator of phonons with wave vector $\vec{q}$ and polarization
%$\lambda$. 
With 
$B\binom{\lambda}{\vec{q}}=\Big(b_{\vec{q}}^{\lambda}+b_{-\vec{q}}^{\lambda \dagger}\Big)$.
The third order anharmonic contribution in Eq.~(\ref{eq3}) is given by
%\begin{eqnarray}
\begin{widetext}
\begin{equation}
\Phi^{(3)}=\frac{\hbar^{3/2}}{3!N^{1/2}}
\sum_{\vec{q}_1\vec{q}_2\vec{q}_3}
\sum_{\lambda_1 \lambda_2 \lambda_3}
\Phi^{(3)}\binom{\lambda_1 \lambda_2 \lambda_3}{\vec{q_1}\vec{q_2}\vec{q_3}} 
B\binom{\lambda_1}{\vec{q}_1}
B\binom{\lambda_2}{\vec{q}_2}
B\binom{\lambda_3}{\vec{q}_3},
\end{equation}
%\end{eqnarray}
%%
%\end{widetext}
with
%\begin{eqnarray}
%\begin{widetext}
%\[
\begin{equation}
\Phi^{(3)}\binom{\lambda_1 \lambda_2 \lambda_3}{\vec{q}_1\vec{q}_2\vec{q}_3}=
%\sum_{\kappa_1 i,\kappa_2 j,\kappa_3 k}
\sum_{\kappa_1 i} \sum_{\kappa_2 j} \sum_{\kappa_3 k}
\frac{e_i^{\kappa_1}(\vec{q}_1,\lambda_1)
e_j^{\kappa_2}(\vec{q}_2,\lambda_2)
e_k^{\kappa_3}(\vec{q}_3,\lambda_3)}
{\sqrt{8\omega(\vec{q}_1,\lambda_1)
\omega(\vec{q}_2,\lambda_2) \omega(\vec{q}_3,\lambda_3)}} 
\Phi_{ijk}^{(3)}\binom{\kappa_1 \kappa_2 \kappa_3}{\vec{q}_1\vec{q}_2\vec{q}_3}.
\label{phi3}
\end{equation}
%\end{widetext}
%\end{eqnarray}
%where
%\begin{widetext}
%\begin{equation}
%\begin{eqnarray}
%\[
%\Phi^{(3)}_{ijk}\binom{\kappa_1 \kappa_2 \kappa_3}{\vec{q}_1\vec{q}_2\vec{q}_3}=
%%\frac{1}{(NM_C)^{3/2}} 
%\frac{1}{\sqrt{N^3M_{\kappa_1}M_{\kappa_2}M_{\kappa_3}}}
%\sum_{\vec{n}_1\vec{n}_2\vec{n}_3}
%\Phi_{ijk}^{(3)}(\vec{n}_1\kappa_1;\vec{n}_2\kappa_2;\vec{n}_3\kappa_3) 
%e^{i[\vec{q}_1\cdot\vec{X}(\vec{n}_1\kappa_1)+
%\vec{q}_2\cdot\vec{X}(\vec{n}_2\kappa_2)+
%\vec{q}_3\cdot\vec{X}(\vec{n}_3\kappa_3)
%]}
%.
%\end{equation}
%\end{eqnarray}
%\end{widetext}
Invariance of the crystal by a displacement through a lattice 
translation vector implies that
%\begin{widetext}
%\begin{subequations}
\begin{align}
%%\begin{equation}
%\hspace{0.8cm}
%\Phi_{ijk}^{(3)}(\vec{n}_1 \kappa_1; \vec{n}_2 \kappa_2; \vec{n}_3 \kappa_3)&=
%\Phi_{ijk}^{(3)}(\vec{n}_1-\vec{n}_3 \kappa_1; 
%\vec{n}_2-\vec{n}_3 \kappa_2; \vec{0} \kappa_3) \label{eq18a},\\
%%\end{equation}
%\intertext{and hence}
%&\nonumber \\
%\begin{eqnarray}
%\begin{equation}
%\begin{widetext}
%\[
\Phi^{(3)}_{ijk}
\binom{\kappa_1 \kappa_2 \kappa_3}{\vec{q}_1\vec{q}_2\vec{q}_3}=& 
\frac{1}{\sqrt{M^3_C}}
%\frac{1}{\sqrt{M_{\kappa_1}M_{\kappa_2}M_{\kappa_3}}}
%\frac{1}{N^{1/2}M_C^{3/2}}
\sum_{\vec{n}_1\vec{n}_2} \Phi_{ijk}^{(3)}
(\vec{n}_1 \kappa_1; \vec{n}_2 \kappa_2; \vec{0} \kappa_3) \nonumber \\
&\times e^{i[\vec{q}_1\cdot\vec{X}(\vec{n}_1 \kappa_1)+
\vec{q}_2\cdot\vec{X}(\vec{n}_2 \kappa_2)+
\vec{q}_3\cdot\vec{r}(\kappa_3)]}
\times \Delta(\vec{q}_1+\vec{q}_2+\vec{q}_3),
\ \ \ \ \ \ \ \ \ \ \ \ \hspace{0.8cm}
%\]
\label{eq18b}
\end{align}
%\end{equation}
%\end{subequations}
%\end{equation}
\end{widetext}
%\end{eqnarray}
where 
\begin{equation}
\Delta(\vec{q}_1+\vec{q}_2+\vec{q}_3)=\sum_{\vec{G}}
\delta_{\vec{q}_1+\vec{q}_2+\vec{q}_3,\vec{G}}.
\label{deltaq}
\end{equation}
Here $\Delta$ vanishes unless ($\vec{q}_1+\vec{q}_2+\vec{q}_3$)
is equal to a lattice vector $\vec{G}$ in 2D reciprocal space. 
In the latter case $\Delta$=$1$.
A phonon scattering process with $\vec{G}=0$ is called Normal while when a 
non-zero $\vec{G}$ is needed to bring back the scattered phonon inside the first BZ the 
process is called Umklapp~\cite{peirels}. 

The fourth order anharmonic term reads
\begin{widetext}
\begin{equation}
\Phi^{(4)}=\frac{\hbar^{2}}{4!N}
\sum_{\vec{q}_1\vec{q}_2\vec{q}_3\vec{q}_4}
\sum_{\lambda_1 \lambda_2 \lambda_3\lambda_4}
\Phi^{(4)}\binom{\lambda_1 \lambda_2 \lambda_3\lambda_4}
{\vec{q_1}\vec{q_2}\vec{q_3}\vec{q_4}} 
B\binom{\lambda_1}{\vec{q}_1}
B\binom{\lambda_2}{\vec{q}_2}
B\binom{\lambda_3}{\vec{q}_3}
B\binom{\lambda_4}{\vec{q}_4},
\end{equation}
with
%\begin{eqnarray}
%\begin{widetext}
%\[
\begin{equation}
\Phi^{(4)}\binom{\lambda_1 \lambda_2 \lambda_3\lambda_4}
{\vec{q}_1\vec{q}_2\vec{q}_3\vec{q}_4}=
%\sum_{\kappa_1 i,\kappa_2 j,\kappa_3 k}
\sum_{\kappa_1 i} \sum_{\kappa_2 j} \sum_{\kappa_3 k}\sum_{\kappa_4 l}
\frac{e_i^{\kappa_1}(\vec{q}_1,\lambda_1)
e_j^{\kappa_2}(\vec{q}_2,\lambda_2)
e_k^{\kappa_3}(\vec{q}_3,\lambda_3)
e_l^{\kappa_4}(\vec{q}_4,\lambda_4)
}
{\sqrt{16\omega(\vec{q}_1,\lambda_1)
\omega(\vec{q}_2,\lambda_2) 
\omega(\vec{q}_3,\lambda_3)
\omega(\vec{q}_3,\lambda_3)}}
\Phi_{ijkl}^{(4)}\binom{\kappa_1 \kappa_2 \kappa_3 \kappa_4}
{\vec{q}_1\vec{q}_2\vec{q}_3\vec{q}_4},
\label{phi4}
\end{equation}
%\end{widetext}
%\end{eqnarray}
where
%\begin{widetext}
%\begin{equation}
\begin{eqnarray}
%\[
\Phi^{(4)}_{ijkl}\binom{\kappa_1 \kappa_2 \kappa_3 \kappa_4}
{\vec{q}_1\vec{q}_2\vec{q}_3\vec{q}_4}=
\frac{1}{M^2_C} 
%\frac{1}{\sqrt{M_{\kappa_1}M_{\kappa_2}M_{\kappa_3}M_{\kappa_4}}}
\sum_{\vec{n}_1\vec{n}_2\vec{n}_3}
\Phi_{ijkl}^{(4)}(\vec{n}_1\kappa_1;\vec{n}_2\kappa_2;\vec{n}_3\kappa_3;
\vec{0}\kappa_4) \ \ \ \ \ \  \ \ \ \ \ \ \ \ \ \ \ \ \ \nonumber\\ 
\ \ \ \ \ \ \ \ \ \ \ \ \ \ \ \ \ 
\times e^{i[\vec{q}_1\cdot\vec{X}(\vec{n}_1\kappa_1)+
\vec{q}_2\cdot\vec{X}(\vec{n}_2\kappa_2)+
\vec{q}_3\cdot\vec{X}(\vec{n}_3\kappa_3)+\vec{q}_4\cdot\vec{r}(\kappa_4)]}
\times \Delta(\vec{q}_1+\vec{q}_2+\vec{q}_3+\vec{q}_4).
\label{eq22}
\end{eqnarray}
\end{widetext}

Invariance of the potential energy against infinitesimal translations of the crystal implies
for $\nu\ge 2,3,...\ $
\begin{equation}
%\begin{subequations}
%\begin{align}
%\begin{equation}
\sum_{\kappa_1\vec{n}_1}\Phi_{ij..}^{(\nu)}(\vec{n}_1\kappa_1,\vec{n}_2\kappa_2,...\ ,\vec{n}_{\nu}\kappa_{\nu})=0.
\label{eq19-aa}
\end{equation}
%\end{equation}
%\intertext{ where $\nu\ge 2,3,...\ $. Equivalently} 
%\begin{equation}
%\lim_{\vec{q}_1\rightarrow 0}\sum_{\kappa_1}\sqrt{M_{\kappa_1}}
%\Phi^{(\nu)}_{ij...}\binom{\kappa_1 \kappa_2 \ ...\ \kappa_{\nu}}
%{\vec{q}_1\vec{q}_2 \ ...\ \vec{q}_{\nu}}
%&=0.
%\end{equation}
%\end{align}
%\end{equation}
%\end{subequations}
% \sum_{\kappa_1} \sqrt{M_{\kappa_1}\Phi_{ij..}^{(\nu)}

%%%%%%%%%%%%%%%%%%%%%%%%%%%%%%%%%%%%%%%%%%%%%%%%%%%%%%%%%%
\section{Physical Quantities}

For the study of thermal expansion and phonon renormalization effects we
restrict ourselves to acoustic modes LA, TA and ZA which we denote by L, T and Z.
%
%We restrict ourselves to temperartures not higher than room $T$. 
%Hence we retain only LA, TA and ZA acoustic modes which we denote by L, T and Z
%respectively, and neglect optical phonons.  
%
In the long wavelength regime the in-plane mode frequencies 
read $\omega(\vec{q},\lambda)=c_{\lambda}q$,
where for $\lambda\in\{$L, T$\}$, $c_{\mbox{{\scriptsize L}}}$ and $c_{\mbox{{\scriptsize T}}}$ are the 
longitudinal and transverse sound velocities, respectively.
%where $c_{\lambda}$ is the longitudinal or transverse sound velocity for $\lambda$=L or T respectively.
%
The out-of-plane mode  (also called flexural mode)
has frequency 
%(also called flexural mode) 
$\omega(\vec{q},\mbox{Z})=\sqrt{\kappa_0}q^2$,
where $\kappa_0=\kappa_B/\rho_{2D}$, $\kappa_B$ is the bending rigidity 
coefficient and $\rho_{2D}$ the density.
With the harmonic force constant model~\cite{Karl} for graphene we obtain 
$\kappa_0=42.48\times10^{-6}$ cm$^4$ s$^{-2}$, which corresponds to $\kappa_B=3.23\times10^{-12}$ erg $\approx$
2.01 eV. Due to a numerical error the value 1.12 eV quoted 
in Ref.~\onlinecite{Karl} is wrong.
One finds a broad range of values for $\kappa_B$ for graphene in the literature
(units eV): 1.68 (Ref.~\onlinecite{lindsay1}), 1.1 (Ref.~\onlinecite{amorin}).
From out of plane phonon dispersions measured by neutron 
scattering on graphite~\cite{Nickow} we estimate $\kappa_B=$2.4~eV.

%%%%%%%%%%%%%%%%%%%%%%%%%%%%%%%%%%%%%%%%%%%%%%%%%%%%%%%%%%%%%%%%%%%
\subsection{Thermoelastic Phenomena}

We want to calculate thermodynamic quantities such as thermal tension and thermal expansion
which depend on lattice anharmonicities.
We start from an undeformed graphene crystal at an initial temperature $T$ and in the absence
of external forces. We recall that the elastic properties of the 2D %two-dimensional 
hexagonal crystal reflect the symmetry of an isotropic solid.
% with two elastic constants.
A small temperature change 
%$\theta$, where $T=T_0$+$\theta$ is the final temperature, 
will cause isotropic lattice deformations which are described by the thermal expansion coefficient
$\alpha_{T}=\sum_{i}d\epsilon_{ii}/dT$.
%
%
%The corresponding thermoelastic 
%free energy per unit volume (area in 2D) reads
%\begin{equation}
%F(\epsilon,T)=-\beta_{T}\theta\sum_{i=1}\epsilon_{ii}+\frac{B_{2D}}{2}\left(\sum_{i=1}\epsilon_{ii}\right)^2.
%\end{equation}
Here the strain $\sum_{i}\epsilon_{ii}\equiv\epsilon_{xx}+\epsilon_{yy}$ characterizes the  change
of the unit cell area.
%, $B_{2D}$ is the bulk modulus. We recall that for graphene $B_{2D}=\lambda_{2D}+\mu_{2D}$,
%where $\lambda_{2D}=\gamma_{12}$, $\mu_{2D}=\gamma_{66}$ are the elastic tension coefficients\cite{Karl}. 
%The quantity $\beta_{T}$ is the thermodynamic tension coefficient, which in case of the 2D 
%hexagonal crystal is a scalar. 
The thermal expansion is related to the thermal tension $\beta_T$ by
\begin{equation}
\alpha_{T}=\beta_{T}B_{2D}^{-1}.
\label{therm-ex-eq}
\end{equation}
Here $B_{2D}$=$\lambda_{2D}+\mu_{2D}$ is the bulk modulus, 
$\lambda_{2D}=\gamma_{12}$ and $\mu_{2D}=\gamma_{66}$ are the elastic tension coefficients\cite{Karl}.
For graphene we use $B_{2D}$=24.89$\times$10$^{4}$~dyn/cm.

Starting from the vibrational energy per unit cell in the quasi-harmonic approximation\cite{Leibfried},
one obtains the tension coefficient as
\begin{equation}
\beta_{T}=\frac{1}{v_{2D}N}\sum_{\vec{q}\lambda}
\gamma(\vec{q},\lambda)
\frac{\partial{E(\omega(\vec{q},\lambda),T)}}{\partial{T}}.
\label{eq_ten}
\end{equation}
Here
\begin{equation}
E(\omega(\vec{q},\lambda),T)=\hbar\omega(\vec{q},\lambda)\left[n(\vec{q},\lambda)+\frac{1}{2}\right],
\label{eq_ten1}
\end{equation}
is the vibrational energy of phonons with harmonic frequency $\omega(\vec{q},\lambda)$ and
$n(\vec{q},\lambda)=(e^{\hbar\omega(\vec{q},\lambda)/k_B T}-1)^{-1}$ is the phonon thermal density at temperature $T$. 
The generalized Gr\"uneisen coefficient reads 
\begin{equation}
\gamma(\vec{q},\lambda)=-\frac{1}{\omega(\vec{q},\lambda)}
\sum_i\frac{\partial{\omega(\vec{q},\lambda)}}{\partial{\epsilon_{ii}}},
\label{eq_gamma}
\end{equation}
where $\lambda\in\{$T,L,Z$\}$ characterize the relative changes of the acoustic phonon frequencies by
strains.
%and obtain for the thermal tension
%\begin{equation}
%\beta_{T}=\frac{1}{v_{2D}N}\sum_{\vec{q}\lambda}
%\gamma(\vec{q},\lambda)
%\frac{\partial{E(\omega(\vec{q},\lambda),T)}}{\partial{T}}.
%\label{eq_ten}
%\end{equation}

The evaluation of the Gr\"uneisen coefficient requires the calculation of 
$\partial{\omega(\vec{q},\lambda)}/\partial{\epsilon_{ii}}$ which involves anharmonic 
interactions.
Details of the calculation are given in Appendix A and the results are discussed in Sect.~V.

%%%%%%%%%%%%%%%%%%%%%%%%%%%%%%%%%%%%%%%%%%%%%%%%%%%%%%%%%%%%%%%%%%%%%%%%%%%%%%%%%%%
%\subsection{Phonon shifts and linewidths}
\subsection{Phonon Resonances}

The anharmonic potentials $\Phi^{(3)}$ and $\Phi^{(4)}$ 
change the harmonic phonon frequencies 
$\omega(\vec{q},\lambda)$. We study the resonances
of the frequency dependent displacement-displacement Green's 
function $D(\vec{q},\lambda;z)$.

%%%%%%%%%%%%%%%%%%%%%%%%
One derives the Dyson equation~\cite{gotze}
%Using Green's functions techniques one derives the equation of motion~\cite{gotze}
%The equation of motion reads 
\begin{equation}
[z^2-\omega^2(\vec{q},\lambda)-\Sigma(\vec{q},\lambda;z)]D(\vec{q},\lambda;z)=\hbar,
\end{equation}
where $z=\omega+i\epsilon$, $\epsilon\rightarrow 0^+$, is the frequency.
The self-energy $\Sigma(\vec{q},\lambda;z)$ reads~\cite{gotze} 
%is given in lowest order of perturbation theory by
\begin{equation}
\Sigma(\vec{q},\lambda;z)=\Sigma'(\vec{q},\lambda;\omega)+i\Sigma''(\vec{q},\lambda;\omega)
\end{equation}
where
\begin{equation}
\Sigma'(\vec{q},\lambda;\omega)=\Sigma^{(3)'}(\vec{q},\lambda;\omega)+
\Sigma^{(4)}(\vec{q},\lambda),
\label{new29-old}
\end{equation}
with
%\begin{eqnarray}
\begin{widetext}
\begin{align}
%\begin{equation}
\Sigma^{(3)'}(\vec{q},\lambda;\omega)=&
\frac{\hbar\omega(\vec{q},\lambda)}{N}
%\Sigma(\vec{q},\lambda;z)&=\hbar \omega(\vec{q},\lambda) 
P\sum_{\vec{q}_2\vec{q}_3}
\sum_{\lambda_2 \lambda_3}
\left|\Phi^{(3)}\binom{\lambda \lambda_2 \lambda_3}
{-\vec{q}\vec{q}_2\vec{q}_3}\right|^2  \ \nonumber \\
&\times \Bigg\lbrace
\frac{1+n(\vec{q}_2,\lambda_2)+n(\vec{q}_3,\lambda_3)}
{\omega-\omega(\vec{q}_2,\lambda_2)-\omega(\vec{q}_3,\lambda_3)}- 
\frac{1+n(\vec{q}_2,\lambda_2)+n(\vec{q}_3,\lambda_3)}
{\omega+\omega(\vec{q}_2,\lambda_2)+\omega(\vec{q}_3,\lambda_3)}
%
%\qquad {} 
+\frac{2[n(\vec{q}_2,\lambda_2)-n(\vec{q}_3,\lambda_3)]}
{\omega+\omega(\vec{q}_2,\lambda_2)-\omega(\vec{q}_3,\lambda_3)}
%
%\frac{n(\vec{q}_2,\lambda_2)-n(\vec{q}_3,\lambda_3)}
%{z-\omega(\vec{q}_2,\lambda_2)+\omega(\vec{q}_3,\lambda_3)}
%
%\bigg)
\Bigg\rbrace,
\label{new29}
\end{align}
%\end{widetext}
%\begin{widetext}
with $P$ standing for the principal part,
\begin{equation}
% \vec{X}^s(\vec{n})=\sum_{\kappa}\frac{M_{\kappa}}{M}X_j(\vec{n}\kappa),
\Sigma^{(4)}(\vec{q},\lambda)=\frac{\hbar\omega(\vec{q},\lambda)}{N}
\sum_{\vec{q}_1\lambda_1}\Phi^{(4)}
\binom{\lambda \lambda_1 \lambda_1 \lambda}
{-\vec{q}\vec{q}_1-\vec{q}_1\vec{q}} %\nonumber\\
\bigg[ 1+2n(\vec{q}_1,\lambda_1) \bigg],
\label{new30}
\end{equation}
%\end{widetext}
%\end{eqnarray}
and where
%\begin{widetext}
\begin{align}
\Sigma''(\vec{q},\lambda;\omega)=
%\begin{equation}
&-\frac{\pi\hbar\omega(\vec{q},\lambda)}{N}
\sum_{\vec{q}_2\vec{q}_3}
\sum_{\lambda_2 \lambda_3}
\left|\Phi^{(3)}\binom{\lambda \lambda_2 \lambda_3}{-\vec{q}\vec{q}_2\vec{q}_3}\right|^2 \times 
\Bigg\lbrace \bigg[1+n(\vec{q}_2,\lambda_2)+n(\vec{q}_3,\lambda_3)\bigg] \nonumber \\ 
& \bigg[ \delta\big(\omega-\omega(\vec{q}_2,\lambda_2)-\omega(\vec{q}_3,\lambda_3)\big) 
-\delta\big(\omega+\omega(\vec{q}_2,\lambda_2)+\omega(\vec{q}_3,\lambda_3)\big)  \bigg]
\nonumber\\ 
& + 2 \bigg[n(\vec{q}_2,\lambda_2)-n(\vec{q}_3,\lambda_3)\bigg]
\delta\big(\omega+\omega(\vec{q}_2,\lambda_2)-\omega(\vec{q}_3,\lambda_3)\big)
%delta\bigg(\omega(\vec{q},\lambda)-\omega(\vec{q}_2,\lambda_2)+\omega(\vec{q}_3,\lambda_3)\bigg)\bigg]
%%
\Bigg\rbrace.
%\end{equation}
%\label{sigmapp}
\label{new31}
\end{align}
\end{widetext}
\noindent 
Within lowest order perturbation theory $\Sigma^{(4)}(\vec{q},\lambda)$ is real.
In expressions (\ref{new29})-(\ref{new31}) the summations in $\vec{q}$-space 
run over the 2D Brillouin zone.
% and the harmonic 
%phonon frequencies $\omega(\vec{q},\lambda)$ are those of the 2D crystal.

The resonances of $D(\vec{q},\lambda,z)$ near $\omega$=$\omega(\vec{q},\lambda)$ are given by
%, writing
% We call $\tilde\omega(\vec{q},\lambda)$ the resonances of $D(\vec{q},\lambda,z)$ near
%\omega$=$\omega(\vec{q},\lambda)$, writing 
\begin{equation}
z=\pm
%\tilde\omega(\vec{q},\lambda)=
\Big\lbrace\big[\omega^2(\vec{q},\lambda)+\Sigma'(\vec{q},\lambda)\big]^2+
\Sigma^{''2}(\vec{q},\lambda)\Big\rbrace^{1/4}
e^{\pm i\Psi(\vec{q},\lambda)}
\label{w-tilde}
\end{equation}
where
\begin{equation}
\Psi(\vec{q},\lambda)=\frac{1}{2}tan^{-1}\Bigg\lbrace\frac{\Sigma''(\vec{q},\lambda)}
{\omega^2(\vec{q},\lambda)+\Sigma'(\vec{q},\lambda)}\Bigg\rbrace.
\end{equation}
Here $\Sigma'(\vec{q},\lambda)$ and $\Sigma''(\vec{q},\lambda)$ stand for 
$\Sigma'(\vec{q},\lambda;\omega)$ and $\Sigma''(\vec{q},\lambda;\omega)$ 
with $\omega=\omega(\vec{q},\lambda)$.

For the in-plane modes $\lambda=\lbrace$L, T$\rbrace$, where the harmonic phonon dispersion is linear 
in the long wavelength regime, 
%it follows from Eqs.~(\ref{new20}),
%(\ref{new21}), (\ref{new29}) and (\ref{new30}) that $\Sigma'(\vec{q},\lambda)$
%for $\lambda \in \lbrace \mbox{L, T}\rbrace$ is quadratic in q and hence 
we obtain the renormalized phonon frequency
\begin{equation}
\Omega(\vec{q},\lambda)=
\omega(\vec{q},\lambda)+\Delta(\vec{q},\lambda)
%\omega(\vec{q},\lambda)+\Delta(\vec{q},\lambda)-i\Gamma(\vec{q},\lambda),
\label{34new}
\end{equation}
where
\begin{equation}
\Delta(\vec{q},\lambda)=\frac{\Sigma'(\vec{q},\lambda)}{2\omega(\vec{q},\lambda)},
\label{35new}
\end{equation}
is the phonon frequency shift.

The phonon damping (line-width) is given by 
\begin{equation}
\Gamma(\vec{q},\lambda)=-\frac{\Sigma''(\vec{q},\lambda)}{2\omega(\vec{q},\lambda)}.
\label{36new}
\end{equation}
%the phonon line-width (damping). 
Expressions corresponding to Eqs.~(\ref{34new})-(\ref{36new}) for $\Delta(\vec{q},\lambda)$
and $\Gamma(\vec{q},\lambda)$ have been obtained originally by 
diagrammatic techniques for 3D anharmonic
crystals~\cite{Kokkedee,Maradudin1}.

For the out-of-plane mode ($\lambda$=Z),  $\Sigma'(\vec{q},$Z$)$  is quadratic in $q$
(see Sect.~VI) and cannot be treated as a 
perturbation to $\omega^2(\vec{q},$Z$)$=$\kappa_0q^4$ in the 
long wavelength regime. We write
%
%Eqs.~(\ref{new20}), (\ref{new21}) 
%and (\ref{new29}), (\ref{new30})
%imply that in the long wavelength regime $\Sigma'(\vec{q},$Z$)$ is quadratic in $q$ too and we write 
\begin{equation}
\Sigma'(\vec{q},\mbox{Z})\equiv q^2 c^2_{\mbox{{\scriptsize Z}}}  
\label{sigmita}
\end{equation}
where $c_{\mbox{{\scriptsize Z}}}$, to be determined later, has 
the dimension of a velocity.
In case that $\Sigma''(\vec{q},$Z$)$ in Eq.~(\ref{w-tilde}) can be neglected, we 
define the renormalized flexural mode frequency
\begin{equation}
\Omega(\vec{q},\mbox{Z})=\sqrt{\omega^2(\vec{q},\mbox{Z})+q^2c^2_{\mbox{{\scriptsize Z}}}}.
\label{renorm}
\end{equation}
For short wavelengths $q>>q_c$ where 
\begin{equation}
q_c=c_{\mbox{\scriptsize Z}}
\sqrt{\frac{2}{\kappa_0}},
\end{equation}
$\Omega(\vec{q},$Z$)$ reduces to 
$\omega(\vec{q},\mbox{Z})=\sqrt{\kappa_0}q^2$.
At long  wavelengths $q<<q_c$ the dispersion becomes linear
\begin{equation}
\Omega(\vec{q},\mbox{Z})=c_{{\scriptsize\mbox{Z}}} q.
\end{equation}
In the intermediate regime where $q_c/2<q<q_c$ we obtain from Eq.~(\ref{w-tilde})
$\Omega(\vec{q},\mbox{Z})=\sqrt{\kappa_oq_c}q^{3/2}$.
These results are familiar from first order perturbation theory 
in crystalline membranes~\cite{membrane}.
Concepts from membrane theory have been applied to 
the continuum theory of graphene~\cite{fasolino,amorin,Mariani}
and graphene nanoribbons~\cite{pablo2}.
%and graphene nanoribbons~\cite{costa1,pablo2}.

In Sect.~VI~B., we will show that $c_{\mbox{{\scriptsize Z}}}$ depends 
on temperature and on the size of the system.

%%%%%%%%%%%%%%%%%%%%%%%%%%%%%%%%%%%%%%%%%%%%%%%%%%%%%%%%%%
\section{Interaction parameters}

%%%%%%%%%%%%%%%%%%%%%%%%%%%%%%%%%%%%%%%%%%%%%%%%%%%%%%%%%%%
{\setlength{\tabcolsep}{1em}
{\renewcommand{\arraystretch}{2}
\begin{table*}[!t]
%begin{minipage}{.5\linewidth}
%     \caption{}
%     \centering
%       \begin{tabular}{ll}
%           1 & 2
%       \end{tabular}
%   \end{minipage}%y
%   \begin{minipage}{.5\linewidth}
%     \centering
%       \caption{}
%       \begin{tabular}{ll}
%           3 & 4
%       \end{tabular}
%   \end{minipage} 
%
%
\vspace{0.2cm}
\begin{tabular}{ c c c c } \hline \hline
{$B_{\alpha}$} & {$B_1$} & {$B_2$} & {$B_3$} \\
\hline
\hline
{$\varphi_{xxx}^{(3)}(A;B_{\alpha})$} & {$f^{(3)}=124.12$} & {$\frac{-1}{8}(f^{(3)}+9g^{(3)})$} & {$\frac{-1}{8}(f^{(3)}+9g^{(3)})$} \\
\hline
{$\varphi_{xyy}^{(3)}(A;B_{\alpha})$} & {$g^{(3)}=40.43$} & {$\frac{-1}{8}(3f^{(3)}-5g^{(3)})$} & {$\frac{-1}{8}(3f^{(3)}-5g^{(3)})$} \\
\hline
{$\varphi_{xxy}^{(3)}(A;B_{\alpha})$} & {$/$} & {$\frac{\sqrt{3}}{8}(f^{(3)}+g^{(3)})$} & {$\frac{-\sqrt{3}}{8}(f^{(3)}+g^{(3)})$} \\
\hline
{$\varphi_{yyy}^{(3)}(A;B_{\alpha})$} & {$/$} & {$\frac{3\sqrt{3}}{8}(f^{(3)}+g^{(3)})$} & {$\frac{-3\sqrt{3}}{8}(f^{(3)}+g^{(3)})$} \\
\hline
{$\varphi_{zzx}^{(3)}(A;B_{\alpha})$} & {$h^{(3)}=-3.35$} & {$\frac{-1}{2}h^{(3)}$} & {$\frac{-1}{2}h^{(3)}$} \\
\hline
{$\varphi_{zzy}^{(3)}(A;B_{\alpha})$} & {$/$} & {$\frac{\sqrt{3}}{2}h^{(3)}$} & {$\frac{-\sqrt{3}}{2}h^{(3)}$} \\
\hline
\hline
\label{table1}
\end{tabular}
\caption{Third order anharmonic force constants for nearest neighbor atoms.
%Numerical values in units $10^{12}$ erg/cm$^3$.} 
Numerical values are for graphene and are in units of $10^{12}$ erg/cm$^3$.
}
\end{table*}}}

%The evaluation of the phonon energy shift and line-width requires the knowledge
%of the harmonic phonon dispersion relations and of the anharmonic coupling coefficients.
We will use phonon dispersions~\cite{Karl} calculated by means of an harmonic 
force constant model (Fig.~\ref{fig1}).
Such a force constant model has been suggested from in-plane inelastic 
X-ray scattering experiments in graphite~\cite{Mohr}.

In the absence of an empirical model for the anharmonic coupling parameters 
$\Phi^{(3)}_{ijk}(\vec{n_1}\kappa_1;\vec{n_2}\kappa_2;\vec{n_3}\kappa_3)$ and
$\Phi^{(4)}_{ijkl}(\vec{n_1}\kappa_1;\vec{n_2}\kappa_2;\vec{n_3}\kappa_3;\vec{n_4}\kappa_4)$,
we take an heuristic approach. 
%In order to specify the third order anharmonic force constants on the hexagonal lattice,
We assume a central force inter-atomic potential,
where the potential function of interaction 
$\varphi(\vec{n_1}\kappa_1;\vec{n_2}\kappa_2)$ between an atom $\kappa'$
at site $\vec{X}(\vec{n'}\kappa')$ and an atom $\kappa$ at site
$\vec{X}(\vec{n}\kappa)$ depends only on the interatomic distance 
$r=|\vec{X}(\vec{n}\kappa)-\vec{X}'(\vec{n}'\kappa')|$.
One has\cite{maradudin}
\begin{subequations}
\begin{align}
%\begin{eqnarray}
\Phi_{ijk}^{(3)}(\vec{n}\kappa;\vec{n}\kappa;\vec{n'}\kappa')&=
-\varphi_{ijk}^{(3)}(\vec{n}\kappa;\vec{n'}\kappa')\ \ \ 
(\vec{n}\kappa)\neq(\vec{n'}\kappa') \label{eq_30a}\\
\Phi_{ijk}^{(3)}(\vec{n}\kappa;\vec{n}\kappa;\vec{n}\kappa)&=
\sum^{'}_{\vec{n'}\kappa'}\varphi_{ijk}^{(3)}(\vec{n}\kappa;\vec{n'}\kappa')
\label{eq_30b}
%\end{eqnarray}
\end{align}
\end{subequations}
\noindent 
Here $\varphi^{(3)}_{ijk}$ are the third order derivatives of the potential $\varphi(r)$.
Similarly one has
\begin{subequations}
\begin{align}
%\begin{eqnarray}
\Phi_{ijkl}^{(4)}(\vec{n}\kappa;\vec{n}\kappa;\vec{n}\kappa;\vec{n'}\kappa')&=
-\varphi_{ijkl}^{(4)}(\vec{n}\kappa;\vec{n'}\kappa')\ \ \ 
(\vec{n}\kappa)\neq(\vec{n'}\kappa') \label{35a}\\
\Phi_{ijkl}^{(4)}(\vec{n}\kappa;\vec{n}\kappa;\vec{n'}\kappa';\vec{n'}\kappa')&=
\varphi_{ijkl}^{(4)}(\vec{n}\kappa;\vec{n'}\kappa')\ \ \ 
(\vec{n}\kappa)\neq(\vec{n'}\kappa') \label{35b}\\
\Phi_{ijkl}^{(4)}(\vec{n}\kappa;\vec{n}\kappa;\vec{n}\kappa;\vec{n}\kappa)&=
\sum^{'}_{\vec{n'}\kappa'}\varphi_{ijkl}^{(4)}(\vec{n}\kappa;\vec{n'}\kappa')\label{35c}.
%\end{eqnarray}
\end{align}
\end{subequations}
All these quantities are invariant with respect to a permutation of the indices $i,j,...$ .

Using Eqs.~(\ref{eq_30a}), (\ref{eq_30b}) and (\ref{eq18b}) we get 
\begin{widetext}
%\begin{eqnarray}
\begin{subequations}
\begin{align}
\Phi_{ijk}^{(3)}\binom{AAA}{\vec{q}_1\vec{q}_2\vec{q}_3}&=
\frac{1}{\sqrt{M_C^3}}\sum_{\alpha}
\varphi_{ijk}^{(3)}(A;B_{\alpha})
\Delta(\vec{q}_1+\vec{q}_2+\vec{q}_3), \\
\Phi_{ijk}^{(3)}\binom{BAA}{\vec{q}_1\vec{q}_2\vec{q}_3}&=
\frac{-1}{\sqrt{M_C^3}}\sum_{\alpha}
\varphi_{ijk}^{(3)}(A;B_{\alpha})
e^{i\vec{q_1}\cdot\vec{r}(B_{\alpha})}
\Delta(\vec{q}_1+\vec{q}_2+\vec{q}_3),
\label{eq_28}
%
%\end{eqnarray}
\end{align}
\end{subequations}
\end{widetext}
%%\end{widetext}
%\noindent $\Phi_{ijk}^{(3)}\binom{ABA}{\vec{q}_1\vec{q}_2\vec{q}_3}$ and 
%$\Phi_{ijk}^{(3)}\binom{AAB}{\vec{q}_1\vec{q}_2\vec{q}_3}$ are obtained
%from Eq.~(\ref{eq_28}) by permutation of indices.
%
%
\noindent etc. for $\Phi_{ijk}^{(3)}\binom{BBA}{\vec{q}_1\vec{q}_2\vec{q}_3}$,
$\Phi_{ijk}^{(3)}\binom{BBB}{\vec{q}_1\vec{q}_2\vec{q}_3}$.

%%%%%%%%%%%%%%%%%%%%%%%%%%%%%%%%%%%%%%%%%%%%%%%%%%%%%%%%%%%
{\setlength{\tabcolsep}{1em}
{\renewcommand{\arraystretch}{2}
\begin{table*}[!t]
\vspace{0.4cm}
\begin{tabular}{ c c c  } \hline \hline
{$B_{\alpha}$} & {$B_1$} & {$B_2$} \\
\hline
\hline
{$\varphi_{xxxx}^{(4)}(A;B_{\alpha})$} & {$j^{(4)}=f_r/r_a^2=20.56$} & {$\frac{1}{16}(j^{(4)}+9k^{(4)}+18m^{(4)})$}  \\
\hline
{$\varphi_{yyyy}^{(4)}(A;B_{\alpha})$} & {$k^{(4)}=f_i/r_a^2=6.69$} & {$\frac{1}{16}(9j^{(4)}+k^{(4)}+18m^{(4)})$}  \\
\hline
{$\varphi_{xxyy}^{(4)}(A;B_{\alpha})$} & {$m^{(4)}=\sqrt{f_if_r}/r_a^2=7.32$} & {$\frac{1}{16}(3j^{(4)}+3k^{(4)}-2m^{(4)})$}  \\
\hline
{$\varphi_{xxzz}^{(4)}(A;B_{\alpha})$} & {$n^{(4)}=\sqrt{f_rf_o}/r_a^2=6.27$} & {$\frac{1}{4}(n^{(4)}+3p^{(4)})$}  \\
\hline
{$\varphi_{yyzz}^{(4)}(A;B_{\alpha})$} & {$p^{(4)}=\sqrt{f_if_o}/r_a^2=3.58$} & {$\frac{1}{4}(p^{(4)}+3n^{(4)})$}  \\
\hline
{$\varphi_{zzzz}^{(4)}(A;B_{\alpha})$} & {$l^{(4)}=f_o/r_a^2=3.06$} & {$l^{(4)}$}  \\
\hline
\hline
\label{table2}
\end{tabular}
\caption{Fourth order anharmonic force constants for nearest neighbor atoms.
Due to symmetry $\varphi_{...}^{(4)}(A;B_3)=\varphi_{...}^{(4)}(A;B_2)$.
Numerical values are for graphene and are in units of $10^{20}$ erg/cm$^4$.
} 
\vspace{0.4cm}
\end{table*}}}
%%%%%%%%%%%%%%%%%%%%%%%%%%%%%%%%%%%5

Similarly we obtain from Eqs.~(\ref{eq22}) and (\ref{35a})-(\ref{35c}) 
\begin{widetext}
%\begin{eqnarray}
\begin{subequations}
\begin{align}
\Phi_{ijkl}^{(4)}\binom{AAAA}{\vec{q}_1\vec{q}_2\vec{q}_3\vec{q}_4}&=
\frac{1}{M_C^2}\sum_{\alpha}
\varphi_{ijk}^{(4)}(A;B_{\alpha})
\Delta(\vec{q}_1+\vec{q}_2+\vec{q}_3+\vec{q}_4), \label{eq_36a}\\
\Phi_{ijkl}^{(4)}\binom{BAAA}{\vec{q}_1\vec{q}_2\vec{q}_3\vec{q}_4}&=
\frac{-1}{M_C^2}\sum_{\alpha}
\varphi_{ijk}^{(4)}(A;B_{\alpha})
e^{i\vec{q_1}\cdot\vec{r}(B_{\alpha})}
\Delta(\vec{q}_1+\vec{q}_2+\vec{q}_3+\vec{q}_4)\label{eq_36b}.
%\Phi_{ijkl}^{(4)}\binom{BBBB}{\vec{q}_1\vec{q}_2\vec{q}_3\vec{q}_4}&=
%\frac{1}{M_C^2}\sum_{\alpha}
%\varphi_{ijkl}^{(4)}(A;B_{\alpha})
%e^{i(\vec{q}_1+\vec{q}_2+\vec{q}_3+\vec{q}_4)\cdot\vec{r}(B_{\alpha})}
%\Delta(\vec{q}_1+\vec{q}_2+\vec{q}_3+\vec{q}_4).
%
\end{align}
\end{subequations}
\end{widetext}
%\end{eqnarray}
%
%\intertext{
%
%%%%%%%%%%%%%%%%%%%%%%%%%%%%%%%%%%%%%%%%%%%%
Replacement of $\Phi^{(n)}\binom{. . A . .}{ . . \vec{q} . .}$ by 
$\Phi^{(n)}\binom{. . B . .}{ . . \vec{q} . .}$ leads to an additional phase factor 
$e^{i\vec{q}\cdot\vec{r}(\mbox{\scriptsize B}_{\alpha})}$ on the right hand side.

\begin{figure}[h]
\vspace{0.25cm}
\includegraphics[width=0.32\textwidth]{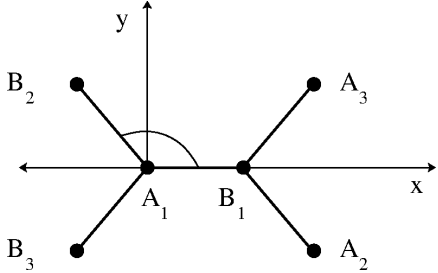}
\caption{Schematic plot of $A_1$ and $B_1$ atoms belonging to the unit cell 
with its corresponding first-nearest neighbors.
}
\label{schem}
\end{figure}

For interactions between nearest neighbor atoms $A$ and $B_1$, at equilibrium positions $(0,0)$
and $a/\sqrt{3},0$ respectively (Fig.~\ref{schem})  we retain 
$\varphi_{xxx}^{(3)}(A;B_1)=f^{(3)}$; 
$\varphi_{xyy}^{(3)}(A;B_1)=g^{(3)}$; 
$\varphi_{xzz}^{(3)}(A;B_1)=h^{(3)}$.
%with $\varphi_{yxy}^{(3)}=\varphi_{yyx}^{(3)}=\varphi_{xyy}^{(3)}$;
%$\varphi_{zzx}^{(3)}=\varphi_{zxz}^{(3)}=\varphi_{xzz}^{(3)}$.
%
The interactions between $A$ and $B_2$ at $(-a/2\sqrt{3},a/2)$ as well as
between $A$ and $B_3$ at $(-a/2\sqrt{3},-a/2)$ are obtained by using the transformation laws
of third rank tensors under rotations by $\pm 120^\circ$, respectively. The results are summarized 
in Table I.
%in Table~\ref{table1}.
%
The numerical values of $f^{(3)}$, $g^{(3)}$ and $h^{(3)}$ are for the 2D crystal graphene and 
are determined from the acoustic mode Gr\"uneisen parameters (see Sect.~V).
%The numerical values of $f^{(3)}$, $g^{(3)}$ and $h^{(3)}$ are estimated by dividing the values of the
%nearest neighbor second order force constants~\cite{Mohr}
%$f_r$, $f_i$, $f_o$ by the C-C bond distance $r_a=1.42 \AA$ (See Table I). 
%
%The interactions between $A$ and $B_2$ at $(-a/2\sqrt{3},a/2)$ as well as
%between $A$ and $B_3$ at $(-a/2\sqrt{3},-a/2)$ are obtained by using the transformation laws
%of third rank tensors under rotations by $\pm 120^\circ$ respectively. The results are summarized 
%in Table I.
%in Table~\ref{table1}.
%$f_r=25.88$, $f_i=8.42$, $f_o=6.18$, all in eV/$\AA^2$, by the C-C bond distance $r_a=1.42 \AA$.
%$f_r$, $f_i$, $f_o$ by the C-C bond distance $r_a=1.42 \AA$ (See Table I). 
%
%Then, $f^{(3)}=29.20 \times 10^{12}$, $g^{(3)}=9.50\times 10^{12}$, 
%$h^{(3)}=-6.98\times 10^{12}$ in units erg/cm$^3$. 
 
The negative value of $h^{(3)}$ is motivated by an argument originally put forward by 
I. M. Lifshitz (Ref.~\onlinecite{Lifshitz}) in formulating the dispersion law for layered 
structures in the long wavelength limit. 
%In thin plates or membranes in-plane stress leads to a frequency 
%increase of bend waves (out-of-plane displacements) which implies
%a negative coefficient of thermal expansion. 
In the present case of a discrete crystal structure the force in the $x$-direction
on atom B$_1$ due to an out-of-plane displacement of atom A$_1$ (Fig.~\ref{schem})
reads $M_C \Ddot{u}_x(B_1)=-\Phi^{(3)}_{zzx}(A_1;A_1;B_1) u^2_z(A_1)$.
Since this force has to be attractive, $-\Phi^{(3)}_{zzx}(A_1;A_1;B_1)=\varphi^{(3)}_{zzx}
\equiv h^{(3)}<0$.
In Sect.~V we will see that the negative value of $h^{(3)}$ is related to a negative value of the
out-of-plane Gr\"uneisen coefficient $\gamma(\vec{q}, $Z$)$ and hence, as has been emphasized
by Mounet and Marzari (Ref.~\onlinecite{mounet}), favors a negative contribution to the thermal expansion. 
%

%The force constants are related to the coupling parameters 
%$\Phi_{ijk}^{(3)}(\vec{n}_1\kappa_1;\vec{n}_2\kappa_2;\vec{n}_3\kappa_3)$ by~\cite{note-1}
%\begin{subequations}
%\begin{align}
%%\begin{eqnarray}
%\Phi_{ijk}^{(3)}(\vec{n}\kappa;\vec{n}\kappa;\vec{n'}\kappa')&=
%-\varphi_{ijk}^{(3)}(\vec{n}\kappa;\vec{n'}\kappa')\ \ \ 
%(\vec{n}\kappa)\neq(\vec{n'}\kappa') \label{eq_30a}\\
%\Phi_{ijk}^{(3)}(\vec{n}\kappa;\vec{n}\kappa;\vec{n}\kappa)&=
%\sum^{'}_{\vec{n'}\kappa'}\varphi^{(3)}(\vec{n}\kappa;\vec{n'}\kappa')
%\label{eq_30b}
%\end{eqnarray}
%\end{align}
%
%\end{subequations}
%\noindent
%where the prime at the summation sign indicates that terms with 
%$\vec{n}\kappa$=$\vec{n'}\kappa'$ are excluded.
%
%Eq.~(\ref{eq_28}) by replacing $(\vec{q}_2+\vec{q}_3)\cdot\vec{r}(B_{\alpha})$
%by $(\vec{q}_1+\vec{q}_3)\cdot\vec{r}(B_{\alpha})$ and $(\vec{q}_1+\vec{q}_2)\cdot\vec{r}(B_{\alpha})$
%respectively.
%In a similar way we proceed with the fourth order 
%coupling parameters $\Phi_{ijkl}^{(4)}$. We first consider the force constants
%defined in Table II.

Numerical values of the fourth order force constants $j^{(4)},k^{(4)}, ..., l^{(4)}$
are estimated as shown in Table II
%by dividing the third order force constants by the C-C bond distance $r_a=1.42~\AA$. 
by dividing the second order force constant of graphene~\cite{Mohr} by
the square of the C-C bond distance.
In analogy with the reasoning about $\varphi_{zzx}^{(3)}$ we consider the force 
$-\Phi_{zzxx}^{(4)}(A_1 ; A_1 ; B_1 ; B_1)\times u_z^2(A_1)u_x(B_1)<0$ where 
$u_x(B_1)>0$. By means of Eq.~(\ref{35b}) follows $\varphi_{zzxx}^{(4)}(A_1 ; A_1 ; B_1 ; B_1)=n^{(4)}>0$.
The same holds for $p^{(4)}$.

%We obtain
%\begin{eqnarray}
%j^{(4)}&=&f_r/r_a^2=20.56   \nonumber\\
%k^{(4)}&=&f_i/r_a^2=6.69  \nonumber\\
%m^{(4)}&=&\sqrt{f_if_r}/r_a^2=7.32\nonumber\\
%n^{(4)}&=&\sqrt{f_rf_o}/r_a^2=6.27\nonumber\\
%p^{(4)}&=&\sqrt{f_if_o}/r_a^2=3.58\nonumber\\
%l^{(4)}&=&f_o/r_a^2=3.06\nonumber
%\end{eqnarray} 
%\noindent in units of $10^{20}$ erg/cm$^4$. 
%{\bf NOTE - We need to mention here relations like $\varphi_{xxyy}=\varphi_{yyxx}$ 
%as we did for the 3rd order case before -}.

%In analogy with Eqs.~(\ref{30a}), (\ref{30b}) we have
%\begin{subequations}
%\begin{align}
%%\begin{eqnarray}
%\Phi_{ijkl}^{(4)}(\vec{n}\kappa;\vec{n}\kappa;\vec{n}\kappa;\vec{n'}\kappa')&=
%-\varphi_{ijkl}^{(4)}(\vec{n}\kappa;\vec{n'}\kappa')\ \ \ 
%(\vec{n}\kappa)\neq(\vec{n'}\kappa') \label{35a}\\
%\Phi_{ijkl}^{(4)}(\vec{n}\kappa;\vec{n}\kappa;\vec{n'}\kappa';\vec{n'}\kappa')&=
%\varphi_{ijkl}^{(4)}(\vec{n}\kappa;\vec{n'}\kappa')\ \ \ 
%(\vec{n}\kappa)\neq(\vec{n'}\kappa') \label{35b}\\
%\Phi_{ijkl}^{(4)}(\vec{n}\kappa;\vec{n}\kappa;\vec{n}\kappa;\vec{n}\kappa)&=
%\sum^{'}_{\vec{n'}\kappa'}\varphi_{ijkl}^{(4)}(\vec{n}\kappa;\vec{n'}\kappa')\label{35c}.
%%\end{eqnarray}
%\end{align}
%\end{subequations}
%\noindent Using then Eq.~(\ref{eq22}), we get

%%%%%%%%%%%%%%%%%%%%%%%%%%%%%%%%%%%%%%%%%%%%%%%%%%%%%%%%%%%
\vspace{0.4cm}
%\section{Analytical Results}
\section{Thermal expansion}

Here we present analytical calculations of the generalized Gr\"uneisen coefficients
and the thermal expansion.
Although we explicitly discuss graphene, the analytical results are applicable
to other layered 2D crystals with D$_{3h}$ symmetry by replacing $M_C$ by 
2$\mu$, where $\mu$ is the reduced mass, and by adapting the corresponding 
numerical values for the material constants.

%Next we investigate lineshifts and damping
%of the in-plane modes (subsect. B) and of the flexural mode (subsect. C).
We will need the anharmonic coupling coefficients for acoustic phonons in the long 
wavelength regime. We use $e_i^A(\vec{k},\lambda)=e_i^B(\vec{k},\lambda)$ as well as 
$e_i^A(\vec{k}-\vec{q},\lambda)\sim e_i^A(\vec{k},\lambda)$ for $\vec{q}\rightarrow 0$.
From Eqs.~(\ref{phi3}), (\ref{eq18b}) and (\ref{eq_30a}), (\ref{eq_30b}) we obtain 

\begin{widetext}
\begin{align}
%\begin{equation}
\Phi^{(3)}\binom{\lambda\ \ \lambda_1\  \lambda_2}{{-\vec{q}}\ \vec{k}\ \vec{q}{-\vec{k}}}=&
\frac{i\sqrt{2}}
{4\sqrt{M_C^3\omega(\vec{q},\lambda)\omega(\vec{k},\lambda_1)\omega(\vec{q}-\vec{k},\lambda_2)}} 
\nonumber \\ 
&\times \sum_{\alpha}\sum_{ijk}
e_i^{A*}(\vec{q},\lambda)e_j^{A}(\vec{k},\lambda_1)e_k^{A*}(\vec{k},\lambda_2)
\ \varphi_{ijk}^{(3)}(A;B_{\alpha})
(\vec{q}\cdot\vec{r}(B_{\alpha})) 
(\vec{k}\cdot\vec{r}(B_{\alpha})) 
((\vec{k}-\vec{q})\cdot \vec{r}(B_{\alpha})).
%sin^2
%\left(\frac{\vec{k}\cdot\vec{r}(B_{\alpha})}{2}\right).
\label{eq49}
\end{align}
%\intertext{
We have performed a series expansion in $\vec{q}$ and in $\vec{k}$ of the exponentials.
%and only retained the linear term. 
Similarly we proceed with Eqs.~(\ref{phi4}), (\ref{eq22}) and (\ref{35a})-(\ref{35c}) and obtain
%}
%
\begin{align}
\Phi^{(4)}\binom{\ \lambda\ \lambda'\ \lambda'\ \lambda}{{-\vec{q}}\ \vec{k}\ {-\vec{k}}\ \vec{q}}=&
\frac{1}{4M_C^2\omega(\vec{q},\lambda)\omega(\vec{k},\lambda')} 
\nonumber \\ 
&\times \sum_{\alpha}\sum_{ijkl}
e_i^{A*}(\vec{q},\lambda)e_j^{A}(\vec{k},\lambda')e_k^{A*}(\vec{k},\lambda')e_l^{A}(\vec{q},\lambda)
\ \varphi_{ijkl}^{(4)}(A;B_{\alpha})
(\vec{q}\cdot\vec{r}(B_{\alpha}))^2 %sin^2
(\vec{k}\cdot\vec{r}(B_{\alpha}))^2. 
\label{eq-45-new-Ok}
\end{align}
%\end{equation}
\end{widetext}
%These expressions agree with Eqs.~(\ref{new20}) and (\ref{new21}).

%Here the lowest order term is quadratic in $\vec{q}$.

%\subsection{Gr\"uneisen coefficients and thermal expansion}
 
In Sect.~III we have seen that the thermal tension $\beta_T$ 
and equivalently the thermal expansion $\alpha_T$ depend linearly on the third order 
anharmonicities through the generalized Gr\"uneisen coefficient $\gamma(\vec{q},\lambda)$, 
Eq.~(\ref{eq_gamma}).
From Eq.(\ref{eq50}) we obtain in the long wavelength regime
\begin{widetext}
\begin{align}
\frac{\partial{\omega(\vec{q},\lambda)}}{\partial{\epsilon_{ii}}}=&
-\frac{1}{4\omega(\vec{q},\lambda)M_C}
\sum_{\alpha}\sum_{kl}
\varphi^{(3)}_{kli}(A;B_{\alpha})r_i(B_{\alpha})
e^{A*}_k(\vec{q},\lambda)e^{A}_l(\vec{q},\lambda)
(\vec{q}\cdot\vec{r}(B_{\alpha}))^2.
%
%\times \nonumber \\ 
%&\bigg[e^{A*}_k(\vec{q},\lambda)e^{A}_l(\vec{q},\lambda)+
%e^{B*}_k(\vec{q},\lambda)e^{B}_l(\vec{q},\lambda)-
%e^{A*}_k(\vec{q},\lambda)e^{B}_l(\vec{q},\lambda)e^{i\vec{q}.\vec{r}(B_{\alpha})}-
%%e^{B*}_k(\vec{q},\lambda)e^{A}_l(\vec{q},\lambda)e^{-i\vec{q}.\vec{r}(B_{\alpha})}\bigg]
\label{eq_grun}
\end{align}
\end{widetext}
\noindent
Here $\vec{r}(B_{\alpha})$, $\alpha=1,2,3$ runs over the three nearest neighbor atoms 
$B_{\alpha}$ of A$_1$ (see Fig.~(\ref{schem})).

%Since at low temperature long wave-length acoustic phonons play a dominant role, 
%we have studied $\gamma(\vec{q},\lambda)$ for the polarizations ZA (out-of-plane acoustic),
%TA (in-plane transversal acoustic) and LA (longitudinal in-plane acoustic) 
%in/the long wave-legnht regime. 
%
In order to obtain quantitative results for $\beta_T$ one has to evaluate the $\vec{q}$-sum in 
Eq.~(\ref{eq_ten}). 
We have used analytical methods which allow us to investigate the limit cases
of high and low $T$ and to discuss singularities in $\vec{q}$-space. 
%Secondly we have performed the summation over the Brillouin zone by numerical integration techniques,
%thereby obtaining results in a broad $T$ range.

We start from Eq.~(\ref{eq_grun}) with the out-of-plane mode $\lambda$=Z.
The polarization vectors in the long wavelength regime are 
$e^{\kappa}_k(\vec{0},$Z$)=\sqrt{1/2}\delta_{kz}$
for $\kappa=$A, B. 
%Expansion of Eq.~(\ref{eq_grun}) with respect 
%to small values of $\vec{q}=(q_x,q_y)$ gives
%
Carrying out the summation over neighbor atoms using Table I, we obtain
% s and consider Eq.~(\ref{eq50}) with $\vec{q}=(q_x,q_y)$.
%The relevant polarization vector components at $\vec{q}=\vec{0}$ are 
%$e^{\kappa}_k(\vec{0},$ZA$)=\sqrt{1/2}\delta_{kz}$ for $\kappa=$A, B. 
%Expanding Eq.~(\ref{eq_grun}) with respect to small values of $\vec{q}$
%and using the anharmonic force constants of Table I, we obtain
\begin{equation}
\frac{\partial \omega(\vec{q},\mbox{Z})}{\partial \epsilon_{xx}}=
\frac{-a^3h^{(3)}}{64\sqrt{3} M_C \omega(\vec{q},\mbox{Z})}
\bigg[3q^2_x+q^2_y\bigg].
\label{eq_ZA_gru}
\end{equation}

%We take the acoustic polarization vectors in Eq.(\ref{}) at $\vec{q}=\vec{0}$.
%The relevant components are $e^{\kappa}_k(\vec{0},$LA$)=\sqrt{1/2}\delta_{kx}$,
%$e^{\kappa}_k(\vec{0},$TA$)=\sqrt{1/2}\delta_{ky}$,
%$e^{\kappa}_k(\vec{0},$ZA$)=\sqrt{1/2}\delta_{kz}$, for $\kappa=$A, B.
%Restricting ourselves to central force nearest neighbor anharmonic force constants (Table I),
%we obtain from Eq.(\ref{}) in the long wave-length regime 
%\begin{equation}
%\frac{\partial \omega(\vec{q},ZA)}{\partial \epsilon_{ii}}=
%-\frac{a^3h}{32 \omega(\vec{q},ZA)}
%\bigg[\sqrt{3}p^2_x+\frac{p^2_y}{\sqrt{3}}\bigg]
%\end{equation}

Since $h^{(3)}$, the anharmonic force constant $\varphi_{zzx}$ is negative, 
the frequency $\omega(\vec{q},$Z$)$ increases with in-plane strain. 
The corresponding expression for $\partial \omega(\vec{q},$Z$)/\partial\epsilon_{yy}$
is obtained by an interchange of  $q^2_x \leftrightarrow q^2_y$ in Eq.~(\ref{eq_ZA_gru}).
Addition of both contributions and use of Eq.~(\ref{eq_gamma}) leads to
%%%%%%%%%%
%The corresponding expression for $\partial \omega(\vec{q},$ZA$)/\partial \epsilon_{yy}$
%is obtained by interchange of $p^2_x \leftrightarrow p^2_y$ on the 
%rhs of Eq.~(\ref{eq_ZA_gru}). Addition of both terms and use of Eq.~(\ref{eq_gamma})
%leads in the long wave-length regime to  
\begin{equation}
\gamma(\vec{q},\mbox{Z})=\frac{a^3h^{(3)}q^2}{16\sqrt{3}M_C\omega^2(\vec{q},\mbox{Z})},
\label{eq-48-new1}
\end{equation}
\noindent 
where $q^2=q^2_x+q^2_y$. 
Since $\omega^2(\vec{q},$Z$)=\kappa_0 q^4$, 
the wave vector average of $\gamma(\vec{q},$Z$)$ diverges logarithmically with $q\rightarrow 0$.
We then consider a finite 2D crystal with linear dimensions $l$.
The corresponding wave vector $q_l=2\pi/l$ entails a lowest non-zero frequency 
$\omega_l($Z$)=\sqrt{\kappa_0}q^2_l$.
Transforming the $\vec{q}$-sum in a frequency integral, the wave vector average of 
Eq.~(\ref{eq-48-new1}) reads
\begin{equation}
\overline{\gamma(\mbox{Z})}=\frac{\tilde{\gamma}(\mbox{Z})v_{2D}}
{4\pi\kappa_0}ln\Bigg(\frac{\omega_s(\mbox{Z})}{\omega_l(\mbox{Z})}\Bigg),
\label{eq49n}
\end{equation}
where $\tilde{\gamma}(\mbox{Z})=a^3h^{(3)}/(16\sqrt{3} M_C)$.
We take for 
$\overline{\gamma(\mbox{Z})}$ the empirical numerical value $\gamma_{\mbox{\scriptsize ZA}}$=-1.5 from 
Ref.~\onlinecite{nikas}. As upper frequency limit we choose $\omega_s($Z$)$=94.3 THz, which
corresponds to $\tilde{\nu}$=$500$ cm$^{-1}$ for the ZA branch in Fig.~1. 
With $l=10^4 a$ the lower frequency limit is $\omega_l($Z$)$=4.25 MHz. Solving
Eq.~(\ref{eq49n}) with respect to $h^{(3)}$, we obtain the value quoted in Table I. 
%

%where $\kappa_0$ is proportional to the  bending rigidity, 
%$\gamma(\vec{q},$Z$)$ 
%diverges as $q^{-2}$ in the limit 
%$q\rightarrow 0$. We write
%\begin{equation}
%\gamma(\vec{q},\mbox{Z})=\frac{\tilde{\gamma}(\mbox{Z})}{q^{2}\kappa_0},
%\label{eq-gamatilde}
%\end{equation}
%is proportional to $p^2=p^2_x+p^2_y$ too, we see that $\gamma(\vec{q},ZA)$ diverges as $p^{-2}$ for 
%$\vec{q}\rightarrow0$.
%where $\tilde{\gamma}($Z$)=a^3h^{(3)}/(16\sqrt{3} M_C)=18.8\times10^{10}$ cm$^{2}$ s$^{-2}$.
%
%With the force constant model of Ref.~[\onlinecite{Karl}] we have
%$\kappa_0=42.48\times10^{-6}$ cm$^4$ s$^{-2}$ and hence 
%$\tilde{\gamma}($Z$)=-44\times10^{14} $cm$^{-2}$.

Considering the in-plane displacement modes T and L we use in the long wavelength 
regime the polarization vectors $e^{\kappa}_k(\vec{0},$T$)=\sqrt{1/2}\delta_{ky}$ and
$e^{\kappa}_k(\vec{0},$L$)=\sqrt{1/2}\delta_{kx}$, for $\kappa=$A, B.
%Proceeding as before we obtain th 
%
%
%Turning to the in-plane displacement modes TA and LA, we consider Eq.~(\ref{eq50}) with 
%$\vec{q}=(q_x,0)$. In the long wave-length regime the relevant polarization vectors then are 
%$e^{\kappa}_k(\vec{0},$TA$)=\sqrt{1/2}\delta_{ky}$ and  
%$e^{\kappa}_k(\vec{0},$LA$)=\sqrt{1/2}\delta_{kx}$, for $\kappa=$A, B. 
Proceeding as before we obtain %the Gr\"uneisen component 
\begin{eqnarray}
\gamma(\vec{q},\mbox{T})=\frac{a^3}{64\sqrt{3}M_C\omega^2(\vec{q},\mbox{T})}
\ \ \ \ \ \ \ \   \nonumber \\
\ \ \ \ \ \ \ \ \ \ 
\times  \bigg[(f^{(3)}+3g^{(3)})q^2_x+(3f^{(3)}+g^{(3)})q^2_y\bigg].
\label{eq_TA_gru}
\end{eqnarray}
%\noindent and
%\begin{equation}
%\gamma(\vec{q},LA)=\frac{a^3}{64\sqrt{3}M_C\omega^2(\vec{q},LA)}\bigg[(3f+g)p^2_x\bigg]
%\end{equation}

%On the other hand we obtain for the TA and LA cases by a similar calculation
%\begin{eqnarray}
%\gamma(\vec{q},TA)=\frac{a^3}{32\sqrt{3}\omega^2(\vec{q},TA)}((f+3g)p^2_x+(3f+g)p^2_y)\\
%\gamma(\vec{q},LA)=\frac{a^3}{32\sqrt{3}\omega^2(\vec{p},LA)}((3f+g)p^2_x+(f+3g)p^2_y)
%\end{eqnarray}
%

The expression for $\gamma(\vec{q},$L$)$ is obtained from Eq.~(\ref{eq_TA_gru})
by interchange of $q^2_x \leftrightarrow q^2_y$ and 
by replacing $\omega(\vec{q},$T$)$ by $\omega(\vec{q},$L$)$.
The long wavelength acoustic phonons in a 2D crystal have frequencies 
$\omega(\vec{q},$T$)=c_{\mbox{\scriptsize T}}q$ and 
$\omega(\vec{q},$L$)=c_{\mbox{\scriptsize L}}q$,
where $c_{\mbox{\scriptsize T}}$ and $c_{\mbox{\scriptsize L}}$ 
are the transversal and longitudinal sound velocities.
Using the Debye interpolation scheme we define an average sound velocity $\hat{c}$ by
\begin{equation}
\frac{2}{\hat{c}^2}=\frac{1}{c^2_{\mbox{\scriptsize L}}}+\frac{1}{c^2_{\mbox{\scriptsize T}}},
\end{equation}
\noindent 
and replace both $\omega(\vec{q}, $T$)$ and $\omega(\vec{q}, $L$)$ by $\hat{c}q$.
With the model of Ref.~\onlinecite{Karl} we have 
%$\hat{c}=17.2\times10^{5}$ cm s$^{-1}$.
$c_{\mbox{{\scriptsize L}}}=23.1\times10^{5}$ cm s$^{-1}$,
$c_{\mbox{{\scriptsize T}}}=14.3\times10^{5}$ cm s$^{-1}$ and hence
$\hat{c}=17.2\times10^{5}$ cm s$^{-1}$.

Adding the long wavelength expressions for $\gamma(\vec{q}, $T$)$ and
$\gamma(\vec{q},$L$)$ we obtain the in-plane Gr\"uneisen constant 
$\gamma(\perp)$ which is independent of the wave vector:
\begin{equation}
\gamma(\perp)=\frac{a^3(f^{(3)}+g^{(3)})}{16\sqrt{3}M_C\hat{c}^2}.
\label{eq-gamaperp}
\end{equation} 
%
%
%\begin{equation}
%\gamma(\vec{q},\perp)=\gamma(\vec{q},\mbox{T})+\gamma(\vec{q},\mbox{L})
%\end{equation} 
%\noindent with the result
%\begin{equation}
%\gamma(\vec{q},\perp)=\frac{\tilde{\gamma}(\perp)}{\hat{c}^2},
%a^3}{16\sqrt{3}M_C \hat{c}^2}[f^{(3)}+g^{(3)}]
%\gamma(\vec{q},\perp)=\frac{a^3}{16\sqrt{3}M_C \hat{c}^2}[f^{(3)}+g^{(3)}]
%\label{eq-gamaperp}
%\end{equation}
%where $\tilde{\gamma}(\perp)=a^3(f^{(3)}+g^{(3)})/16\sqrt{3}M_C$.
Here and in the following $\perp$ has the meaning of in-plane, i. e. 
normal to the highest symmetry axis. 
We identify $\gamma(\perp)$, Eq.~(\ref{eq-gamaperp}) with the average 
$(\gamma_{\mbox{\scriptsize LA}}+\gamma_{\mbox{\scriptsize TA}})/2$=1.5, 
taken from Ref.~\onlinecite{nikas}.
Comparison with 
Eq.~(\ref{eq-gamaperp}) yields ($f^{(3)}+g^{(3)}$)=164.55$\times$10$^{12}$ erg/cm$^3$.
Assuming that the ratio $f^{(3)}/g^{(3)}$ is equal to 25.88/8.42=3.07 as 
inferred from the second order stretching and shearing force constants~\cite{Mohr},
we obtain the values quoted in Table I.

Conversely we have used the present values of $f^{(3)}$ and $g^{(3)}$ to calculate the E$_{2g}$ in-plane
optical mode Gr\"uneisen parameter (biaxial stress) and obtain  $\gamma($E$_{2g})$=1.51. 
We recall that the in-plane E$_{2g}$ Gr\"uneisen parameter inferred from Raman scattering spectra on graphite
under hydrostatic pressure~\cite{hanfland} leads to $\gamma($E$_{2g})$=1.59, as quoted in Ref.~\onlinecite{Reich}.
Raman spectroscopy on uniaxially strained graphene~\cite{ferrari} leads to  $\gamma($E$_{2g})$=1.99.

\begin{figure}[t]
\vspace{0.25cm}
\includegraphics[width=0.48\textwidth]{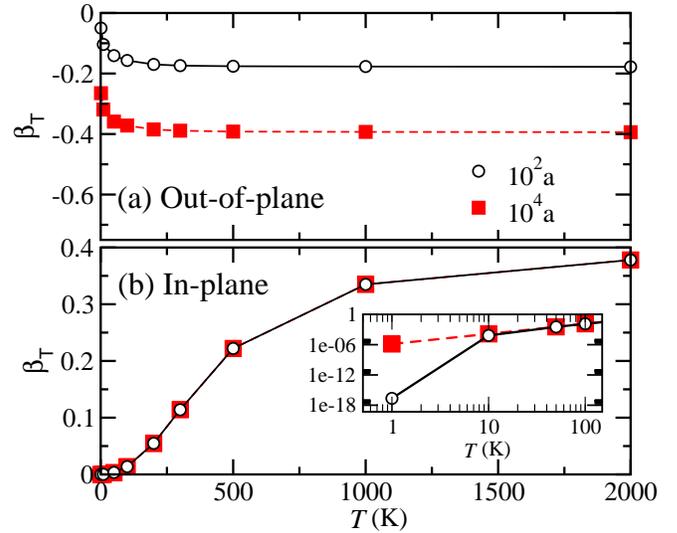}
\caption{(Color online) 
Thermal tension coefficient $\beta_T$ for system sizes $l$=10$^2a$ (circles) and 10$^4a$ (squares). 
The (a) out-of-plane $\beta_T(\mbox{Z})$ and (b) in-plane $\beta_T(\perp)$
components are given in units of dyn cm$^{-1}$ K$^{-1}$. 
Notice in the inset the size dependence at low $T$ of $\beta_T(\perp)$.  
}\label{plot-Karl}
\end{figure}

We turn now to the thermal tension defined by Eq.~(\ref{eq_ten}). 
Given the different analytic behavior of $\gamma(\vec{q},$Z$)$
and $\gamma(\perp)$ we will consider separately the out-of-plane
contributions to $\beta_T$ by writing 
\begin{equation}
\beta_{T}=\beta_T(\mbox{Z})+\beta_T(\perp)
\label{eq-sumperp}
\end{equation}
\noindent with $\beta_T(\perp)=\beta_T(\mbox{T})+\beta_T(\mbox{L})$.
We start %from Eq.~(\ref{eq_ten}) 
with $\lambda$=Z. 
Transforming the 
$\vec{q}$-sum into a frequency integral %by means of %the dispersion 
%$\omega(\vec{q},$Z$)=\sqrt{\kappa_0}q^2$ and using Eqs.~(\ref{eq_ten1}) and (\ref{eq-gamatilde}),
we get 
\begin{equation}
\beta_T(\mbox{Z})=\frac{\tilde{\gamma}(\mbox{Z})\hbar^2}
{4\pi k_B T^2 \kappa_0}
\int^{\omega_{s}(\mbox{{\scriptsize Z}})}_{\omega_{l}(\mbox{{\scriptsize Z}})}
d\omega \frac{\omega e^{\hbar\omega/k_B T}}
{(e^{\hbar\omega/k_B T}-1)^2}.
\label{eq_inte}
\end{equation}
Here and in the following we use the third order anharmonic force constants from Table I.  
Then $\tilde{\gamma}$(Z)=-9.03$\times$10$^{10}$cm$^2$s$^{-2}$
which implies that $\beta_T(\mbox{Z})$ is negative. 
The size dependence is accounted for by $\omega_l$(Z).
%
%Here we take as upper limit $\omega_{s}(\mbox{Z})=103.7$ THz, which corresponds to 
%$\tilde\nu=550$ cm$^{-1}$.
%
%In the case of an infinite crystal $\omega_{l}(\mbox{Z})=0$ and the integral diverges logarithmically.
%
%$\hbar\omega=k_B T$ for room temperature.
%The integral diverges logarithmically for $\omega$=0. 
%and hence the concept of thermal 
%tension (equivalently thermal expansion) becomes meaningless for a 2D crystal with infinite extension.
%We then consider a finite 2D crystal with linear dimensions $l$.
%The corresponding wave vector $q_l=2\pi/l$ entails a lowest non-zero frequency 
%$\omega_l($Z$)=\sqrt{\kappa_0}q^2_l$. Replacing the lower 
%limit of the integral in Eq.~(\ref{eq_inte}) by $\omega_l($Z$)$ we obtain
%
In the high $T$ limit (classical case), $k_B T> \hbar \omega$, this expression reduces to  
\begin{equation}
\beta_T(\mbox{Z})=\frac{\tilde{\gamma}(\mbox{Z})}{\kappa_0}
\frac{k_B}{4\pi}ln\bigg(\frac{\omega_s(\mbox{Z})}
{\omega_l(\mbox{Z})}\bigg),
\label{eq61-lowZA}
\end{equation}
\noindent and in the low $T$ limit (quantum case), $k_BT  <\hbar\omega_l(\mbox{Z}) <\hbar\omega_s(\mbox{Z})$, 
\begin{equation}
\beta_T(\mbox{Z})=\frac{\tilde{\gamma}(\mbox{Z})}{\kappa_0}
\frac{\hbar\omega_l(\mbox{Z})}
{4\pi T}e^{-\hbar\omega_l(\mbox{{\scriptsize Z}})/k_B T}.
\label{eq61-high}
\end{equation}
\noindent 
%Since $\tilde{\gamma}($Z$)<$0, the contribution of the flexural mode to the 
%thermal tension is negative~\cite{Lifshitz,mounet}.
While $\beta_T($Z$)$ is constant at high $T$, it vanishes with $T\rightarrow$ 0,
in accordance with Nernst's theorem~\cite{Landau1}.
In Fig.~\ref{plot-Karl}(a) we plotted $\beta_T$(Z), 
Eq.~(\ref{eq_inte}), as function of temperature for two different
crystal sizes. 
%Notice the agreement with the limit cases of high and low $T$, 
%Eqs.~(\ref{eq61-lowZA}) and (\ref{eq61-high}) respectively.

%#####################
%If we take into account the renormalization of the flexural mode by anharmonic effects, Eq.~(\ref{renorm}),
%we obtain at high $T$ where $k_BT>\hbar\omega_s($Z$)>\hbar\omega_l($Z$)$ for a finite crystal
%\begin{equation}
%\beta_T(\mbox{Z})=\frac{\tilde{\gamma}(\mbox{Z})k_B}{4\pi\kappa_0}
%ln\Bigg[\frac{c^2_{\mbox{{\scriptsize Z}}}+\sqrt{\kappa_0}\omega_s(\mbox{Z})}{c^2_{\mbox{{\scriptsize Z}}}+\sqrt{\kappa_0}\omega_l(\mbox{Z})}\Bigg].
%\end{equation}
%\noindent 
%#####################

%The result remains finite for an infinite crystal, $\omega_l(\mbox{Z})=0$, as long
%as $\xi\neq0$. 
%
%
%At low $T$ where $k_BT<\hbar\omega_l($Z$)<\hbar\omega_s($Z$)$ we have 
%\begin{equation}
%\beta_T(\mbox{Z})=\frac{\tilde{\gamma}(\mbox{Z})\hbar}{2\pi c^4_{\mbox{{\scriptsize Z}}} T}
%\omega^3_s(\mbox{Z})e^{-\hbar\omega_l(\mbox{{\scriptsize Z}})/k_B T,}
%\end{equation}
%for a finite crystal where $\omega_l($Z$)=c^2_{\mbox{{\scriptsize Z}}}q_l$. In case of an infinite crystal with 
%$k_BT<\hbar\omega_s($Z$)$,
%\begin{equation}
%\beta_T(\mbox{Z})=\frac{3\tilde{\gamma}(\mbox{Z})k^3_BT^2\zeta(3)}{\pi\hbar^2 c^4_{\mbox{{\scriptsize Z}}}},
%\end{equation}
%where $\zeta(3)$ is the Riemann zeta function. 

In studying the contribution $\beta_T(\perp)$ due to the in-plane modes
we make use of the linear dispersion and replace $\omega(\vec{q},$T$)$ and $\omega(\vec{q},$L$)$
by $\hat{c}q$.  The upper frequency limit is $\omega_s(\perp)=(4\pi\hat{c}^2/v_{2D})^{1/2}$.
For the sake of consistency we take a crystal with finite size where $\omega_l(\perp)=\hat{c}\ 2\pi/l$.
We then obtain 
\begin{equation}
\beta_T(\perp)=\frac{\gamma(\perp)\hbar^2}{2\pi\hat{c}^2k_BT^2}
\int_{\omega_l(\perp)}^{\omega_s(\perp)}d\omega
\frac{\omega^3e^{\hbar\omega/k_BT}}
{\big(e^{\hbar\omega/k_BT}-1\big)^2},
\label{eq-64-1}
\end{equation}
%we remind that the Gr\"uneisen coefficient $\gamma(\perp)$ is a constant in the long wavelength regime.
%Making use of the linear dispersion $\omega(\vec{q})=\hat{c}q$ and replacing $\omega(\vec{q},$T$)$
%and $\omega(\vec{q},$L$)$ by $\hat{c}q$, we obtain the upper frequency limit
%$\omega_s(\perp)=(4\pi\hat{c}^2/v_{2D})^{1/2}$. For a crystal with finite size the lower frequency reads
%$\omega_l(\perp)=\hat{c}\ 2\pi/l$, where $l$ is the characteristic length. 
%Then $\beta_T(\perp)$ is given by the frequency integral
%\begin{equation}
%\beta_T(\perp)=\frac{\gamma(\perp)\hbar^2}{2\pi\hat{c}^2k_BT^2}
%\int_{\omega_l(\perp)}^{\omega_s(\perp)}d\omega
%\frac{\omega^3e^{\hbar\omega/k_BT}}
%{\big(e^{\hbar\omega/k_BT}-1\big)^2}
%\end{equation} 
\noindent 
%where $\tilde{\gamma}(\perp)$=4.44$\times$10$^{12}$cm$^2$s$^{-2}$.
where $\gamma(\perp)$=1.5, which implies that $\beta_T(\perp)$ is positive.
In contradistinction with Eq.~(\ref{eq_inte}) the integral exists also for 
$\omega_l(\perp)=0$, i. e. for $l\rightarrow\infty$ . 
In the high $T$ regime where $k_BT>\hbar\omega_s(\perp)>\hbar\omega_l(\perp)$ we obtain 
\begin{equation}
\beta_T(\perp)=\frac{\gamma(\perp)k_B}{v_{2D}}\bigg(1-\pi \frac{v_{2D}}{l^2}\bigg)
\label{eq64-lowperp}
\end{equation}
\noindent 
%Since $l^2$ is a measure of the area of the crystal $(v_{2D}/l^2)\approx N^{-1}$,
%where $N$ is the number of unit cells. 
In the low $T$ regime 
$k_BT<\hbar\omega_l(\perp)<\hbar\omega_s(\perp)$ we obtain  
%$\hbar\omega_l(\perp)<k_BT<\hbar\omega_s(\perp)$
\begin{equation}
\beta_T(\perp)=\frac{\gamma(\perp)\hbar}{2\pi \hat{c}^2 T}\omega^3_l(\perp)
e^{-\hbar\omega_l(\perp)/k_BT}.
\label{eq64-highperp}
\end{equation}
%In case of a crystal with infinite size and 
%$k_BT<\hbar\omega_s(\perp)$, we get at low $T$
%\begin{equation}
%\beta_T(\perp)=\frac{3\tilde{\gamma}(\perp)k^3_BT^2\zeta(3)}{\hat{c}^4\hbar^2}
%\end{equation}

\begin{figure}[t]
\vspace{0.25cm}
\includegraphics[width=0.48\textwidth]{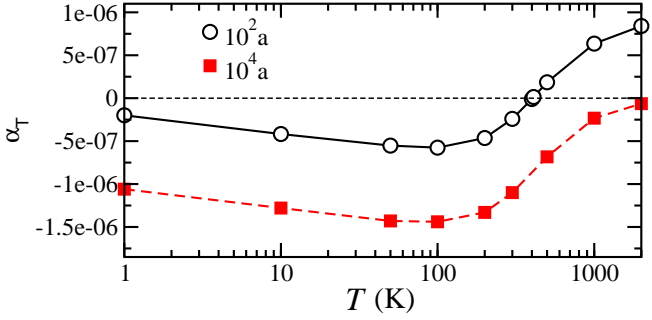}
\caption{(Color online) 
Thermal expansion coefficient $\alpha_T$ for
$l$=10$^2a$ and 10$^4a$ sample sizes.
Units K$^{-1}$.
%Units 10$^{-6}$ K$^{-1}$.
}\label{plot-Karl-2}
\end{figure}

In Fig.~\ref{plot-Karl}~(b) we plotted $\beta_T(\perp)$, 
Eq.~(\ref{eq-64-1}), as function of $T$
for two different crystal sizes. Notice here again the agreement with the limit cases 
Eqs.~(\ref{eq64-lowperp}) and (\ref{eq64-highperp}) 
of high and low $T$, respectively. In particular it follows from 
Eq.~(\ref{eq64-lowperp}) that $\beta_T(\perp)$ is quasi size independent at large $T$.

%Since $|\beta_T$(Z)$|>\beta_T(\perp)$, the total thermal tension 
%(and hence by Eqs.~(\ref{eq-sumperp}) and (\ref{therm-ex-eq}) the thermal expansion) 
%is negative, even at high $T$.
%Plots of the thermal expansion are given in Fig.~\ref{plot-Karl-2}.
%Although the in-plane anharmonic coupling
%is larger than the out-of-plane coupling, indeed $\tilde{\gamma}(\perp)/
%\tilde{\gamma}($Z$)$=5.31,
%the dominance of $\beta_T$(Z) over $\beta_T(\perp)$ is due to the relative smallness of 
%$\kappa_0/(\hat{c}^2 v_{2D})$=2.74$\times 10^{-2}$, which is a measure of ratio of the 
%out-of-plane and in-plane restoring forces.

%While for the $l$=10$^4a$ system $|\beta_T(\mbox{Z})|>\beta_T(\perp)$, always holds,
%we find (Fig.~\ref{plot-Karl}) that for $l$=10$^2a$, $|\beta_T$(Z)$|$ becomes smaller than 
%$\beta_T(\perp)$ above $T\approx 410$~K. Hence the thermal expansion $\alpha_T$ 
%changes sign for smaller systems (Fig.~\ref{plot-Karl-2}).

Evaluation of Eqs.(\ref{eq_inte}) and (\ref{eq-64-1}) shows that $|\beta_T($Z$)|>\beta_T(\perp)$ at low $T$
and hence $\beta_T$, Eq.~(\ref{eq-sumperp}) is negative. 
A change of sign to positive values becomes possible with increasing temperature.
Solution of the equation
\begin{equation}
|\beta_T(\mbox{Z}\mbox{;}T, l)|=\beta_T(\perp\mbox{;}T, l)
\label{ekun}
\end{equation}
yields pairs of values $\{T_{\alpha},l_{\alpha}\}$ where $\beta_T$, Eq.~(\ref{eq-sumperp}), 
changes sign, (Fig.~\ref{plot-Karl-2}).
Here $T_{\alpha}$ is an implicit function of $l_{\alpha}$.
We obtain $\{l_{\alpha}=10^2a, T_{\alpha}=407$K$\}$; $\{l_{\alpha}=10^3a, T_{\alpha}=697$K$\}$.
In case of an infinite system (thermodynamic limit), $\beta_T$(Z)  diverges 
logarithmically (see Eq.~(\ref{eq61-lowZA})). Then Eq.~(\ref{ekun}) has no solution and $\beta_T$
remains negative up to highest $T$.
With our model parameters we find that for $l=10^4a$, $T_{\alpha}$ becomes already 
unphysically large ($ 10^4$~K).
 
We will show below (Sect.~VI) that the renormalization of the flexural mode however decreases $|\beta_T($Z$)|$
and results in room temperature values of $T_{\alpha}$ for 
graphene samples of macroscopic size.
%At the end of Sect.~VI we show that renormalization of the flexural 
%mode decreases $|\beta_{T}($Z$)|$ and leads to a positive thermal 
%expansion above 320 K also for the $l$=10$^4a$ system. 
Since the ratio $|\beta_T$(Z)$|$ versus $\beta_T(\perp)$ depends on material constants,
it is conceivable that for some finite size monolayer materials  
$|\beta_T$(Z)$|<\beta_T(\perp)$ always holds and the thermal expansion is positive,
as has been concluded~\cite{sevik} for MoS$_2$.

%%%%%%%%%%%%%%%%%%%%%%%%%%%%%%%%%%%%%%%%%%%%%%%%%%%%%%%%%%%%%%%
\section{Frequency Shifts and Linewidths}
%\subsection{In-plane modes}

Phonon lineshifts are due to third and fourth order anharmonicities
and phonon dampings due to third order.
%Next we investigate lineshifts and damping
%of the in-plane modes and of the flexural mode.

\subsection{In-plane modes}

We study the case where a long wavelength in-plane phonon ($\lambda=$L, T) of wavevector $\vec{q}$
decays into two out-of-plane phonons ($\lambda=$Z) with wave vectors $\vec{k}$ and $\vec{q}-\vec{k}$.
Since all wave vectors are small, there are no Umklapp processes. 
The frequency shift due to 
third order anharmonicities, obtained by means of Eqs.~(\ref{35new}), 
(\ref{new29-old}) and (\ref{new29}), reads
%\begin{widetext}
\begin{eqnarray}
\Delta^{(3)}(\vec{q},\lambda)=\frac{\hbar}{2N}
%\Sigma(\vec{q},\lambda;z)&=\hbar \omega(\vec{q},\lambda) 
P
\sum_{\vec{k}}
\left|\Phi^{(3)}\binom{\lambda\ \mbox{Z}\  \ \mbox{Z}}
{ {-\vec{q}}\ \vec{k}\  {\vec{q}-\vec{k}}}\right|^2 \nonumber \\
\times \bigg\lbrace
\frac{1+n(\vec{k},\mbox{Z})+n(\vec{k}-\vec{q},\mbox{Z})}
{\omega(\vec{q},\lambda)-\omega(\vec{k},\mbox{Z})-\omega(\vec{k}-\vec{q},\mbox{Z})}
\bigg\rbrace.
\label{eq67}
\end{eqnarray}
%\end{widetext}
\noindent With $\vec{q}=(q,0)$ taken as polar axis along the x-direction, we have 
$\vec{k}=k(cos\ \varphi, sin\ \varphi)$. We approximate the polarization vectors entering 
$\left|\Phi^{(3)}\binom{\lambda\ \mbox{Z}\  \mbox{Z}}
{{-\vec{q}}\ \vec{k}\  {\vec{q}-\vec{k}}}\right|$ by $e_i^A(\vec{q},\lambda)=\sqrt{1/2} \delta_{i\zeta}$,
with $\zeta=x$ for $\lambda=$L and  $\zeta=y$ for $\lambda=$T.
%
%The coupling between in-plane and out-of-plane mode is then governed by 
%$\varphi^{(3)}_{xzz}(A,B_{\alpha})$ and $\varphi^{(3)}_{yzz}(A,B_{\alpha})$.
Furthermore, we use $e_j^A(\vec{k},$Z$)\approx e_j^A(\vec{k}-\vec{q},$Z$)\approx \sqrt{1/2}\delta_{jz}$,
approximate $\omega(\vec{k}-\vec{q},$Z$)$ by $\omega(\vec{k},$Z$)$ and 
$sin^2(\vec{k}\cdot\vec{r}(B_{\alpha}))$ by $(\vec{k}\cdot\vec{r}(B_{\alpha}))^2$. 
From Eq.~(\ref{eq49}) we then obtain for $\lambda=$L
\begin{equation}
%\begin{equation}
\Phi^{(3)}\binom{\mbox{L}\ \mbox{Z}\  \mbox{Z}}{{-\vec{q}}\ \vec{k}\ \vec{q}{-\vec{k}}}=
\frac{i\sqrt{q}h^{(3)}a^3(2 cos^2 \varphi +1 )}
{64\sqrt{3{M_C^3c_{\mbox{\scriptsize L}}\kappa_0}}}
\label{eq-z}
\end{equation}
The expression for $\lambda=$ T is obtained from Eq.~(\ref{eq-z}) replacing $(2 cos^2 \varphi +1 )$
by $2\ cos \varphi\ sin \varphi$ and $c_{\mbox{\scriptsize L}}$ by $c_{\mbox{\scriptsize T}}$. 
Transforming the $\vec{k}$-sum in Eq.~(\ref{eq67}) to a 2D integral, 
%carrying out the angular integral and transforming the $k$-integral into a frequency integral, 
%using $\omega(\vec{k},$Z$)=\sqrt{\kappa_0}k^2$,
we have
\begin{equation}
\Delta^{(3)}(\vec{q},\lambda)=\frac{\hbar q C^{(3)}(\lambda)}{2} P
\int^{\omega_s(\mbox{{\scriptsize Z}})}_{\omega_l(\mbox{{\scriptsize Z}})}d\omega
\frac{(1+2n(\omega))}{\omega(\vec{q},\lambda)-2\omega}
\end{equation} 
where $n(\omega)=(e^{\hbar\omega/k_BT}-1)^{-1}$.
Here we have defined for $\lambda$=L 
\begin{equation}
C^{(3)}(\mbox{L})=\frac{(h^{(3)}a^4)^2 3\sqrt{3}}{\pi (256)^2 c_{\mbox{\scriptsize L}} M_C^3 \kappa_o^{3/2}},
\end{equation}
%
%$C^{(3)}($L$)=(h^{(3)}a^4)^2 3\sqrt{3}/(\pi (256)^2 c_{\mbox{\scriptsize L}} 
%M_C^3 \kappa_o^{3/2})$, and
while for $\lambda$=T, $c_{\mbox{\scriptsize L}}$ has to be replaced by $c_{\mbox{\scriptsize T}}$ 
and $3\sqrt{3}$ by $1/\sqrt{3}$.
We obtain $C^3\mbox{(L)=7.51}\times 10^{27}\mbox{cm}^{-1}\mbox{g}^{-1}$ and 
$C^3\mbox{(T)=1.35}\times 10^{27}\mbox{cm}^{-1}\mbox{g}^{-1}$. 
%With $n(\omega)=(e^{\hbar\omega/k_B T}-1)^{-1}$ one finds that the frequency integral diverges at $\omega=0$.
%One has to introduce again an infrared cut-off $\omega_l($Z$)$ which corresponds to a crystal of finite size, 
%where $q\ge2\pi/l$ with $\omega_l($Z$)<\omega(\vec{q},\lambda)/2<\omega_s($Z$)$. 
%
%We consider the integral
%\begin{equation}
%I(\vec{q},\lambda)=P\int_{\omega_l(\mbox{{\scriptsize Z}})}^{\omega_s(\mbox{{\scriptsize Z}})}
%d\omega \frac{1+2n(\omega)}
%{w(\vec{q},\lambda)-2\omega}
%\end{equation}
Carrying out the integration we obtain in the quantum case 
%$k_BT\ll\hbar\omega_l(\mbox{{\scriptsize Z}})$
%$\hbar\omega_l($Z$) \gg k_B T$ 
%where $n(\omega)\ll 1$, 
\begin{equation}
\Delta^{(3)}(\vec{q},\lambda)=\frac{\hbar q}{4} C^{(3)}(\lambda)
ln\bigg(
\frac{\omega(\vec{q},\lambda)-2\omega_l(\mbox{Z})}
{2\omega_s(\mbox{Z})-\omega(\vec{q},\lambda)}.
\label{eq70}
\bigg)
\end{equation}
In the classical case %$k_B T \gg \hbar\omega_s($Z$)$
we get
\begin{equation}
\Delta^{(3)}(\vec{q},\lambda)=\frac{k_B T C^{(3)}(\lambda)}{c_{\lambda}}
ln\bigg(
\frac{\omega_s(\mbox{Z})[\omega(\vec{q},\lambda)-2\omega_l(\mbox{Z})]}
{\omega_l(\mbox{Z})[2\omega_s(\mbox{Z})-\omega(\vec{q},\lambda)]}
\bigg).
\label{eq71}
\end{equation}
\noindent For $T$=1000 K and q=$\pi/$10$a$ we get 
$\Delta^{(3)}(\vec{q}, $L$)=6.83\times 10^9\mbox{s}^{-1} 
\ \mbox{i. e.} 3.6\times 10^{-2} \mbox{cm}^{-1}$,
and $\Delta^{(3)}(\vec{q}, $T$)=1.91 \times 10^9 \mbox{s}^{-1}
\ \mbox{i. e.} 1.00\times 10^{-2} \mbox{cm}^{-1}$. 

\begin{figure}[t]
\vspace{0.25cm}
\includegraphics[width=0.45\textwidth]{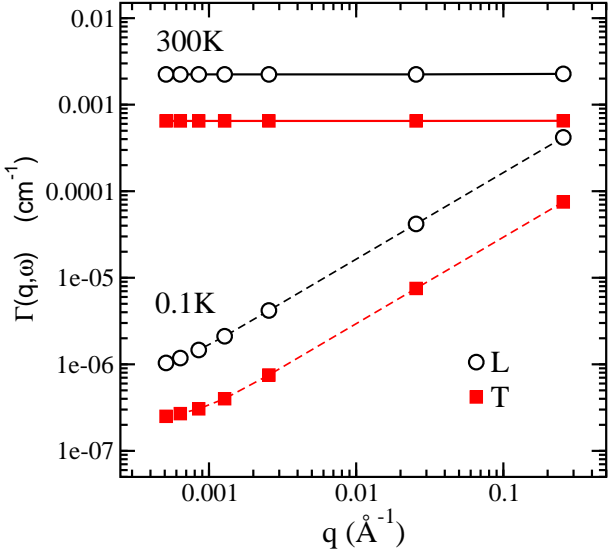}
\caption{
(Color online) Linewidths for $\lambda=$L (circles) and T (squares) at different temperatures as indicated.
The 2D crystal size is $l=10^4a$.
}\label{fig5}
\end{figure}

Turning to the corresponding linewidth process we find 
by means of Eqs.~(\ref{new31}) and (\ref{36new})
%\begin{widetext}
\begin{subequations}
\begin{align}
\Gamma(\vec{q},\lambda)=&\frac{\pi\hbar}{2N}\sum_{\vec{k}}
\left|\Phi^{(3)}
\binom{\lambda \ \mbox{Z}\  \mbox{Z}}{{-\vec{q}}\  \vec{k}\  {\vec{q}-\vec{k}}}
\right|^2   \nonumber \\
&\times \bigg(1+n(\vec{k},\mbox{Z})+n(\vec{k}-\vec{q},\mbox{Z})\bigg)\ \nonumber \\
&\times \delta\bigg(\omega(\vec{q},\lambda)-\omega(\vec{k},\mbox{Z})-\omega(\vec{k}-\vec{q},\mbox{Z})\bigg).
\end{align}
\end{subequations}
%\Gamma(\vec{q},\lambda)&=
%%\frac{\pi\hbar}{2N}\sum_{\vec{k}}
%\left|\Phi^{(3)}\binom{\lambda \mbox{Z} \mbox{Z}}
%{{-\vec{q}} \vec{k} {\vec{q}-\vec{k}}\right|^2
%\bigg(1+n(\vec{k},\mbox{Z})+n(\vec{k}-\vec{q},\mbox{Z})\bigg)
%\delta(\omega(\vec{q},\lambda)-\omega(\vec{k},\mbax{Z})-\omega(\vec{k}-\vec{k},\mbox{Z}))
%\label{eq_73a}\\
%\end{align}
%\end{subequations}
%\end{widetext}
With the same approximations as outlined before we obtain
\begin{subequations}[resume]
\begin{align}
\Gamma(\vec{q},\lambda)=\frac{\pi}{4}\hbar q C^{(3)}(\lambda)
\Bigg[1+2n\bigg(\frac{\omega(\vec{q},\lambda)}{2}\Bigg)\Bigg]
%\Gamma(\vec{q},\lambda)=\frac{\pi}{4}\hbar q C^{(3)}(\lambda)
%\int_{\omega_l(\mbox{{\scriptsize Z}})}^{\omega_s(\mbox{{\scriptsize Z}})}
%& d\omega \ (1+2n(\omega))\ \times \nonumber \\
%&\delta\bigg(\omega-\frac{\omega(\vec{q},\lambda)}{2}\bigg),
\label{linew-b}
\end{align}
\end{subequations}
which in the quantum regime becomes
\begin{equation}
\Gamma(\vec{q},\lambda)=\frac{\pi}{4}\hbar q C^{(3)}(\lambda).
\label{eq73}
\end{equation}
%For $q=\pi/a$, we find $\Gamma(\vec{q},$L$)$=3.43$\times$10$^9$s$^{-1}$
%and  $\Gamma(\vec{q},$T$)$=6.16$\times$10$^8$s$^{-1}$.
%
In the classical regime we get
\begin{equation}
\Gamma(\vec{q},\lambda)=\frac{\pi}{c_{\lambda}}k_B T C^{(3)}(\lambda),
\end{equation}
\noindent the result is independent of the wave vector~\cite{nano-marzari}.
Plots of the linewidths evaluated by means of Eq.~(\ref{linew-b})
for $\lambda=$L and T are given in Fig.~\ref{fig5}.

The frequency shift due to scattering of an in-plane phonon with a flexural mode
(fourth order anharmonic process) reads
%
%The frequency shift due to the fourth order anharmonicities reads for the decay of an in-plane ($\lambda=$L$, $T$)$
%phonon into two flexural (Z) phonons 
%\begin{widetext}
\begin{equation}
\Delta^{(4)}(\vec{q},\lambda)=\frac{\hbar}{2N}
%\Sigma(\vec{q},\lambda;z)&=\hbar \omega(\vec{q},\lambda) 
\sum_{\vec{k}}
\Phi^{(4)}\binom{\lambda\ \mbox{Z}\  \ \mbox{Z}\ \ \lambda}
{ {-\vec{q}}\ \vec{k}\  {-\vec{k}}\ \vec{q}} \big[1+2n(\vec{k},\mbox{Z})\big].
\label{eq-4th_perp}
\end{equation}
\noindent Using long wavelength approximations as before, we obtain from Eq.~(\ref{eq-45-new-Ok})
\begin{eqnarray}
\Phi^{(4)}\binom{\mbox{L} \ \mbox{Z}\  \ \mbox{Z}\ \ \mbox{L}}
{ {-\vec{q}}\ \vec{k}\  {-\vec{k}}\ \vec{q}}=
%\Sigma(\vec{q},\lambda;z)&=\hbar \omega(\vec{q},\lambda) 
\frac{qa^4}{16\times 96 M_C^2 \sqrt{\kappa_0} c_{\mbox{\scriptsize L}}}   \nonumber \\
\times \Big[ (11n^{(4)}+p^{(4)})cos^2\varphi+(n^{(4)}+3p^{(4)})sin^2\varphi\Big].
\label{eq-4th_perp1}
\end{eqnarray}
\noindent  The expression for $\lambda=$T is obtained by replacing $c_{\mbox{\scriptsize L}}$ by 
$c_{\mbox{\scriptsize T}}$ and interchanging $n^{(4)}$ with $p^{(4)}$.
The summation over the Brillouin zone in Eq.~(\ref{eq-4th_perp}) is readily transformed into a frequency integral
\begin{equation}
\Delta^{(4)}(\vec{q},\lambda)=\hbar q C^{(4)}(\lambda)
%\Sigma(\vec{q},\lambda;z)&=\hbar \omega(\vec{q},\lambda) 
\int_{\omega_l(\mbox{\scriptsize Z})}^{\omega_s(\mbox{\scriptsize Z})}
d\omega \ \Big[ 1+2n(\omega)\Big],
\label{eq-4th_perp-int}
\end{equation}
\noindent where for $\lambda=$L
\begin{equation}
C^{(4)}(\mbox{L})=\frac{a^6\sqrt{3}(3n^{(4)}+p^{(4)})}
{128\times96 \pi M_C^2 c_{\mbox{\scriptsize L}}\kappa_o}.
\end{equation}
Again $C^{(4)}$(T) is obtained by the substitutions just mentioned.
In the quantum limit, $k_B T \ll \hbar \omega_l($Z$)$, we have 
\begin{equation}
\Delta^{(4)}(\vec{q},\lambda)=\hbar q C^{(4)}(\lambda)\Big[\omega_s(\mbox{Z})-\omega_l(\mbox{Z})\Big],
\end{equation}
and in the classical limit, $k_B T \gg \hbar\omega_s($Z$)$,
\begin{equation}
\Delta^{(4)}(\vec{q},\lambda)=2 q k_B T C^{(4)}(\lambda)\ ln\Bigg( \frac{\omega_s(\mbox{Z})}{\omega_l(\mbox{Z})}\Bigg).
\label{corri4}
\end{equation}
Notice in both cases the linear dependence on the wave vector of the in-plane mode. Since $n^{(4)}$ and $p^{(4)}$
are positive (see Table II), we conclude that the lineshifts above are positive.

%######################################################
\subsection{Flexural mode}

\begin{figure}[t]
\vspace{0.25cm}
\includegraphics[width=0.45\textwidth]{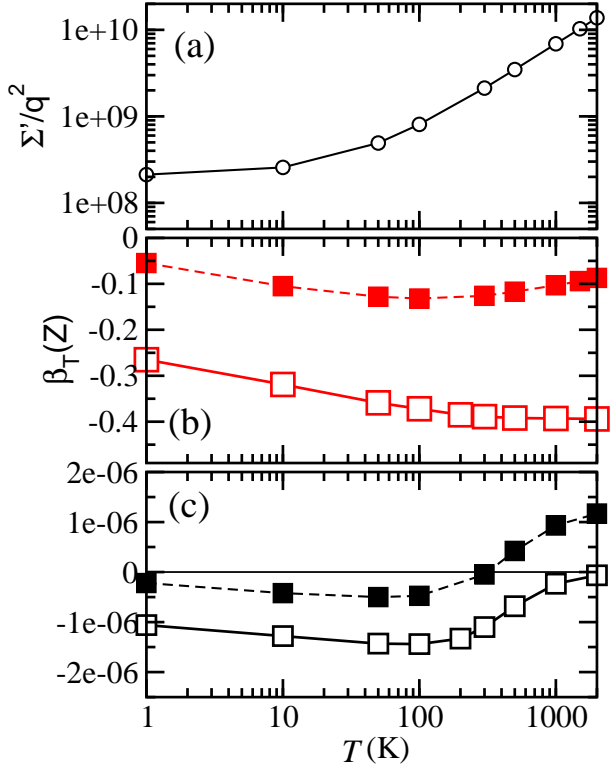}
\caption{
(a) Quantity $c^2_{\mbox{{\scriptsize Z}}}$=$\Sigma'(\vec{q},$Z$)/q^2$ as function of $T$, 
units cm$^2$ s$^{-2}$. % $l$=10$^4a$.
(b) Thermal tension $\beta_T$(Z) and (c) thermal expansion $\alpha_T$ 
evaluated with renormalized flexural mode frequency $\Omega(\vec{q},$Z$)$ (filled squares)
as function of $T$.  $\beta_T$(Z) is given in units of dyn cm$^{-1}$ K$^{-1}$ and 
$\alpha_T$ in K$^{-1}$.
The 2D crystal size is $l=10^4a$.
Compare with Fig.~\ref{plot-Karl-2}.
% $l$=10$^4a$.
}\label{fig6-karlo}
\end{figure}

We first investigate the decay and the lineshift of the flexural mode due to third order 
anharmonicities. As already emphasized~\cite{nano-marzari}, the scattering rate us dominated by the absorption
processes ZA+ZA$\rightarrow$LA(TA). The frequency dependent self-energy (See Sect. III, B)
reads

\begin{eqnarray}
\Sigma^{(3)}(\vec{q}, \mbox{Z};z)=
\frac{2\hbar \omega(\vec{q}, \mbox{Z})}{N}
\sum_{\vec{k}\lambda}\Bigg|\Phi^{(3)}
\binom{\mbox{Z}\  \mbox{Z}\ \lambda}
%{ {-\vec{q}}\ \vec{k}\  {-\vec{k}}\ \vec{q}}&\nonumber \\
{ \vec{q}\ {\vec{k}-\vec{q}}\ {-\vec{k}}}
\Bigg|^2  \ \nonumber \\
\times \frac{n(\vec{k}-\vec{q}, \mbox{Z})-n(\vec{k}, \mbox{Z})}
{z+\omega(\vec{k}-\vec{q}, \mbox{Z})-\omega(\vec{k}, \lambda)}\ \ \ \ \ \ 
\ \ \ \ \ \ \ \ 
\end{eqnarray}
where $z=\omega+i\epsilon$, $\epsilon\rightarrow0^+$, and $\lambda=$L(T).
We investigate this quantity for $\omega=\omega(\vec{q}, $Z$)$.
In order to get an analytically tractable problem, we take $\vec{q}=(q,0,0)$.
In addition we replace $(\vec{k}-\vec{q})\cdot\vec{r}(B_{\alpha})$ in Eq.~(\ref{eq49}) by
its supremum $|\vec{k}-\vec{q}| |\vec{r}(B_{\alpha})|$. 
As consequence of these approximations only the scattering into the L mode 
is different from zero. 
From the study of $\Sigma^{(3)''}$ we obtain the decay rate (compare Eq.~(\ref{36new}))
\begin{equation}
\Gamma(\vec{q}, \mbox{Z})=\hbar q^4\frac{128 \kappa_0^{3/2}}
%{9\sqrt{3}c_{\mbox{\scriptsize L}}^3}
{9c_{\mbox{\scriptsize L}}^3}
C^{(3)}(\mbox{L})
csch\Bigg(
\frac{\hbar \omega(q,\mbox{Z})}{k_B T}
\Bigg)
\end{equation}
In the classical regime we get
\begin{equation}
\Gamma(\vec{q}, \mbox{Z})=q^2k_B T\frac{128\pi \kappa_0}
%\Gamma(\vec{q}, \mbox{Z})=q^2k_B T\frac{128\pi \kappa_0^{3/2}}
%{9\sqrt{3}c_{\mbox{\scriptsize L}}^3}
{9c_{\mbox{\scriptsize L}}^3}
C^{(3)}(\mbox{L})
\end{equation}
and in the quantum regime
\begin{equation}
\Gamma(\vec{q}, \mbox{Z})=\hbar q^4\frac{256 \kappa_0^{3/2}}
{9c_{\mbox{\scriptsize L}}^3}
%{9\sqrt{3}c_{\mbox{\scriptsize L}}^3}
C^{(3)}(\mbox{L})
e^{-\hbar \omega(q,\mbox{\scriptsize Z})/k_B T}
\end{equation}
The $q^2$ dependence in the classical regime has been predicted earlier~\cite{nano-marzari}.

%
%From the study of $\Sigma^{(3)}$ we obtain the lineshift. In the classical regime we find
%\begin{equation}
%\Delta^{(3)}(q, \mbox{Z})=-\frac{16}{9\sqrt{3}c_{\mbox{\scriptsize L}}}
%k_B T C^{(3)}(\mbox{L})
%ln\Bigg(\frac{\omega_s(\mbox{L})}{\omega_l(\mbox{L})}\Bigg),
%\end{equation}
%where $\omega_s$(L)=$c_{\mbox{\scriptsize L}}2\pi/l$ and $\omega_s$(L)$\approx$200 THz.
%The lineshifth is negative and increases logarithmically in absolute value with the 
%size of the system. 
%
From the study of $\Sigma^{(3)'}$ we find in the classical regime
\begin{equation}
\Sigma^{(3)'}(\vec{q}, \mbox{Z})=-\frac{32}{9c_{\mbox{\scriptsize L}}}
q^2 \sqrt{\kappa_0}
%\omega(\vec{q},\mbox{Z})
k_B T C^{(3)}(\mbox{L})
ln\Bigg(\frac{\omega_s(\mbox{L})}{\omega_l(\mbox{L})}\Bigg),
\label{eq1-75}
\end{equation}
where $\omega_l$(L)=$c_{\mbox{\scriptsize L}}2\pi/l$ 
and $\omega_s$(L)=245 THz.
The phonon self-energy is negative and diverges logarithmically with the size
of the system. %the factor $ \omega(\vec{q}, $Z$)$ leads to a $q^2$ dependence.

In the quantum regime the self-energy $\Sigma^{(3)'}(\vec{q}, $Z$)$ vanishes
exponentially with lowering $T$.
%In the quantum regime the lineshift decreases exponentially with lowering $T$.

We next investigate the renormalization %$\Sigma^{(4)'}(\vec{q}, $Z$)$ 
of the flexural mode due to fourth order anharmonicities.
%We investigate the renormalization $\Sigma'(\vec{q},$Z$)$ of the flexural mode due to fourth order 
%anharmonicities. 
%
%Here we neglet the contributions due to third order anharmonicities since they are of higher order in $\vec{q}$
%and hence negligible in the long wavelenght limit in comparison to the fourth order terms. 
%
From Eqs.~(\ref{new29-old}) and (\ref{new30}) we get
%\begin{eqnarray}
\begin{equation}
\Sigma^{(4)}(\vec{q},\mbox{Z})=\frac{\hbar \omega(\vec{q}, \mbox{Z})}{N}
\sum_{\vec{k},\lambda}
\Phi^{(4)}\binom{\mbox{Z}\  \lambda\  \lambda\ \mbox{Z}}
%{ {-\vec{q}}\ \vec{k}\  {-\vec{k}}\ \vec{q}}&\nonumber \\
{ {-\vec{q}}\ \vec{k}\  {-\vec{k}}\ \vec{q}}
\Big[1+2n(\vec{k},\lambda)\Big],
\label{eq-4th-0}
\end{equation}
%\end{eqnarray}
\noindent where $\lambda$=$\lbrace$L, T, Z$\rbrace$. 
Separating in-plane and out of plane scattering modes $\lambda$ we write
\begin{equation}
\Sigma^{(4)}(\vec{q},\mbox{Z})=\Sigma^{(4)}
(\vec{q},\mbox{Z};\perp)+\Sigma^{(4)}(\vec{q},\mbox{Z};\mbox{Z})
\end{equation}
where $\perp$ stands for $\lbrace$L, T$\rbrace$.
We readily transform to frequency integrals and obtain
\begin{equation}
\Sigma^{(4)}(\vec{q},\mbox{Z};\perp)=q^2 \hbar \sum_{\lambda}C^{(4)}(\mbox{Z};\lambda)
\int_{\omega_l(\lambda)}^{\omega_s(\lambda)}
d\omega \ \omega^2\ \Big[ 1+2n(\omega)\Big],
\label{eq4-4th-1}
\end{equation}
where for $\lambda$=L
\begin{equation}
C^{(4)}(\mbox{Z;L})=\frac{a^6\sqrt{3}(3n^{(4)}+p^{(4)})}
{32\times96 \pi M_C^2 c^4_{\mbox{\scriptsize L}}},
\end{equation}
while for $\lambda$=T, $n^{(4)}$ and $p^{(4)}$ are interchanged and $c_{\mbox{\scriptsize L}}$
is replaced by $c_{\mbox{\scriptsize T}}$. 
In Eq.~(\ref{eq4-4th-1}) the integration limits are 
$\omega_s($L$)$=245 THz, $\omega_s($T$)$=151 THz, and for the sample with 
$l$=10$^4a$, $\omega_l($L$)$=5.9$\times10^{10}$ Hz, $\omega_l($T$)$=3.65$\times10^{10}$ Hz.
\noindent In the classical case we have 
\begin{equation}
\Sigma^{(4)}(\vec{q},\mbox{Z};\perp)=q^2 k_B T \sum_{\lambda} C^{(4)}(\mbox{Z};\lambda)
\Big[\omega^2_s(\lambda)-\omega^2_l(\lambda)\Big]
\label{eq4-4th-121}
\end{equation}

In the quantum regime zero point motion gives at $T=0$:
\begin{equation}
\Sigma^{(4)}(\vec{q},\mbox{Z};\perp)=
\frac{q^2\hbar}{3}\sum_{\lambda}
C^{(4)}(\mbox{Z};\lambda)
[\omega_s^3(\lambda)-\omega_l^3(\lambda)].
\end{equation}
 
Likewise we obtain 
\begin{equation}
\Sigma^{(4)}(\vec{q},\mbox{Z};\mbox{Z})= q^2 \hbar C^{(4)}(\mbox{Z};\mbox{Z})
\int_{\omega_l(\mbox{{\scriptsize Z}})}^{\omega_s(\mbox{{\scriptsize Z}})}
d\omega \ \ \Big[ 1+2n(\omega)\Big],
\label{eq4-4th-12}
\end{equation}
where
\begin{equation}
C^{(4)}(\mbox{Z;Z})=\frac{a^6\sqrt{3}\ l^{(4)}}
{16\times96 \pi M_C^2 \kappa_0}
\end{equation}
with $\omega_s($Z$)$=94.25 THz and $\omega_l($Z$)$=4.25 MHz.
In the classical limit we get
\begin{equation}
\Sigma^{(4)}(\vec{q},\mbox{Z};\mbox{Z})= 2 q^2 k_B T C^{(4)}(\mbox{Z};\mbox{Z})
ln\Bigg(\frac{\omega_s(\mbox{Z})}{\omega_l(\mbox{Z})}\Bigg),
\label{label15}
\end{equation}
which diverges for an infinite system where 
$\omega_l($Z$)$=0. 
Here zero point motion gives at $T$=0:
\begin{equation}
\Sigma^{(4)}(\vec{q},\mbox{Z};\mbox{Z})= 
q^2 \hbar C^{(4)}(\mbox{Z};\mbox{Z}) 
[\omega_s(\mbox{Z})-\omega_l(\mbox{Z})].
\end{equation}

From Eqs.~(\ref{eq1-75}), (\ref{eq4-4th-1}) and (\ref{eq4-4th-121}) 
we see that the contributions to the flexural mode self-energy are due to third and 
fourth order anharmonicities and are proportional to $q^2$ at long wavelengths.
%
%While at low $T$ the contributions due to third order anharmonicities become
%negligible, zero point motion effects due to fourth order anharmonicities
%contribute to the self-energy and are still proportional to $q^2$.
%
We then identify $c^2_{{\scriptsize Z}}$ introduced in Eq.~(\ref{sigmita}) as:
\begin{equation}
c^2_{\mbox{{\scriptsize Z}}}=\frac{\Sigma^{(3)'}(\vec{q},\mbox{Z})
+\Sigma^{(4)}(\vec{q},\mbox{Z})}{q^2}
\end{equation}

Notice that the in-plane and out-of-plane contributions to 
$\Sigma'(\vec{q},\mbox{Z})$ are proportional to $q^2$, as anticipated in Eq.~(\ref{sigmita}).
In the language of membrane theory~\cite{Nelson-Peliti} this result corresponds to a 
$q^{-2}$ singularity due to the first anharmonic correction to the bare bending rigidity.
We have evaluated expressions (\ref{eq1-75}), 
(\ref{eq4-4th-1}) and (\ref{eq4-4th-12}) as function of temperature in the interval 
0.1-2000~K.
We find that the negative term $\Sigma^{(3)'}(\vec{q},\mbox{Z})$ is more 
than two orders of magnitude smaller than the positive term $\Sigma^{(4)}(\vec{q},\mbox{Z})$
and hence negligible.
Both $\Sigma'(\vec{q},$Z;$\perp)$ and $\Sigma'(\vec{q},$Z;Z) are monotone increasing functions with increasing 
$T$, with $\Sigma'(\vec{q},$Z;Z$)>\Sigma'(\vec{q},$Z;$\perp)$. 
Above $T$=100~K, $\Sigma'(\vec{q}$,Z;Z$)$ is more than one order of magnitude larger.
In Fig.~\ref{fig6-karlo}(a) we have plotted the sum  
$c^2_{\mbox{{\scriptsize Z}}}=\Sigma'(\vec{q}$,Z$)/q^2$ for the case $l$=10$^4a$.
We find that for the case $l$=10$^2a$ the renormalization effect is about a factor 2 smaller.

%\begin{figure*}
%\vspace{0.25cm}
%\includegraphics[width=0.42\textwidth]{figs/gruneinsen_all.eps}
%\hspace{0.25cm}
%\includegraphics[width=0.45\textwidth]{figs/therm-exp-oct14.eps}
%%\includegraphics[width=0.48\textwidth]{figs/therm_exp.eps}
%\caption{(Color online)
%(Left panel) Generalized Gr\"uneisen parameters $\gamma(\vec{q},\omega)$ 
%for (a) ZA, (b) TA and (c) LA acoustic phonon modes.  
%$\gamma_x$ ($\gamma_y$) contribution is indicated by a dashed-red (dotted-blue) line. 
%(Right panel) Temperature dependence of the thermal expansion coefficient $\alpha_T$.
%}\label{Grune}
%\end{figure*}
\begin{figure}[t]
\vspace{0.25cm}
\includegraphics[width=0.47\textwidth]{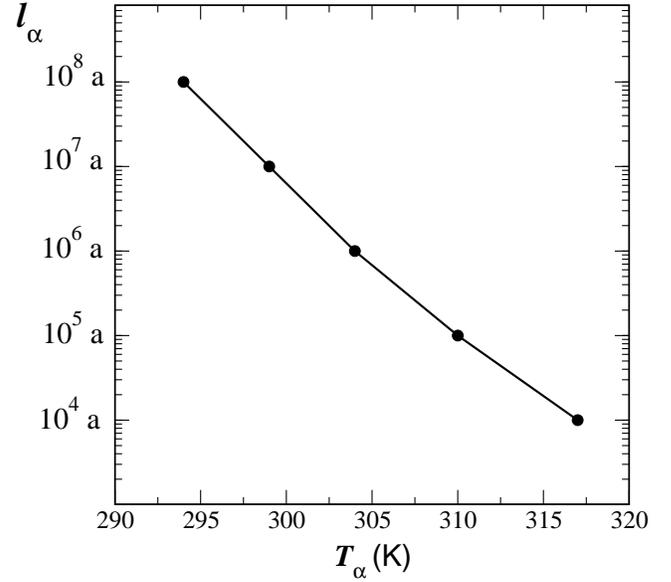}
\caption{
System size $l_{\alpha}$ as function of temperature $T_{\alpha}$ for change from negative to positive thermal
expansion. Discrete points are calculated self-consistently with renormalization of flexural mode.
%$\{l_{\alpha}, T_{\alpha}\}$, Renormalized case.
}\label{fig7n}
\end{figure}

Finally we have studied the effect of renormalization on the thermal expansion. 
Therefore we have evaluated $\beta_{T}$(Z) by means of Eqs.~(\ref{eq_ten}) and (\ref{eq-48-new1}),
with $\omega(\vec{q},$Z$)$ replaced by the renormalized frequency 
$\Omega(\vec{q},$Z$)$, Eqs.~(\ref{sigmita}) and (\ref{renorm}). 
Thereby we take into account self-consistently that $c_{\mbox{{\scriptsize Z}}}^2$ depends on $l$ and $T$. 
%Transforming the $\vec{q}$-sum into an integral 
We obtain
\begin{equation}
\beta_T(\mbox{Z})=\frac{\tilde{\gamma}(\mbox{Z})\hbar^2}{2\pi k_B T^2}
\int_{q_l}^{q_s}dq\ q^3\
\frac{e^{\hbar\Omega(\vec{q},\mbox{{\scriptsize Z}})/k_BT}}
{\big(e^{\hbar\Omega(\vec{q},\mbox{{\scriptsize Z}})/k_BT}-1\big)^2},
\label{eq4th-neww}
\end{equation}
where $q_s$=$(\omega_s$(Z)$/\sqrt{\kappa_0})^{1/2}$=1.20$\times$10$^8$cm$^{-1}$ and 
$q_l$=$(\omega_l$(Z)$/\sqrt{\kappa_0})^{1/2}$=2.56$\times$10$^4$cm$^{-1}$.
In Fig.~\ref{fig6-karlo}(b) we have plotted $\beta_T$(Z) as function of $T$. 
Notice that the renormalized  $\beta_T$(Z) is in absolute value smaller 
than the unrenormalized quantity (empty squares).
%(see Fig.~\ref{plot-Karl}(a), case $l$=10$^4a$). 
%
Hence the renormalization of the flexural mode 
favors the transition from negative to positive thermal expansion. 
The transition temperature $T_{\alpha}$ decreases with increasing size of the system. 
This is shown in Fig.~\ref{fig7n}, obtained by solving Eq.~(\ref{ekun}) self-consistently 
for the renormalized case.
%decreases the tendency of negative thermal expansion. 
%Within the present model of interactions we then obtain for the case $l$=10$^4a$,
%$\beta_T(\perp)>|\beta_T($Z$)|$, i. e. a. positive thermal expansion, for $T$
%above $\approx$320~K.
%Solving again numerically Eq.~(\ref{ekun}) we obtain 
%$\{l_{\alpha}=10^4a, T_{\alpha}=317$~K$\}$; $\{l_{\alpha}=10^5a, T_{\alpha}=310$~K$\}$, and  $\{l_{\alpha}=10^6a, T_{\alpha}=304$~K$\}$.
%
%Notice that here $T_{\alpha}$ decreases with increasing system size since the concomittant
%increase of the renormalization parameter $c_{\mbox{{\scriptsize Z}}}^2$, Eq.~(\ref{sigmita}),
%leads to a decrease of $|\beta_T(T$, Z$)|$. However we are not able to reach the thermodynamic 
%limit since then $c_{\mbox{{\scriptsize Z}}}^2$ 
%or equivalently the selfenergy $\sum^{(4)}(\vec{q},$Z;Z$)$ diverges 
%logarithmically (see Eq.~(\ref{label15})). In fact system sizes larger than
%$l=10^6a$ are already beyond our numerical means. 
Although $c_{\mbox{{\scriptsize Z}}}^2$ diverges in the thermodynamic limit,
the weak logarithmic divergence allows one in fact to consider systems of macroscopic size 
($l=10^8a$).

For a recent discussion of the experimental situation, which includes analysis of substrate
corrections, see Ref.~\onlinecite{linas}.

%
%In the high $T$ limit Eq.~(\ref{eq4th-neww}) leads to the result
%\begin{equation}
%\beta_T(\mbox{Z})=\frac{\tilde{\gamma}(\mbox{Z})k_B}{4\pi \kappa_0}
%ln\bigg[\frac{c^2_{\mbox{{\scriptsize Z}}}+\sqrt{\kappa_0}\omega_s(\mbox{Z})}
%{c^2_{\mbox{{\scriptsize Z}}}+\sqrt{\kappa_0}\omega_l(\mbox{Z})}\bigg].
%\end{equation}
%\noindent Notice that in case of an infinite sample $\omega_l$(Z)=0, the right hand side remains
%finite while the unrenormalized $\beta_T$(Z), Eq.~(\ref{eq61-lowZA}), diverges.

%%%%%%%%%%%%%%%%%%%%%%%%%%%%%%%%%%%%%%%%%%%%%%%%%%%%%%%%%%
\section{Numerical Results}

\begin{figure}[t]
\vspace{0.25cm}
\includegraphics[trim=0.5cm 1.2cm 4.35cm 1.4cm, clip, width=0.49\textwidth]{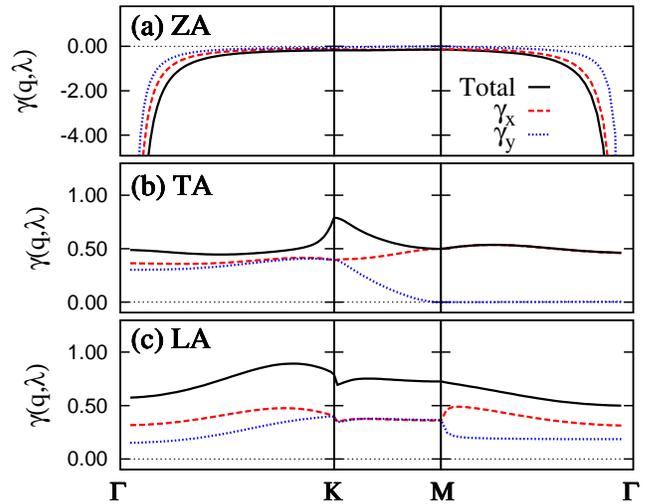}
\caption{(Color online)
Generalized Gr\"uneisen parameters $\gamma(\vec{q},\lambda)$ 
for (a) ZA, (b) TA and (c) LA acoustic phonon modes of graphene.  
$\gamma_x(\vec{q},\lambda)$ ($\gamma_y(\vec{q},\lambda)$), where $x$ ($y$) refer to strain 
$\epsilon_{xx}$ ($\epsilon_{yy}$),
contribution is indicated by a dashed-red (dotted-blue) line.  
}\label{Grune}
\end{figure}

\begin{figure}[t]
\vspace{0.20cm}
\includegraphics[width=0.41\textwidth]{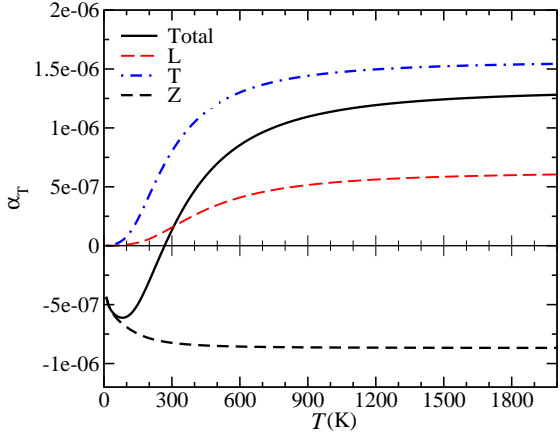}
\caption{(Color online)
Temperature dependence of the thermal expansion coefficient $\alpha_T$ of graphene.
Units K$^{-1}$. Sample size $l=80a$.
}\label{thermfig}
\end{figure}

\begin{figure}[t]
\includegraphics[trim=0.5cm 1.1cm 4.35cm 1.2cm, clip, width=0.49\textwidth]{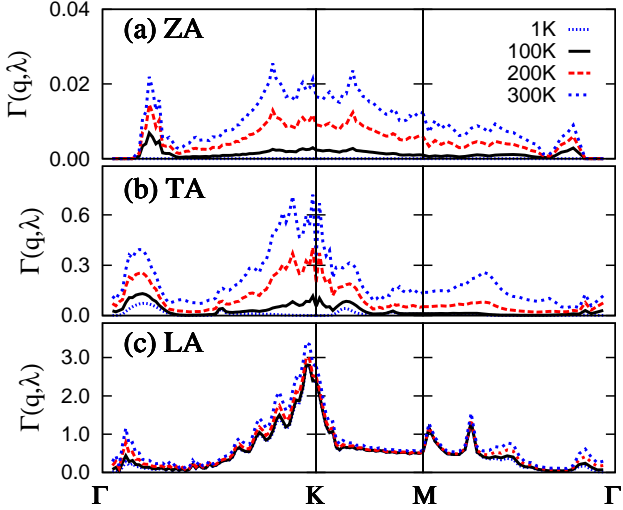}
\caption{(Color online)
Temperature dependence of $\Gamma(\vec{q},\lambda)$ 
for (a) ZA, (b) TA and (c) LA acoustic phonon modes of graphene. Units cm$^{-1}$. 
}\label{damp1}
\end{figure}

The physical quantities outlined above were calculated independently by numerical techniques.
This approach serves for verifying the analytical predictions 
and for obtaining valid results in an extended range of temperatures and wave vectors. 
We investigate Umklapp-processes at the edges of the BZ.
The calculation was realized through a discrete mesh of $\vec{q}$-points in the BZ 
which was designed following the description presented in Ref.~\onlinecite{santoro2}.
%where anharmonicities on surface phonons of Al were investigated.
The basic steps of the algorithm %, with comparison of different meshes, 
are given in Appendix B.

The generalized Gr\"uneisen parameters $\gamma(\vec{q},\lambda)$ (Eq.~(\ref{eq_gamma}))
obtained for $\vec{q}$ along the $\Gamma$-K-M-$\Gamma$ high symmetry crystallographic 
path are shown in Fig.~\ref{Grune}.
The results are to be compared with their analytical counterparts Eq.~(\ref{eq-48-new1}) for 
$\gamma(\vec{q},$~Z$)$ and Eq.~(\ref{eq-gamaperp}) for $\gamma(\perp)$. 
In particular in Eq.~(\ref{eq-48-new1}) we have shown that $\gamma(\vec{q},$~Z$)$
is negative and diverges as $q^{-2}$ in the thermodynamic limit. Notice that 
even in the case of a relatively dense mesh of $\vec{q}$-values 
we are limited in the numerical approach as will be discussed below.
%by the smallest value $\approx 0.025~\AA^{-1}$ which corresponds to a system 
%size $\le~ 10^2\AA$. 
For the in-plane contributions $\gamma($T$)$ and $\gamma($L$)$,
we find that both are positive and approach finite values, in agreement with Eq.~(\ref{eq-gamaperp}).
%
%The contribution of the flexural mode $\gamma(\vec{q},$~Z$)$, in agreement with the discussion 
%given in Eq.~(\ref{eq-48-new1}), 
%is negative and diverges logarithmically for $q\rightarrow 0$. 
%In-plane contributions $\gamma(\vec{q},$~T$)$ and $\gamma(\vec{q},$~L$)$, 
%on the contrary, are both positive and approach finite values in the long wavelength regime
%in agreement with Eq.~(\ref{eq-gamaperp}).
%

The thermal expansion coefficient $\alpha_T$ (Eq.~(\ref{therm-ex-eq})), 
displayed in Fig.~\ref{thermfig}, behaves also as predicted 
(see Fig.~\ref{plot-Karl-2}). 
The flexural (in-plane) mode(s) contribution is negative 
(positive) in the whole range of $T$.
However, here the crossover from negative to positive thermal 
expansion takes place at $\approx 275$~K. 
This value is expected from the discrete mesh adopted in the calculation where 
the smallest values of $\vec{q}$ considered
correspond to a system size $l\approx 80\ a$ (see Appendix B).
Inclusion of smaller values of $\vec{q}$ (larger system sizes) is 
out of reach due to numerical inaccuracy in the diagonalization of the dynamical matrix.  
Notice that this result does not include the renormalization of the flexural mode discussed 
at the end of Sect.~VI~B.

We proceed now with the study of the frequency shifts and linewidths. 
In the previous analysis (Sect.~VI) we restricted ourselves, 
for simplicity, to the most important scattering processes at low $\vec{q}$.  
The numerical treatment of the problem, nevertheless, allows a complete 
survey of every possible scattering mechanism.
We start with the frequency linewidths which, using the same notation as before,
can be calculated through: 
%\begin{equation}
%\Delta(\vec{q},\lambda)=\Delta^{(3)}(\vec{q},\lambda)+\Delta^{(4)}(\vec{q},\lambda),
%In Eq.~(\ref{eq26}) $P$ denotes the principal part.
%The phonon line-width reads
%\begin{widetext}
\begin{widetext}
\begin{align}
%\begin{equation}
\Gamma(\vec{q},\lambda)&=\frac{\pi\hbar}{2N}
\sum_{\vec{q}_2\vec{q}_3}
\sum_{\lambda_2 \lambda_3}
\left|\Phi^{(3)}\binom{\lambda \lambda_2 \lambda_3}{-\vec{q}\vec{q}_2\vec{q}_3}\right|^2 
\Bigg\lbrace \bigg[1+n(\vec{q}_2,\lambda_2)+n(\vec{q}_3,\lambda_3)\bigg] 
\delta\bigg(\omega(\vec{q},\lambda)-\omega(\vec{q}_2,\lambda_2)-\omega(\vec{q}_3,\lambda_3)\bigg) \nonumber \\
&\qquad {}
+ 2 \bigg[n(\vec{q}_2,\lambda_2)-n(\vec{q}_3,\lambda_3)\bigg]
\delta\bigg(\omega(\vec{q},\lambda)+\omega(\vec{q}_2,\lambda_2)-\omega(\vec{q}_3,\lambda_3)\bigg)
%delta\bigg(\omega(\vec{q},\lambda)-\omega(\vec{q}_2,\lambda_2)+\omega(\vec{q}_3,\lambda_3)\bigg)\bigg]
%%
\Bigg\rbrace.
%\end{equation}
\label{eq28-last}
\end{align}
\end{widetext}
%\noindent where $\delta(...)$ stands for the Dirac delta function.
The $\delta$-function was represented as 
$\delta(\omega)=\lim_{\xi\rightarrow 0} e^{-\omega^2/\xi^2}/(\xi \sqrt{\pi})$.
After having analyzed a broad range of parameters we found  
good convergence in the results by adopting %$\xi=5$~cm$^{-1}$ and 
$\xi=5$~cm$^{-1}$ (see Appendix B).  
This value is comparable to the one used in related studies\cite{santoro2,greco,lorenzo}. 

\begin{figure*}
\vspace{0.10cm}
\includegraphics[trim=0.5cm 1.1cm 4.35cm 1.2cm, clip, width=0.49\textwidth]{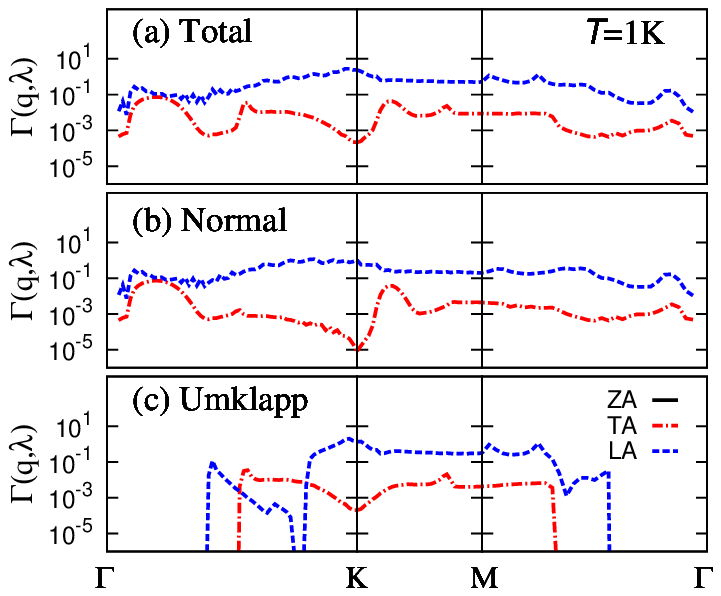}
\hspace{0.15cm}
%includegraphics[width=0.44\textwidth]{figs-paper/ZA12_relevant_all-lg-q3ZA.eps}
\includegraphics[trim=0.5cm 1.1cm 4.35cm 1.2cm, clip, width=0.49\textwidth]{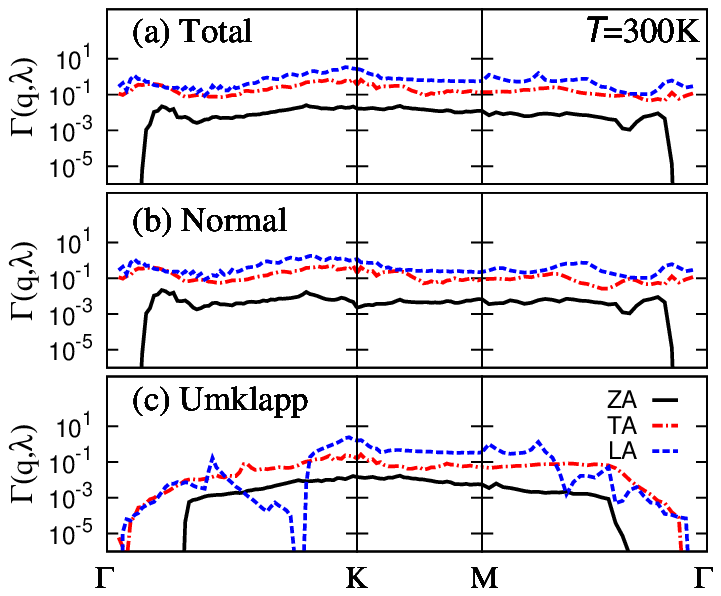}
\vspace{0.10cm}
\caption{(Color online)
Phonon linewidths $\Gamma(\vec{q},\lambda)$ for $T$=1 (left) and 300~K (right).
The log-scale for the y-axis is used to allow comparison of 
Normal and Umklapp contributions. Units cm$^{-1}$.
%in the low-$\vec{q}$ limit.
}\label{1damp_10K}
\end{figure*}

%{\it Phonon linewidths}.
In Fig.~\ref{damp1} we show the total phonon linewidths $\Gamma(\vec{q},\lambda)$
obtained for the three acoustical modes $\lambda$=ZA, TA and LA, at different temperatures as indicated.
Notice the different scales that have been adopted on the $\Gamma$-axes.
This behaviour is a result of the larger possibilities for damping, satisfying conservation laws,  
available for LA, and then, subsequently for TA and ZA.  
Similar to what happens in normal 3D metals, such as Cu, Ag and Au, 
peaked structures located at intermediate values of $\vec{q}$ are present~\cite{fultz1, fultz2}. 
As we show below, they are associated with 
different active scattering channels.
%There is also an entire hierarchy 
%$\Gamma(\vec{q},$~L$)>\Gamma(\vec{q},$~T$)>\Gamma(\vec{q},$~Z$)$
%which is due to the higher magnitude of LA and, subsequently, TA 
%phonon frequencies as compared to ZA what increases the  
%scattering possibilities satisfying with the energy conservation. 
%
With the raise of temperature, higher phonon occupations 
in Eq.~(\ref{eq28-last}), 
produce an increase in the phonon linewidths
due to the thermal activation of additional scattering processes for every value of $\vec{q}$. 

%Normnal - Umklapp
The total linewidths, together with its contributions from Normal and Umklapp 
processes for $T$=1 and 300~K, are displayed in Fig.~\ref{1damp_10K}. 
%Here we adopted a log-scale that permits a better comparison.
%
%Notice that the Normal processes dominate the scattering in 
%the vicinity of the $\Gamma$-point and 
Umklapp processes become comparable to Normal processes only near the 
border of the BZ, close to the K- and the M-point~\cite{balandin1}.
%, what 
%is expected since they require the participation of a phonon located 
%outside the BZ which is less probable for small values of $\vec{q}$. 
%
For the flexural phonon mode at $T$=1~K  $\Gamma(\vec{q},$~Z$)$  
is null ($<10^{-7}$) irrespective of the value of $\vec{q}$. 
Then at 300~K, $\Gamma(\vec{q},$~Z$)$ is non zero in a large region of the 
$\Gamma$-K-M-$\Gamma$ path but it still vanishes in the limit $q\rightarrow 0$. 
On the other hand, in-plane mode linewidths, $\Gamma(\vec{q},$~T$)$ and $\Gamma(\vec{q},$~L$)$,
experience larger changes with $T$ in the long wavelength regime.
In this limit, they both vary about two orders of magnitude with increasing $T$ from 1 to 300~K 
(from $\sim 10^{-3}$ to $\sim 10^{-1}$) and, as we discussed before, 
different behaviors for low- and high-$T$ can be identified. 
%The long wavelength regime will be analyzed in further detail below.

\begin{figure*}[h]
\vspace{0.10cm}
\includegraphics[trim=0.5cm 1.1cm 4.35cm 0.9cm, clip, width=0.49\textwidth]{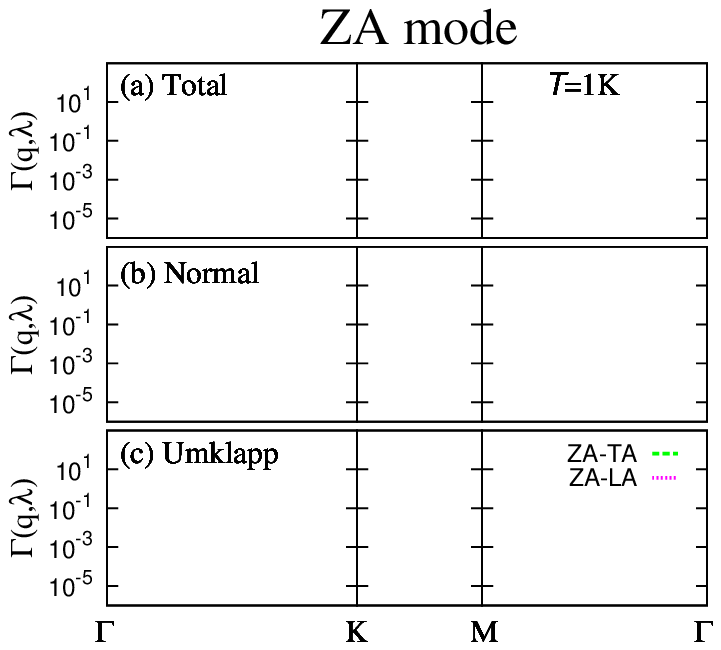}
\hspace{0.10cm}
\includegraphics[trim=0.5cm 1.1cm 4.35cm 0.9cm, clip, width=0.49\textwidth]{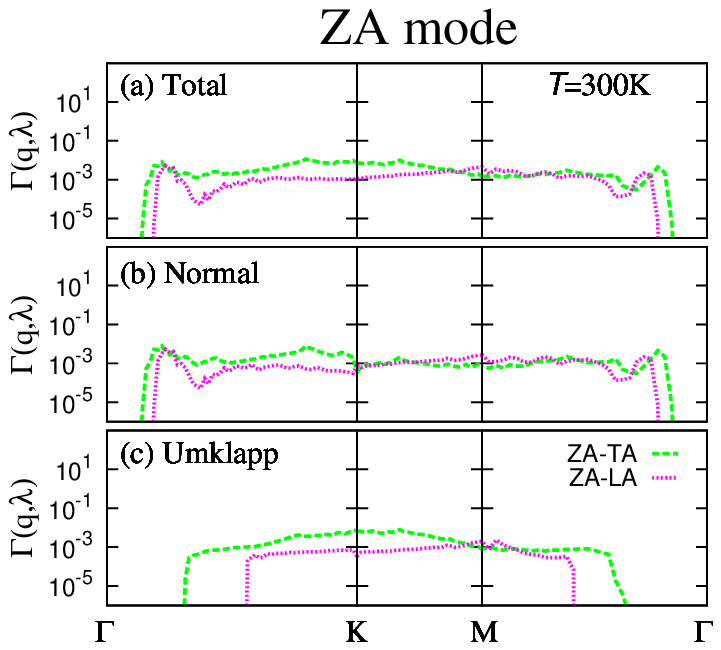}
\vspace{0.30cm}

\includegraphics[trim=0.5cm 1.1cm 4.35cm 0.9cm, clip, width=0.49\textwidth]{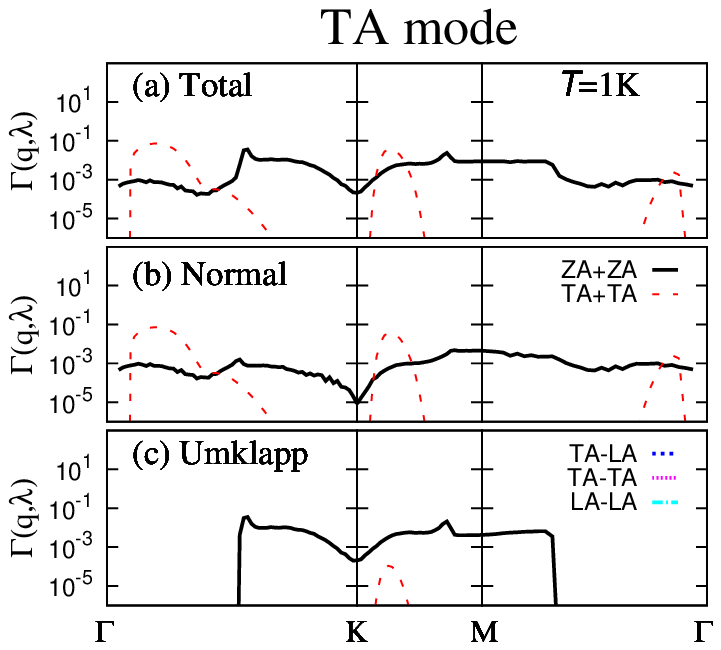}
\hspace{0.10cm}
\includegraphics[trim=0.5cm 1.1cm 4.35cm 0.9cm, clip, width=0.49\textwidth]{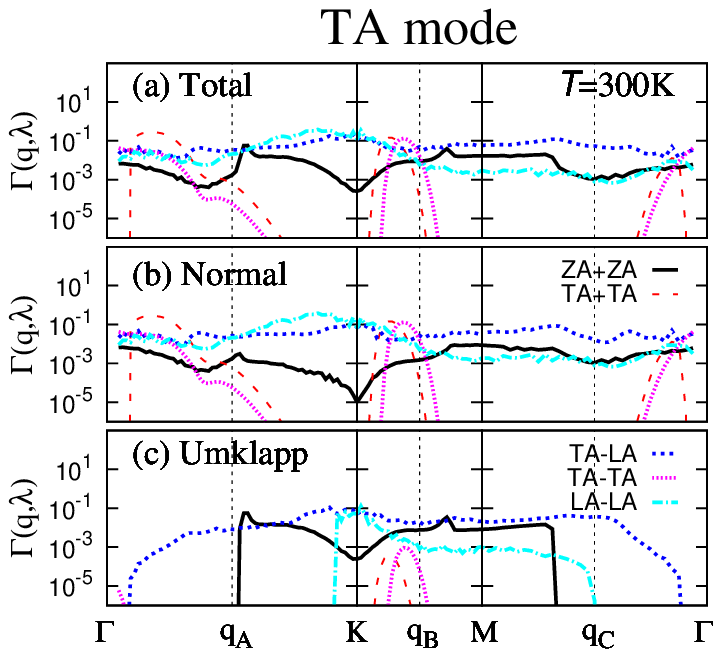}
\vspace{0.30cm}

\includegraphics[trim=0.5cm 1.1cm 4.35cm 0.9cm, clip, width=0.49\textwidth]{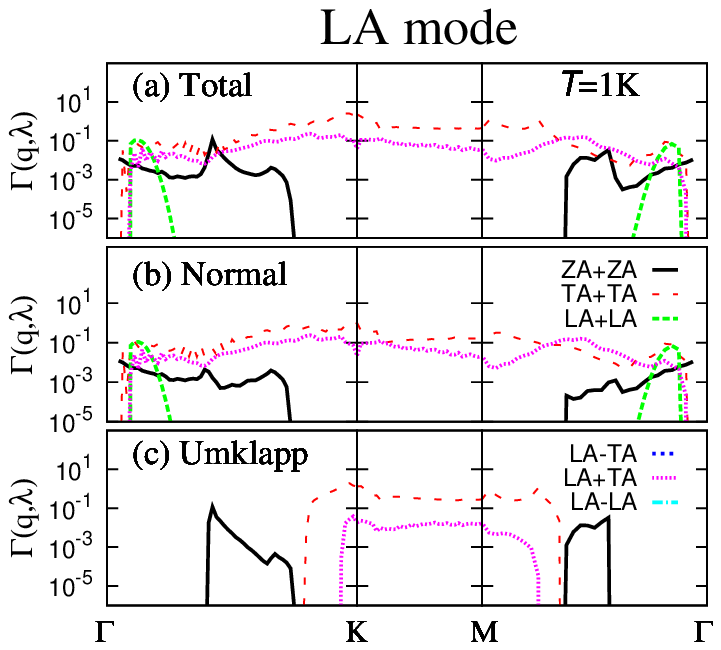}
\hspace{0.10cm}
\includegraphics[trim=0.5cm 1.1cm 4.35cm 0.9cm, clip, width=0.49\textwidth]{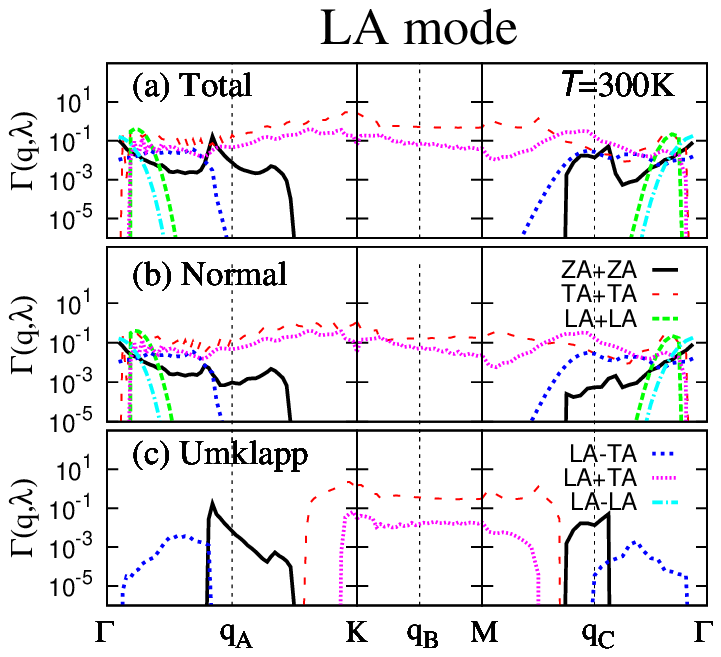}
\caption{(Color online)
Contribution of the different scattering channels to the phonon linewidths 
at $T$=1~K (left) and 300~K (right).
The total linewidth (a) is composed of Normal (b) 
and Umklapp (c) processes. Units cm$^{-1}$.
}\label{channels_1}
\end{figure*}

\begin{figure*}[th]
\vspace{0.25cm}
\includegraphics[trim=0.5cm 0.8cm 1.55cm 0.9cm, clip, width=0.49\textwidth]{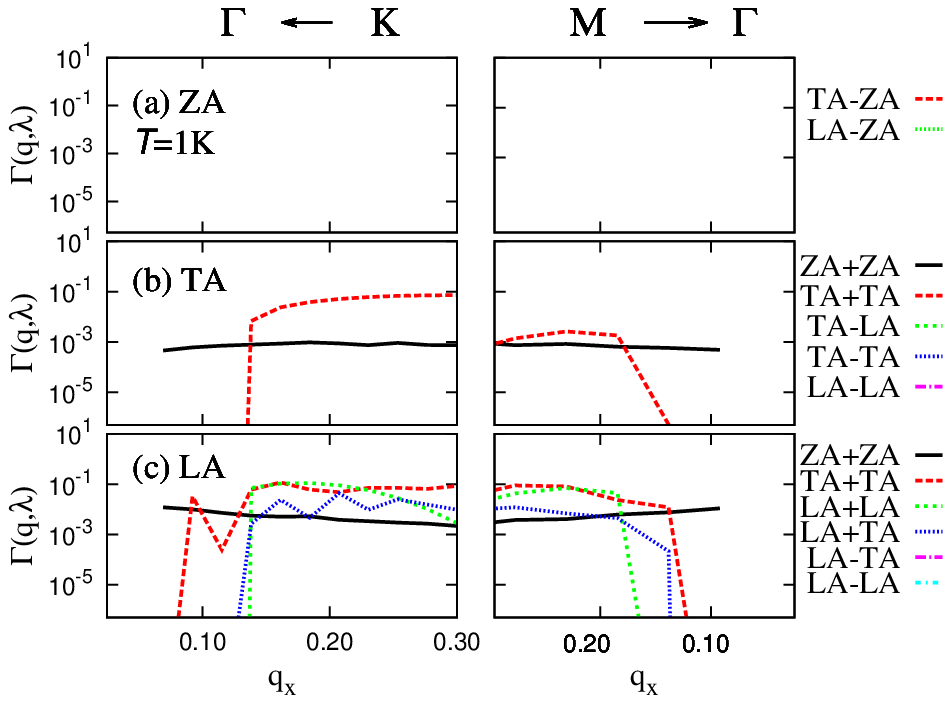}
\hspace{0.05cm}
\includegraphics[trim=0.5cm 0.8cm 1.55cm 0.9cm, clip, width=0.49\textwidth]{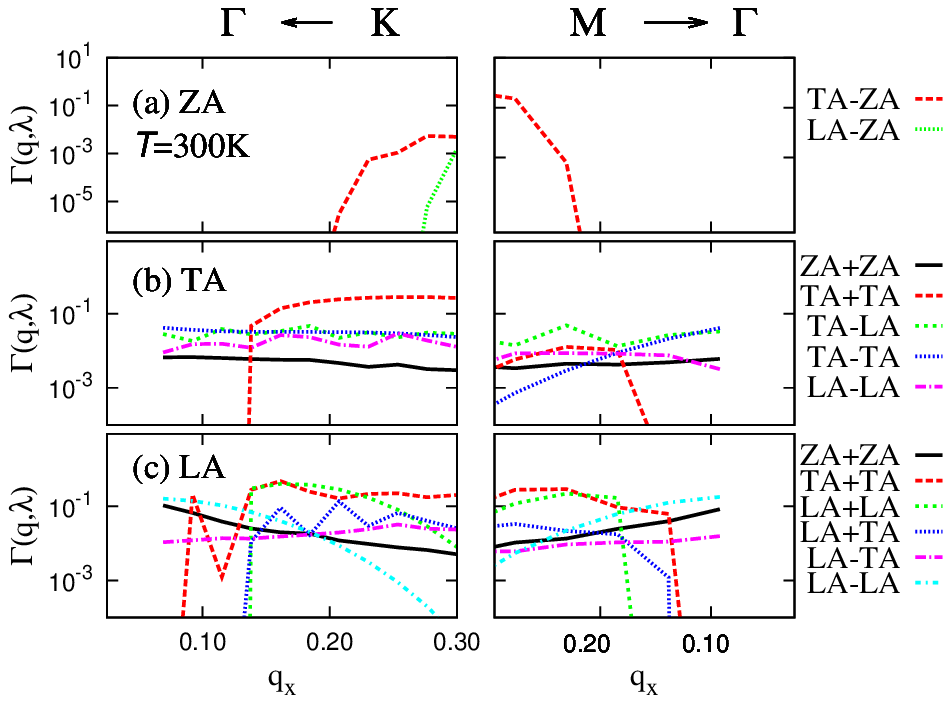}
\caption{(Color online) Contribution of the different 
scattering channels to the phonon linewidths at 1~K (left panels) and 300~K 
(right panels) in the long wavelength regime. Units cm$^{-1}$
($q_x$ (x-axis) is given in units of $\AA^{-1}$).
}\label{lowq-partial}
\end{figure*}

\begin{figure*}[th]
\vspace{0.25cm}
\includegraphics[trim=0.5cm 0.1cm 0.40cm 0.1cm, clip, width=0.32\textwidth]{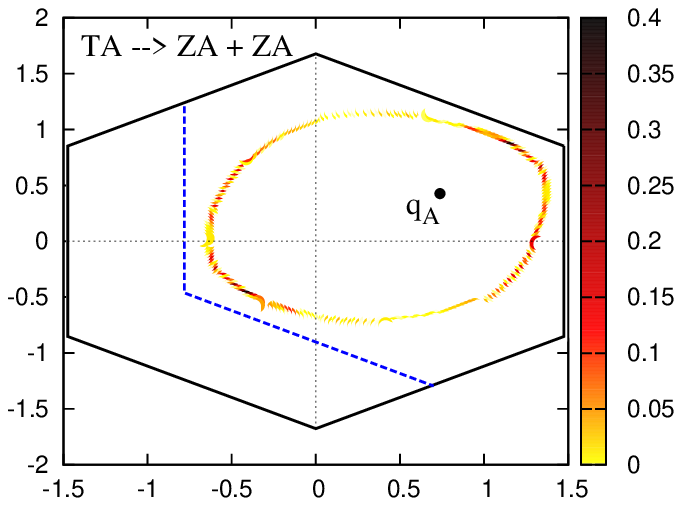}
\hspace{0.1cm}
\includegraphics[trim=0.5cm 0.1cm 0.40cm 0.1cm, clip, width=0.32\textwidth]{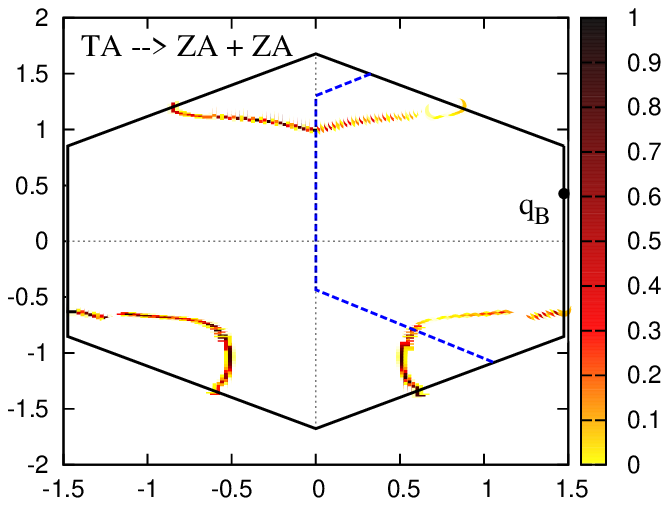}
\hspace{0.1cm}
\includegraphics[trim=0.5cm 0.1cm 0.40cm 0.1cm, clip, width=0.32\textwidth]{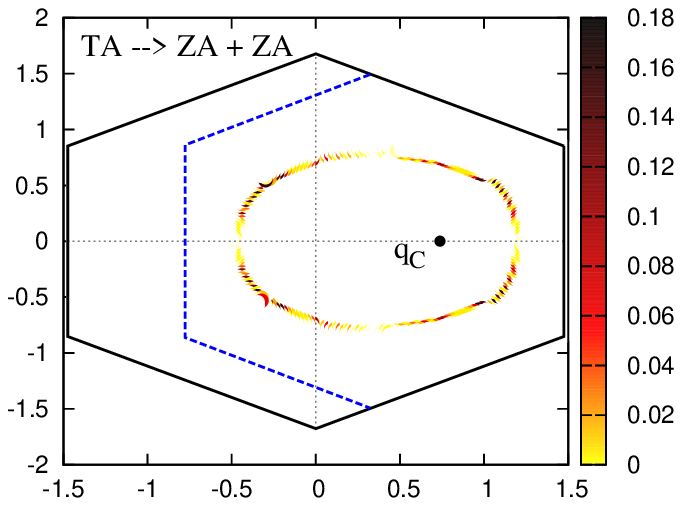}

\includegraphics[trim=0.5cm 0.1cm 0.40cm 0.1cm, clip, width=0.32\textwidth]{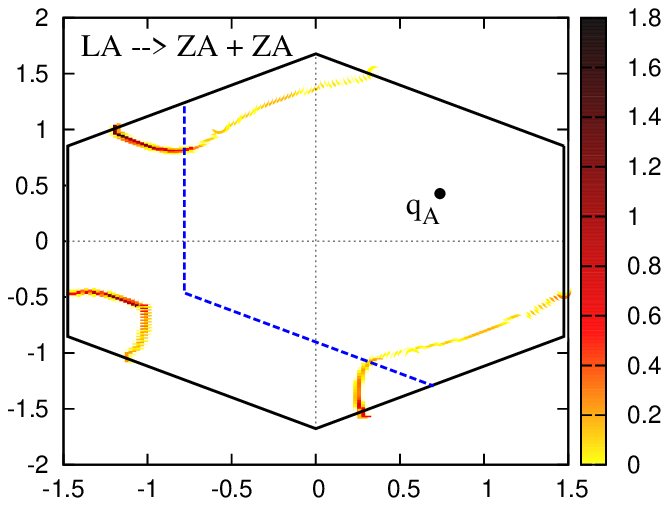}
\hspace{0.1cm}
\includegraphics[trim=0.5cm 0.1cm 0.40cm 0.1cm, clip, width=0.32\textwidth]{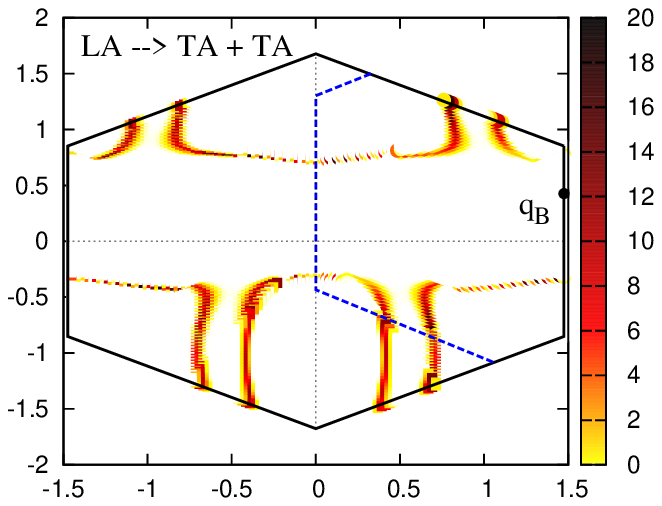}
\hspace{0.1cm}
\includegraphics[trim=0.5cm 0.1cm 0.40cm 0.1cm, clip, width=0.32\textwidth]{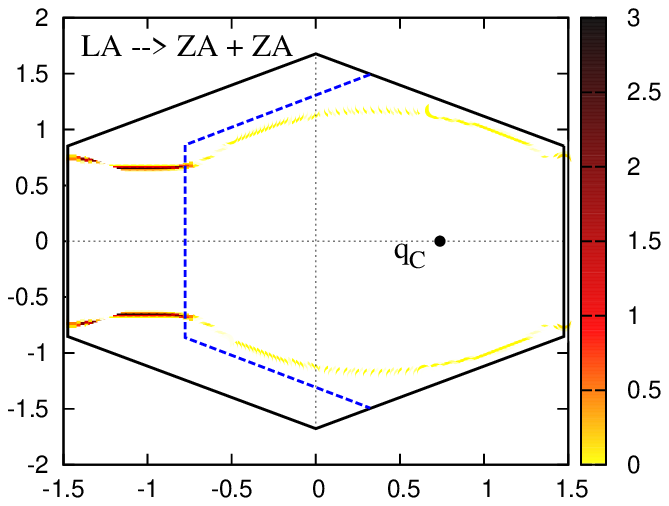}
\caption{(Color online)
Contour plots of the contribution of selected scattering channels inside the BZ 
for the TA and LA acoustic phonon modes of graphene at $\vec{q}_A, \vec{q}_B$ and $\vec{q}_C$ 
(i. e. the middle points  of $\Gamma-$K$-$M$-\Gamma$). 
Units cm$^{-1}$ ($q_x$ (x-axis) and $q_y$ (y-axis) are given in units of $\AA^{-1}$). 
}\label{damp_ZB-1}
\end{figure*}

%{\it Damping pathways}.
The relative importance of the scattering mechanisms is obtained 
by considering separately each of the summands in Eq.~(\ref{eq28-last}).
The dependence with the wave vector $\vec{q}$ for  
all of the active scattering channels at $T$=1 and 300~K are shown in Fig.~\ref{channels_1}. 
Partial contributions to the linewidth of the flexural mode ZA are finite 
in a large sector of the $\Gamma$-K-M-$\Gamma$ path only at 300~K 
(the case $T$=1~K is included to emphasize that $\Gamma(\vec{q}, $ Z$)$ is null).
Here the main important active channels are ZA$\leftrightarrow$ZA-TA and ZA-LA (top right panels).
%with the process involving a TA phonon slightly larger.
% 
It has been reported that processes where ZA phonons are present 
can exist only if two of them participate simultaneously in the scattering.
This result, which has been referred as a selection rule~\cite{lindsay1,lindsay2}, 
is verified by our findings. 

The scattering of in-plane phonons presents more complicated characteristics than 
that of the flexural phonons discussed above. In this case 
a larger number of channels are already open at low-$T$. 
For the TA mode at $T$=1~K, the active scattering channels are 
TA$\leftrightarrow$ZA+ZA and TA+TA (middle left).
%and its relative importance depends on $\vec{q}$ (top right plot). 
%This is mainly because of the conservation of energy ($\delta$-function in Eq.~(\ref{eq28-last})) 
%which is satisfied by a mayority of ZA phonons due to their lower frequencies.
Then, at 300~K, TA$\leftrightarrow$TA-LA, TA-TA and LA-LA become also activated by $T$
and the relative dominance of each them depends strongly on 
$\vec{q}$ in a non trivial way (middle right).
For the LA mode the picture is more complex. 
Due to its larger frequency, already at low-$T$,  
LA$\leftrightarrow$ZA+ZA, TA+TA, LA+LA and LA+TA are all active (bottom left). 
Then, at 300~K additionally LA$\leftrightarrow$LA-LA becomes important, 
particularly in the limit $q\rightarrow 0$ (bottom right).
%particularly in the long wavelength regime (bottom right).
 
The behaviour of the scattering channels in the long wavelength regime 
is analyzed in more detail in Fig.~\ref{lowq-partial}. 
In agreement with the analytical description, we observe that 
at low-$T$ ($T$=1~K) the dominant scattering processes 
are TA$\leftrightarrow$ZA+ZA and LA$\leftrightarrow$ZA+ZA (left panels). 
At 300~K however, TA$\leftrightarrow$TA-TA and LA$\leftrightarrow$LA-LA
become more important (right panels). 
Notice that while in the analytical approach presented before we were able to study 
larger system sizes ($l=10^4\ a$) with the current numerical treatment 
we are limited to $l\approx 80\ a$. Therefore, in the long wavelength regime 
the results shown in Fig.~\ref{fig5} for the ZA+ZA scattering channel of in-plane phonon modes
are complementary to those in Fig.~\ref{lowq-partial}.
We remark the good agreement obtained between both independent calculations
(see the values of $q$ in the $x$-axis of both figures).

\begin{figure*}[t]
\vspace{0.25cm}
\includegraphics[trim=0.5cm 1.1cm 4.25cm 1.5cm, clip, width=0.48\textwidth]{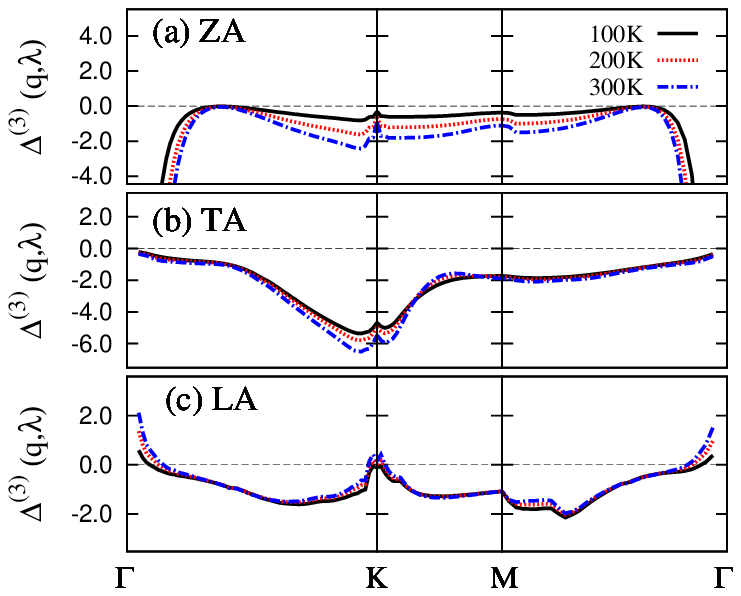}
\hspace{0.1cm}
\includegraphics[trim=0.5cm 1.1cm 4.25cm 0.9cm, clip, width=0.48\textwidth]{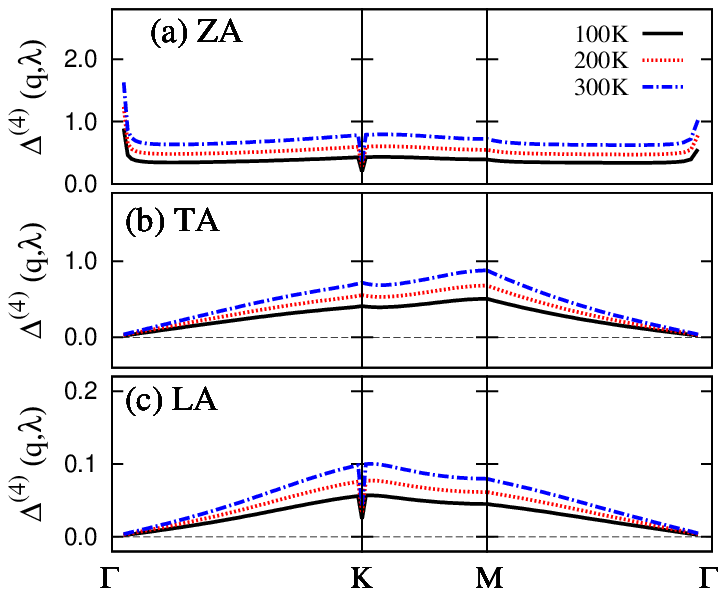}
\caption{(Color online)
Temperature dependence of the 3rd- (left panel) and 4th-order (right panel) bandshifts for the 
(a) ZA, (b) TA and (c) LA acoustic phonon modes of graphene. 
Units cm$^{-1}$.
}\label{shift_10K}
\end{figure*}

Further interesting insights on the features of the anharmonic scattering 
for the in-plane phonon modes are presented in Fig.~\ref{damp_ZB-1}. 
Here we show the spatial distribution inside the BZ for a few selected relevant 
scattering channels with $\vec{q}$ equal to $\vec{q}_A$, $\vec{q}_B$ and $\vec{q}_C$
(middle points of $\Gamma-$K$-$M$-\Gamma$) at 300~K. 
These curves can be interpreted as follows.
The plot displayed on the top left panel describes the processes where an excited TA phonon, 
with fixed momentum $\vec{q}=\vec{q}_A$, is scattered 
through every possible $\vec{q}_2$ by the channel ZA+ZA 
(remember that $\vec{q}_3=-\vec{q}_A-\vec{q}_2\pm\vec{G}$).
Notice that due to the conservation of energy and momentum only a reduced number of processes
are active. The resulting circular shaped line (around $\vec{q}_A$) indicates 
that ZA phonons propagating in every possible direction 
will be generated from the original unperturbed in-plane TA $\vec{q}_A$ phonon.
The blue-dashed line separates Normal (region enclosing $\vec{q}_A$) 
from Umklapp processes.
Here the absence of Umklapp contributions is consistent 
with the result of Fig.~\ref{channels_1} (middle-right panel) 
where the Umklapp TA$\leftrightarrow$ZA+ZA process vanishes 
for values of $\vec{q}$ right above $\vec{q}_A$.
  
Similar analysis can be performed for the remaining panels of Fig.~\ref{damp_ZB-1}.
Observe in particular that for $\vec{q}=\vec{q}_B$, 
with $\vec{q}_B$ located at the border of the BZ,
Umklapp processes become more important. Additional similar plots 
are presented in Appendix B.
These results are relevant for the microscopical understanding 
of the thermal conductivity since they determine to which extent 
any excited phonon, in any given initial direction, will be scattered by phonons  
propagating in any other direction. 
%Here, only Umklapp processes account  
%for the momentum transfer required for the effective thermal conductivity~\cite{book,maradudin}. 
%The study of the thermal conductivity is however beyond the scope of the present work.

%\begin{figure}[t]
%\vspace{0.25cm}
%\includegraphics[width=0.48\textwidth]{figs/low-q-limit.eps}
%\includegraphics[width=0.45\textwidth]{figs/proc_all_sum_each_final.eps}
%\label{low-q-temp}
%\end{figure}

%%%% SHIFTS %%%%%%%%%%%%%
Finally, we study the 3rd- and 4th-order~\cite{haro,seol} frequency shifts which 
%again using the same notation as before, 
can be calculated through:
\begin{widetext}
\begin{align}
%\begin{equation}
\Delta^{(3)}(\vec{q},\lambda)=\frac{\hbar}{2N}
%\Sigma(\vec{q},\lambda;z)&=\hbar \omega(\vec{q},\lambda) 
\sum_{\vec{q}_2\vec{q}_3}
\sum_{\lambda_2 \lambda_3}
\left|\Phi^{(3)}\binom{\lambda \lambda_2 \lambda_3}
{-\vec{q}\vec{q}_2\vec{q}_3}\right|^2 
&P
\bigg\lbrace
\frac{1+n(\vec{q}_2,\lambda_2)+n(\vec{q}_3,\lambda_3)}
{\omega(\vec{q},\lambda)-\omega(\vec{q}_2,\lambda_2)-\omega(\vec{q}_3,\lambda_3)}-
\frac{1+n(\vec{q}_2,\lambda_2)+n(\vec{q}_3,\lambda_3)}
{\omega(\vec{q},\lambda)+\omega(\vec{q}_2,\lambda_2)+\omega(\vec{q}_3,\lambda_3)} \nonumber \\
&\qquad {} 
+\frac{2[n(\vec{q}_2,\lambda_2)-n(\vec{q}_3,\lambda_3)]}
{\omega(\vec{q},\lambda)+\omega(\vec{q}_2,\lambda_2)-\omega(\vec{q}_3,\lambda_3)}
%
%\frac{n(\vec{q}_2,\lambda_2)-n(\vec{q}_3,\lambda_3)}
%{z-\omega(\vec{q}_2,\lambda_2)+\omega(\vec{q}_3,\lambda_3)}
%
\bigg\rbrace,
\label{eq26}
\end{align}
\noindent 
\begin{equation}
\Delta^{(4)}(\vec{q},\lambda)=\frac{\hbar}{2N}
\sum_{\vec{q}_1,\lambda_1}
\Phi^{(4)}\binom{\lambda \lambda_1 \lambda_1 \lambda}
{-\vec{q}\vec{q}_1-\vec{q}_1\vec{q}} 
\Big[
%1+2n\binom{\lambda_1}{\vec{q}_1}
1+2n(\vec{q}_1,\lambda_1)
\Big].
\label{eq27}
\end{equation}
\end{widetext}
In this case the principal part $P$ was represented as 
$P\lbrace1/\omega\rbrace=\omega/(\omega^2+\eta^2)$ 
with $\eta=1$~cm$^{-1}$ (Appendix B).  
%An analysis the dependence of $\Gamma(\vec{q},\lambda)$ as a function of $\xi$ is presented in

The obtained results are displayed in Fig.~\ref{shift_10K}. 
%
%Once more, they coincide with the analytical predictions.
In agreement with the analytical prediction,
the 3rd-order frequency shift of the flexural ZA mode is negative
and experiences a logarithmic divergence in the long wavelength regime.
In this limit, for the in-plane modes on the contrary,
$\Delta^{(3)}(\vec{q}, $T$)$ tends to zero with almost 
no appreciable variation versus $T$ and $\Delta^{(3)}(\vec{q},$~L$)$ 
is negligibly small at low-$T$ but it increases smoothly with $T$.

% 
%ere, the negative 3rd-order shift which is stabilized by the 4th-order processes (See below). 
%

The strong 3rd-order negative bandshift $\Delta^{(3)}(\vec{q},$~Z$)$ near $\Gamma$ 
means, in principle, 
that another configuration different from a flat sample, will be more stable.
As we show in Fig.~\ref{shift_10K} (right), 4th-order anharmonic interactions
counterbalances the 3rd-order bandshift, at $\vec{q}\rightarrow 0$, for the ZA
mode and stabilize the layer.
Other characteristics in this regime are the linear dependence with $\vec{q}$
of $\Delta^{(4)}(\vec{q}, $T$)$ and $\Delta^{(4)}(\vec{q}, $L$)$ which are 
in agreement with the conclusions derived in Eq.~(\ref{corri4}).

%%%%%%%%%%%%%%%%%%%%%%%%%%%%%%%%%%%%%%%%%%%%%%%%%%%%%
\section{Conclusions}

We have studied thermal expansion and phonon broadenings and lineshifts of non-ionic 2D 
crystals by means of anharmonic lattice dynamics, thereby implementing analytical 
and numerical methods.
We have used a semi-empirical model for the interatomic force constants, comprising in-plane
acoustic phonons and out-of-plane flexural modes.

Analytical techniques allow to investigate the long wavelength limit which is 
beyond the reach of numerical calculations and ab-initio methods.
Since the out-of-plane Gr\"uneisen constant $\gamma$(Z) diverges logarithmically in the 
$\vec{q}\rightarrow$0 limit, we have considered crystals of finite size and 
investigated finite size effects.
We have taken graphene as an example for quantitative evaluations. 
Thereby we have found a change of sign of the thermal expansion as a 
function of crystal size.
%In particular we have studied the change of sign of the thermal expansion as a 
%function of crystal size. 
A change of sign has been found earlier by 
Monte Carlo calculations~\cite{faso3}, there the finite size of the system is 
inherent in the method.

We have investigated analytically the wave vector dependence of the decays
and lineshifts of in-plane and out-of-plane phonons in the classical (high $T$)
and in the quantum regime. In the classical regime we confirm earlier results 
about the decay rates of an in-plane phonon into two flexural modes and of the inverse 
absorption process~\cite{nano-marzari}.
In the quantum regime the wave vector dependence and the $T$ dependence of these processes 
are different.
We have studied the lineshifts (equivalently the real part of the self energy)
due to third and fourth order anharmonicities. 
In the classical regime the real part of the self energy of the flexural mode in a 
third order absorption process is found to be negative, proportional to $T$
and $q^2$, and diverging logarithmically with the size of the system.
In the quantum regime it vanishes exponentially with decreasing $T$. 
Self energy corrections of the flexural mode due to fourth order anharmonic
processes are positive and proportional to $q^2$ in the classical as well as in the 
quantum regime.
Similar results have been obtained for crystalline membranes 
in the classical~\cite{Nelson-Peliti} and in the quantum~\cite{amorin} regimes.
As a consequence of the corresponding change of the dispersion of the flexural mode at long
wavelengths~\cite{Mariani, amorin}, we find that the temperature $T_{\alpha}$  of the change 
from negative to positive thermal expansion is lowered 
and close to 300~K for systems of macroscopic size.

The numerical analysis (Sect. VII and Appendix B) is complementary to the analytical treatment 
of Sect. VI. Special care has been devoted to obtain a dense grid of $\vec{q}$ points 
covering the BZ. 
Various scattering channels that contribute to decays 
and lineshifts of the in-plane and out-of-plane modes have been investigated in a systematic way.
In the nearest numerically attainable neighborhood of the $\Gamma$-point we find 
agreement between analytical and numerical results.
In addition, beyond the long wavelength regime, we have studied 
Umklapp processes and compared their contributions with Normal processes
for various scattering channels. 
We find that in a broad $T$ range below room temperature the decay rate of flexural 
modes is much less affected by Umklapp processes than the decay rate of in-plane modes.
This result supports earlier theoretical conclusions that flexural modes are responsible
for an anomalous large intrinsic thermal 
conductivity~\cite{nano-marzari,broido1,fisher1,lorenzo,fisher2}.

\section*{Acknowledgments}
\label{agradecimientos} 
We thank B. Verberck, D. Lamoen and A. Dobry for useful comments. 
We acknowledge funding from the FWO~(Belgium)-MINCyT~(Argentina) collaborative research  
project. This work is supported by the Euro GRAPHENE project CONGRAN.

%%%%%%%%%%%%%%%%%%%%%%%%%%%%%%%%%%%%%%%%%%%%%%%%%%%%%%%%%%%
\section*{Appendix A}
\numberwithin{equation}{section}
\numberwithin{figure}{section}
\renewcommand{\theequation}{A-\arabic{equation}}
\renewcommand{\thefigure}{A-\arabic{equation}}
% redefine the command that creates the equation no.
\setcounter{equation}{0}  % reset counter 
\setcounter{figure}{0}  % reset counter 
%\section*{Appendix A} 

We calculate the change  
of the phonon frequency $\omega(\vec{q},\lambda)$ under homogeneous strains
$\epsilon_{ij}$ ($i,j\in\{1,2\}$) in two dimensions for a non primitive crystal. 
Homogeneous strains are related to the center of mass displacement of the unit cell $\vec{n}$ by
\begin{equation}
s_i(\vec{n})=\sum_j \epsilon_{ij} X^s_j(\vec{n}),
\end{equation}
where the center of mass equilibrium position reads
\begin{equation}
\vec{X}^s(\vec{n})=\sum_{\kappa}\frac{M_{\kappa}}{M}X_j(\vec{n}\kappa),
\end{equation}
and where $M=\sum_{\kappa}M_{\kappa}$ is the total mass per unit cell. 
Hence
\begin{equation}
s_i(\vec{n})=\sum_{\kappa} u_i(\vec{n}\kappa),
\end{equation}
with
\begin{equation}
u_i(\vec{n}\kappa)=\sum_j \epsilon_{ij} \frac{M_{\kappa}}{M}X_j(\vec{n}\kappa).
\label{B4}
\end{equation}
We need to calculate 
\begin{equation}
\delta\omega(\vec{q},\lambda)=\sum_i \frac{\partial{\omega(\vec{q},\lambda)}}{\partial{X_i(\vec{n}\kappa)}} u_i(\vec{n}\kappa),
\end{equation}
where $u_i(\vec{n}\kappa)$ is given by Eq.~(\ref{B4}).
Starting from Eq.~(\ref{eq9}) %and (\ref{eq11}) 
we obtain 
\begin{eqnarray}
\delta\omega(\vec{q},\lambda)=\frac{1}{2\omega(\vec{q},\lambda)}\sum_{kl}\sum_{\kappa_1\kappa_2}
\frac{e^{\kappa_1*}_k(\vec{q},\lambda)e^{\kappa_2}_l(\vec{q},\lambda)}
{\sqrt{M_{\kappa_1}M_{\kappa_2}}} 
\nonumber\\
\times \sum_{\vec{h}_2} \delta \Phi^{(2)}_{kl}(\vec{0}\kappa_1;\vec{h}\kappa_2)
e^{i\vec{q}[\vec{X}(\vec{h}\kappa_2)-\vec{X}(\vec{0}\kappa_1)]}.
\label{B6}
\end{eqnarray}
The change of the second order coupling parameter due to the displacements $u_i(\vec{n}\kappa)$ reads
\begin{equation}
\delta\Phi^{(2)}_{kl}(\vec{0}\kappa_1;\vec{h}\kappa_2)=
\sum_{\vec{n} \kappa i} \Phi^{(3)}_{kli}(\vec{0}\kappa_1;\vec{h}\kappa_2;\vec{n}\kappa)
u_i(\vec{n}\kappa),
\end{equation}
where $\Phi^{(3)}$ is the third order anharmonic coupling. 

Differentiation of Eq.~(\ref{B6}) with respect to the strains yields
\begin{widetext}
\begin{align}
\frac{\partial{\omega(\vec{q},\lambda)}}{\partial{\epsilon_{ij}}}=
\frac{1}{2\omega(\vec{q},\lambda)}
\sum_{kl}\sum_{\kappa_1\kappa_2\kappa}\sum_{\vec{n}\vec{h}}
\frac{e^{\kappa_1*}_k(\vec{q},\lambda)e^{\kappa_2}_l(\vec{q},\lambda)}
{\sqrt{M_{\kappa_1}M_{\kappa_2}}} 
\Phi^{(3)}_{kli}(\vec{0}\kappa_1;\vec{h}\kappa_2;\vec{n}\kappa)
X_i(\vec{n}\kappa)\frac{M_{\kappa}}{M} 
e^{i\vec{q}[\vec{X}(\vec{h}\kappa_2)-\vec{X}(\vec{0}\kappa_1)]}
\label{last_A}
\end{align}
%\end{widetext}
%Applying this general result to the specific case of graphene we obtain the result of Eq.~(\ref{eq50}).
For the case of central forces we obtain
%\begin{widetext}
\begin{align}
\frac{\partial{\omega(\vec{q},\lambda)}}{\partial{\epsilon_{ii}}}=
-\frac{1}{4\omega(\vec{q},\lambda)M_C}
\sum_{kl}\sum_{\kappa\kappa'}\sum_{\vec{n}}&
\varphi^{(3)}_{kli}(\vec{0}\kappa;\vec{n}\kappa')
\bigg[X_i(\vec{n}\kappa')-X_i(\vec{0}\kappa)\bigg]\nonumber\\
&\times 
\bigg[e^{\kappa*}_k(\vec{q},\lambda)e^{\kappa'}_l(\vec{q},\lambda)-
e^{\kappa'*}_k(\vec{q},\lambda)e^{\kappa}_l(\vec{q},\lambda)
e^{i\vec{q}.[\vec{X}(\vec{n}\kappa')-\vec{X}(\vec{0}\kappa)]
}\bigg]
\label{eq50}
\end{align}
\end{widetext}
Using $\varphi^{(3)}_{kli}(\vec{0}\kappa;\vec{n}\kappa')=
-\varphi^{(3)}_{kli}(\vec{n}\kappa';\vec{0}\kappa)$ we see that the right hand side of (\ref{eq50})
is symmetric with respect to an interchange of atoms $(\vec{n}\kappa')\leftrightarrow (\vec{0}\kappa)$.
In case of graphene, acoustic modes satisfy $e_k^A(\vec{q},\lambda)=e_k^B(\vec{q},\lambda)$.

%%%Taking into account the symmetry of the crystal and using
%%%$\varphi^{(3)}_{kli}(\vec{0}\kappa;\vec{n}\kappa')=
%%%-\varphi^{(3)}_{kli}(\vec{n}\kappa';\vec{0}\kappa)$, we find that the right 
%%%hand side of Eq.~(\ref{eq50}) is symmetric respect to an interchange of atoms  
%%%$(\vec{n}\kappa')\leftrightarrow (\vec{0}\kappa)$.

%%%%%%%%%%%%%%%%%%%%%%%%%%%%%%%%%%%%%%%%%%%%%%%%%%%%%
\section*{Appendix B} 
\numberwithin{equation}{section}
\numberwithin{figure}{section}
\numberwithin{table}{section}
\renewcommand{\theequation}{B-\arabic{equation}}
\renewcommand{\thefigure}{B-\arabic{figure}}
\renewcommand{\thetable}{B-\arabic{table}}
% redefine the command that creates the equation no.
\setcounter{equation}{0}  % reset counter 
\setcounter{figure}{0}  % reset counter 
\setcounter{table}{0}  % reset counter 

\begin{figure*}[t]
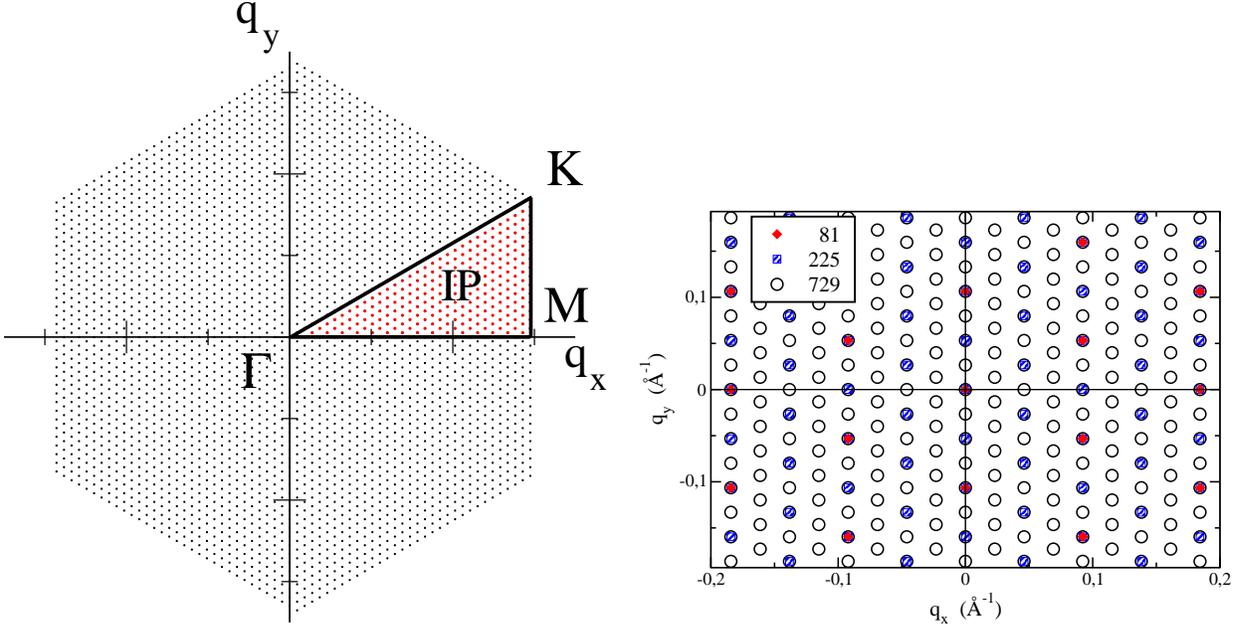

\vspace{0.25cm}
\includegraphics[width=0.45\textwidth]{1BZ.eps}
\hspace{0.2cm}
\includegraphics[width=0.44\textwidth]{grid_comp.eps}
\caption{
Discrete mesh with $225$ $\vec{q}$-points inside the irreducible part (IP) of the 1BZ (left).
Comparison of meshes in the vicinity of the $\Gamma$-point (right). 
}\label{grid}
\end{figure*}

Anharmonic force constants 
$\Phi^{(3)}\binom{\lambda_1 \lambda_2 \lambda_3}{\vec{q}_1\vec{q}_2\vec{q}_3}$ (Eq.~(\ref{phi3}))
and $\Phi^{(4)}\binom{\lambda_1 \lambda_2 \lambda_3 \lambda_4}{\vec{q}_1\vec{q}_2\vec{q}_3 \vec{q}_4}$
(Eq.~(\ref{phi4})) are defined by Fourier transforms 
given by rather complex summations where each term contains multiple factors
such as the energy $\omega(\vec{q},\lambda)$ and the polarization vector
$e^{\kappa}_i(\vec{q},\lambda)$ evaluated at distinct values of $\vec{q}$ and $\lambda$ simultaneously. 
The first step in the construction of an algorithm 
%for  the numerical calculation 
is therefore to obtain and store 
$\omega(\vec{q},\lambda)$ and $e^{\kappa}_i(\vec{q},\lambda)$ 
for the $\vec{q}$-points which will be included in the calculation.

This first step was done here by using a code developed previously 
for the study of harmonic phonons in graphene reported in Ref.~\onlinecite{Karl}.
%Since the unit cell of graphene is composed by two atoms (A and B) we found that 
%$e^{\kappa}_i(\vec{q},\lambda)$ acquire non-zero imaginary components 
%for values of $\vec{q}$ moving away from $\Gamma$-point
%where they have instead only non-zero real components. 
%
Due to the symmetry of the crystal, numerical diagonalization of 
the dynamical matrix $D(\vec{q})$, carried out
through packages from the LAPACK libraries~\cite{lapack},
is performed only inside the irreducible part (IP) 
of the Brillouin zone (1BZ) (Fig~\ref{grid} (left)).
Further phonon frequencies and polarization vectors, on the remaining $\vec{q}$-points 
in the entire 1BZ, are generated by symmetry operations satisfying 
 the properties $\vec{e}(\vec{q},\lambda)$=$\vec{e}^{*}
(-\vec{q},\lambda)$ and $\omega(\vec{q},\lambda)$=$\omega(-\vec{q},\lambda)$, as
well as the orthonormality and closure conditions:
\begin{subequations}
\begin{align}
%\begin{eqnarray}
\sum_{\kappa i} e_i^{\kappa *}(\vec{q},\lambda)
e_i^{\kappa}(\vec{q},\lambda')=\delta_{\lambda \lambda'}, \\
\sum_{\lambda} e_i^{\kappa *}(\vec{q},\lambda)
e_j^{\kappa'}(\vec{q},\lambda)=\delta_{ij} \delta_{\kappa \kappa'}.  
%\end{eqnarray}
\end{align}
\end{subequations}
Because of the band-crossing as function of the phonon momentum $\vec{q}$,  
the use of an auxiliary algorithm, based on the eigenvector orthogonality,
is needed to maintain the sorting of phonon modes after diagonalization~\cite{sort}.

Momentum conservation in Eq.~(\ref{deltaq}) implies that 
$\vec{q_1}$, $\vec{q_2}$ and $\vec{q_3}$ included in the calculation must satisfy
\begin{equation}
\vec{q_1}+\vec{q}_2+\vec{q}_3=\pm\vec{G},
\label{eq_grid}
\end{equation} 
\noindent 
where $\vec{G}$ is a vector of the reciprocal lattice.
For this purpose we adopted a special finite mesh in such a way that $\vec{q_1}$, 
$\vec{q}_2$ and $\vec{q}_3$, satisfying Eq.~(\ref{eq_grid}), belong to the mesh itself
what reduces considerably the number of diagonalizations 
of the dynamical matrix. 
The mesh is defined such that every $\vec{q}$-point is given by 
\begin{equation}
\vec{q}=n_1\vec{\delta}_1+n_2\vec{\delta}_2
\end{equation}
\noindent where $n_1$ and $n_2$ are integers and $\vec{\delta}_1$ and $\vec{\delta}_2$
are the unit vectors that define the mesh.  
The algorithm was designed following closely the steps described in Ref.~\onlinecite{santoro2} 
where anharmonicities of surface phonons in Al were studied~\cite{santoro1}.  
We refer the reader to that work for further details on the  
construction of the mesh.
  
Convergence of numerical results was verified by comparing results for three different meshes
with 81, 225 and 729 distinct $\vec{q}$-points inside the IP. 
Figure~\ref{grid} (left) displays the case with 225 $\vec{q}$-points. 
A comparison of the different meshes near the $\Gamma$ point is 
displayed in Fig.~\ref{grid} (right).   
Note that smaller meshes are 
subsequently contained in the larger ones. 
Thus, by increasing the mesh we keep the existing $\vec{q}$-points and new ones, 
lying at the intermediate distance between two consecutive wave-vectors, are added.

Contributions from Normal and Umklapp processes are identified in a simple way. 
Given a fixed $\vec{q}_1$, the summation over $\vec{q}_2$ in Eqs.~(\ref{eq28-last}) and (\ref{eq26})
runs over the whole BZ.
For each pair $\vec{q}_1$, $\vec{q}_2$, $\vec{q}_3$  
becomes unambiguously defined by $\vec{q}_3=-\vec{q}_1-\vec{q}_2$. 
Then, if the wave-vector $\vec{q}_3$ lies inside the BZ the process is Normal. 
Otherwise, a non-zero $\pm G$ is used to re-map $\vec{q}_3$ 
to its equivalent $\vec{q}$-point inside the BZ and the process is counted as Umklapp.  

\begin{figure*}[t]
\includegraphics[trim=0.5cm 1.1cm 4.25cm 0.9cm, clip, width=0.49\textwidth]{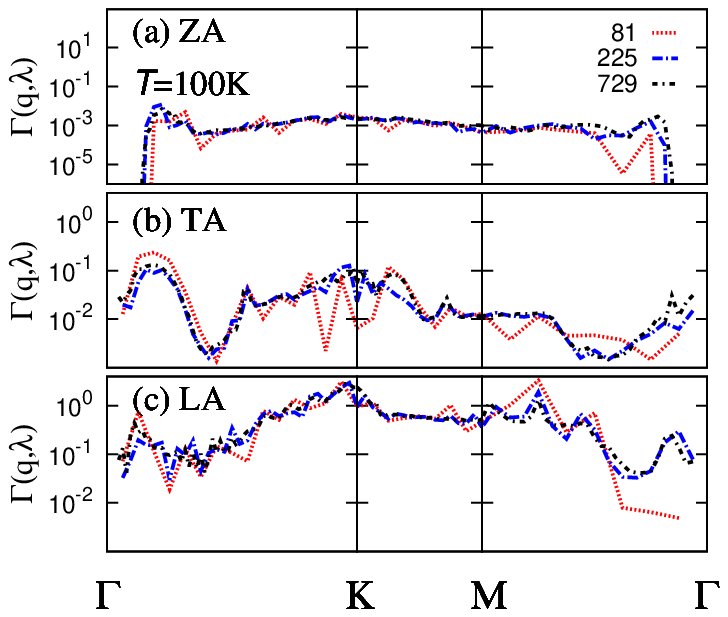}
\hspace{0.10cm}
\includegraphics[trim=0.5cm 1.1cm 4.25cm 0.9cm, clip, width=0.49\textwidth]{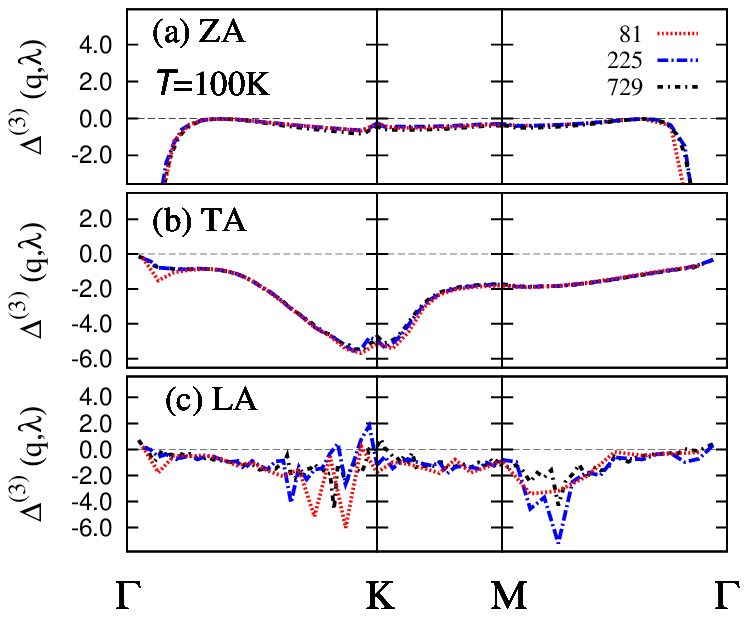}
\caption{(Color online)
Comparison of results for 
the linewidth (left panels) and phonon shift (right panels) for  
different number of $q$-points in the mesh along the BZ. 
}
\label{grids-gamma}
\end{figure*}

\begin{figure*}[t]
\includegraphics[trim=0.5cm 1.1cm 2.55cm 0.9cm, clip, width=0.49\textwidth]{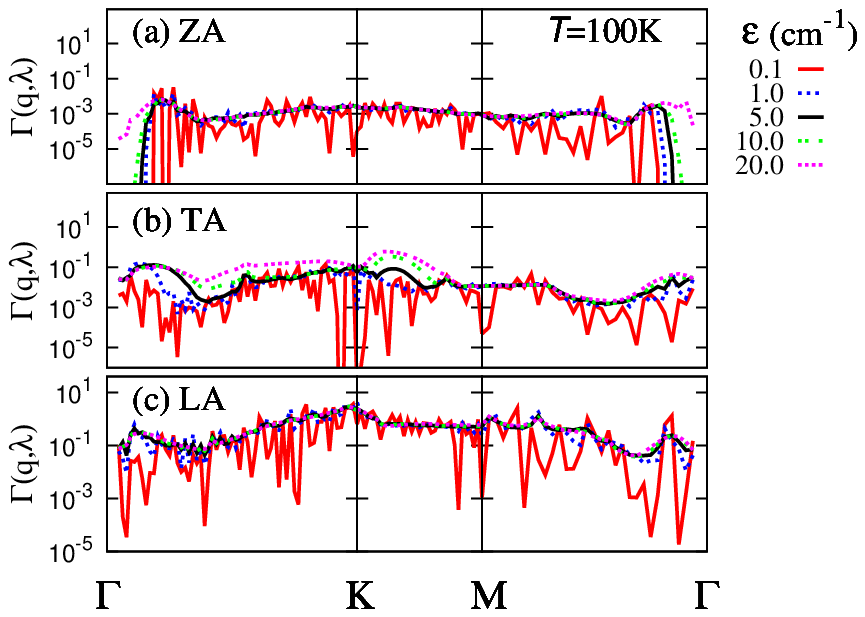}
\hspace{0.10cm}
\includegraphics[trim=0.5cm 1.1cm 2.55cm 0.9cm, clip, width=0.49\textwidth]{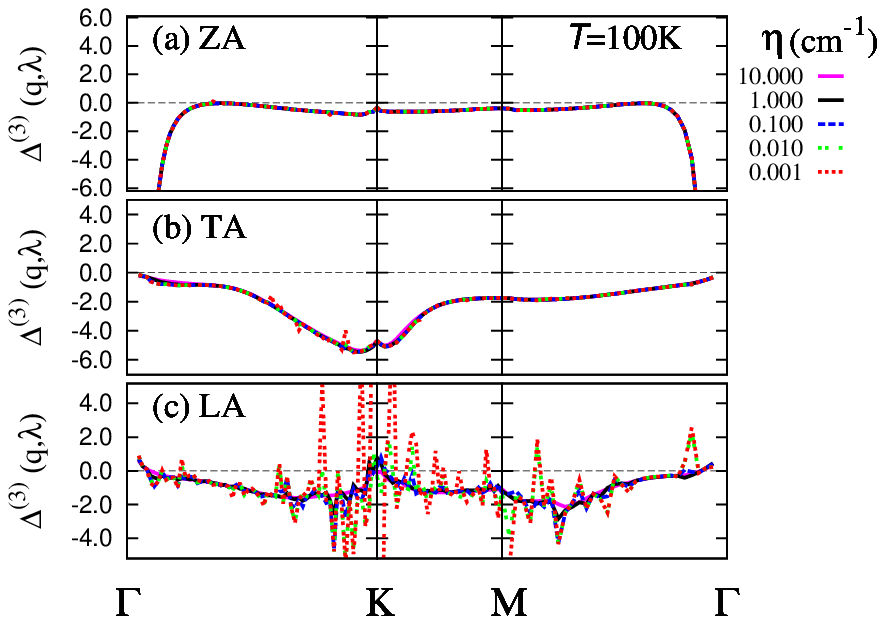}
\caption{(Color online) 
Effects of the auxiliar variables $\eta$ and $\epsilon$ on the 
evaluation of the principal part and the delta function 
in the calculation of the phonon shift (right panels) and linewidths (left panels).
%Mesh with 729 $\vec{q}$-points. T=100~K. 
}\label{eta_comp}
\end{figure*}

The dependence of the numerical results for the 3rd-order linewidths $\Gamma(\omega,\vec{q})$
and bandshifths $\Delta(\omega,\vec{q})$ with the number of $\vec{q}$-points 
of the mesh at T=100~K is analyzed in Fig.~\ref{grids-gamma}. 
The overall agreement as function of the phonon momentum is evident. 
Similar results were found also for other temperatures. 
The total number of individual scattering processes, for every possible scattering channel,
considered in the whole path $\Gamma$-K-M-$\Gamma$ for each mesh is shown in Table B-1.
%considered in the whole path $\Gamma$-K-M-$\Gamma$ for each mesh is shown in Table~\ref{table2_B}.

{\setlength{\tabcolsep}{1em}
{\renewcommand{\arraystretch}{2}
\begin{table}[th]
\vspace{0.25cm}
\begin{tabular}{| c | c | c |} \hline
{Mesh} &  {Normal} & {Umklapp} \\
\hline
\hline
{$729$} & {$922594$} & {$621662$} \\
\hline
{$225$} & {$118407$} & {$77689$} \\
\hline
{$81$} &  {$15581$} & {$9697$} \\
\hline
\end{tabular}
\label{table2_B}
\caption{Total number of scattering processes included in the calculation 
for the whole high-symmetry crystallographic path $\Gamma$-K-M-$\Gamma$.}
\end{table}}}

The auxiliar variables $\xi$ and $\eta$ required to evaluate 
the $\delta$-function and the principal part $P$, respectively, 
are studied on Fig.~\ref{eta_comp}. 
Here we show the case of 729 $\vec{q}$-points inside the IP at T=100~K. 
As is expected, when $\xi$ is too small the phonon linewidths $\Gamma(\vec{q},\lambda)$ 
results in uncorrelated peaks, i. e. $\xi=0.1$ cm$^{-1}$  (solid-red curve). 
In the opposite limit, i. e.,  $\xi=20$ cm$^{-1}$  (dot-dashed light-blue curve), 
the peaks become over-broadened and the phonon linewidth is over-estimated.
Therefore we used $\xi=5$ cm$^{-1}$  (solid black curve), which results in good 
convergence and is consistent with values adopted in 
previous related studies\cite{santoro2,greco,lorenzo}. 
Similar trend is present with the behavior of $\eta$. In this case the 
best choice turned out to be $\xi=1$ cm$^{-1}$.

%%%%%%%%%%%%%%%%%%%%%%%%%%%%%%%%%%%%%%%%%%%%%%%%%%%%%
%\section*{Appendix C} 
A last point which deserves special mention is the determination of the scattering channels.
This point constitutes a mayor test for the accurancy of the code.
Given a particular $\vec{q}$, any scattering process
\begin{eqnarray} 
\lambda_1&\leftrightarrow \lambda_2+\lambda_3\\ 
(\lambda_1&\leftrightarrow \lambda_2-\lambda_3) \nonumber 
\label{equi1}
\end{eqnarray}
should be identically to 
\begin{eqnarray} 
\lambda_1&\leftrightarrow \lambda_3+\lambda_2\\
(\lambda_1&\leftrightarrow \lambda_3-\lambda_2). \nonumber
\label{equi2}
\end{eqnarray}
for arbitrary $\lambda_i$= ZA, TA and LA, with $i=1,2,3$.
These processes, however, are calculated independently and 
the equality 
is valid only after the complete summations (Eqs.~(\ref{eq28-last}) and (\ref{eq26})) 
over the entire BZ (every possible $\vec{q}_2$ is included) are performed.
Thus, every single process involving different $\vec{q}_1$, $\vec{q}_2$ (and $\vec{q}_3$)
has to be taken into account properly, otherwise the equivalence 
will not be satisfied. 

The relationships (\ref{equi1}) and (\ref{equi2}) can be viewed as a consequence of the 
symmetry of the lattice and the properties of the inter-atomic force constants under 
inter-change of $\vec{q}_2$ with $\vec{q}_3$. 
Note for instance that using the equations of the Sect. IV, it can be shown that
\begin{equation}
\Phi_{ijk}^{(3)}\binom{BBB}{\vec{q}_1\vec{q}_2\vec{q}_3}=
\Phi_{ijk}^{(3)}\binom{BBB}{\vec{q}_1\vec{q}_3\vec{q}_2}. 
\label{eq42}
\end{equation}
\noindent The same is valid for $\Phi_{ijk}^{(3)}\binom{AAB}{\vec{q}_1\vec{q}_2\vec{q}_3}$, 
$\Phi_{ijk}^{(3)}\binom{AAA}{\vec{q}_1\vec{q}_2\vec{q}_3}$ 
and $\Phi_{ijk}^{(3)}\binom{BAA}{\vec{q}_1\vec{q}_2\vec{q}_3}$.
The remaining anharmonic force constant satisfy instead the following conditions 
where, in addition to $\vec{q}_2$ and $\vec{q}_3$, the order of the A and B atoms
must be also interchanged
\begin{eqnarray}
\Phi_{ijk}^{(3)}\binom{BAB}{\vec{q}_1\vec{q}_2\vec{q}_3}&=&
\Phi_{ijk}^{(3)}\binom{BBA}{\vec{q}_1\vec{q}_3\vec{q}_2} \\
\Phi_{ijk}^{(3)}\binom{ABA}{\vec{q}_1\vec{q}_2\vec{q}_3}&=&
\Phi_{ijk}^{(3)}\binom{AAB}{\vec{q}_1\vec{q}_3\vec{q}_2} \\
\Phi_{ijk}^{(3)}\binom{AAB}{\vec{q}_1\vec{q}_2\vec{q}_3}&=&
\Phi_{ijk}^{(3)}\binom{ABA}{\vec{q}_1\vec{q}_3\vec{q}_2} \\
\Phi_{ijk}^{(3)}\binom{AAB}{\vec{q}_1\vec{q}_2\vec{q}_3}&=&
\Phi_{ijk}^{(3)}\binom{ABA}{\vec{q}_1\vec{q}_3\vec{q}_2}. 
\label{eq46}
\end{eqnarray}
\vspace{0.11cm}

\noindent With the replacement of expressions in Eqs.~(\ref{eq42}) to (\ref{eq46}) 
into the Eq.~(\ref{eq26}) for 
$\Phi^{(3)}\binom{\lambda_1 \lambda_2 \lambda_3}{\vec{q}_1\vec{q}_2\vec{q}_3}$
together with the interchange of $\lambda_2$ and $\lambda_3$ it can be shown   
that 
%\begin{equation}
%\Phi^{(3)}\binom{\lambda_1 \lambda_2 \lambda_3}{\vec{q}_1\vec{q}_2\vec{q}_3}=
%%\sum_{\kappa_1 i,\kappa_2 j,\kappa_3 k}
%\sum_{\kappa_1 i} \sum_{\kappa_2 j} \sum_{\kappa_3 k}
%\frac{e_i^{\kappa_1}(\vec{q}_1,\lambda_1)
%e_j^{\kappa_2}(\vec{q}_2,\lambda_2)
%e_k^{\kappa_3}(\vec{q}_3,\lambda_3)}
%{\sqrt{8\omega(\vec{q}_1,\lambda_1)
%\omega(\vec{q}_2,\lambda_2) \omega(\vec{q}_3,\lambda_3)}} 
%\Phi_{ijk}^{(3)}\binom{\kappa_1 \kappa_2 \kappa_3}{\vec{q}_1\vec{q}_2\vec{q}_3}.
%\label{phi3}
%\end{equation}
\begin{eqnarray}
\Phi^{(3)}_{AAA}\binom{\lambda_1 \lambda_2 \lambda_3}{\vec{q}_1\vec{q}_2\vec{q}_3}&=&
\Phi^{(3)}_{AAA}\binom{\lambda_1 \lambda_3 \lambda_2}{\vec{q}_1\vec{q}_2\vec{q}_3} \\
\Phi^{(3)}_{ABB}\binom{\lambda_1 \lambda_2 \lambda_3}{\vec{q}_1\vec{q}_2\vec{q}_3}&=&
\Phi^{(3)}_{ABB}\binom{\lambda_1 \lambda_3 \lambda_2}{\vec{q}_1\vec{q}_2\vec{q}_3} \\
\Phi^{(3)}_{BBB}\binom{\lambda_1 \lambda_2 \lambda_3}{\vec{q}_1\vec{q}_2\vec{q}_3}&=&
\Phi^{(3)}_{BBB}\binom{\lambda_1 \lambda_3 \lambda_2}{\vec{q}_1\vec{q}_2\vec{q}_3} \\
\Phi^{(3)}_{BAA}\binom{\lambda_1 \lambda_2 \lambda_3}{\vec{q}_1\vec{q}_2\vec{q}_3}&=&
\Phi^{(3)}_{BAA}\binom{\lambda_1 \lambda_3 \lambda_2}{\vec{q}_1\vec{q}_2\vec{q}_3}
\end{eqnarray}
and, for example, that
\begin{eqnarray}
\Phi^{(3)}_{BBA}\binom{\lambda_1 \lambda_2 \lambda_1}{\vec{q}_1\vec{q}_2\vec{q}_3}&=&
\Phi^{(3)}_{BAB}\binom{\lambda_1 \lambda_1 \lambda_2}{\vec{q}_1\vec{q}_2\vec{q}_3} \\
\Phi^{(3)}_{AAB}\binom{\lambda_1 \lambda_2 \lambda_3}{\vec{q}_1\vec{q}_2\vec{q}_3}&=&
\Phi^{(3)}_{ABA}\binom{\lambda_1 \lambda_2 \lambda_1}{\vec{q}_1\vec{q}_2\vec{q}_3} 
\nonumber\\
\end{eqnarray}
and other similar ones, where we have used the notation
\begin{eqnarray}
\Phi^{(3)}_{\kappa_1 \kappa_2 \kappa_3}
\binom{\lambda_1 \lambda_2 \lambda_1}{\vec{q}_1\vec{q}_2\vec{q}_3}=
\frac{e_i^{\kappa_1}(\vec{q}_1,\lambda_1) e_i^{\kappa_2}(\vec{q}_2,\lambda_2)
e_i^{\kappa_3}(\vec{q}_3,\lambda_3)}
{\sqrt{8\omega(\vec{q}_1,\lambda_1)
\omega(\vec{q}_2,\lambda_2) \omega(\vec{q}_3,\lambda_3)}} \nonumber\\
\times \Phi_{iii}^{(3)}\binom{\kappa_1 \kappa_2 \kappa_3}{\vec{q}_1\vec{q}_2\vec{q}_3}.
\ \ \ \ \ \ \ \ \ \ 
\end{eqnarray}

In addition, the calculation of any given $\Phi_{ijk}^{(3)}$ requires the 
evaluation of exponential functions where the scalar product $\vec{q}\cdot\vec{X}$ 
may result in an integer fraction of $\pi$. 
Any small error or loss of precision in the sum over the wave vectors
$\vec{q}_1$, $\vec{q}_2$, $\vec{q}_3$ (Normal) and $\pm\vec{G}$ (Umklapp) 
can prevent the numerical equivalence between Eqs.~(\ref{equi1}) and (\ref{equi2}).

\begin{figure*}[t]
\vspace{0.25cm}
\includegraphics[trim=0.5cm 0.1cm 0.40cm 0.1cm, clip, width=0.32\textwidth]{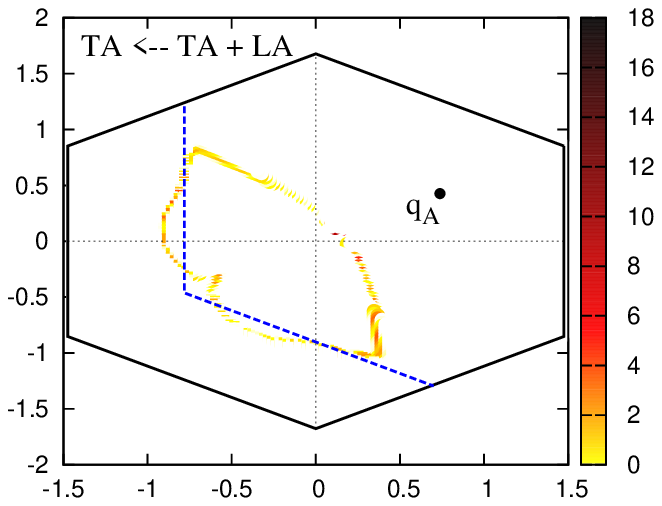}
\hspace{0.1cm}
\includegraphics[trim=0.5cm 0.1cm 0.40cm 0.1cm, clip, width=0.32\textwidth]{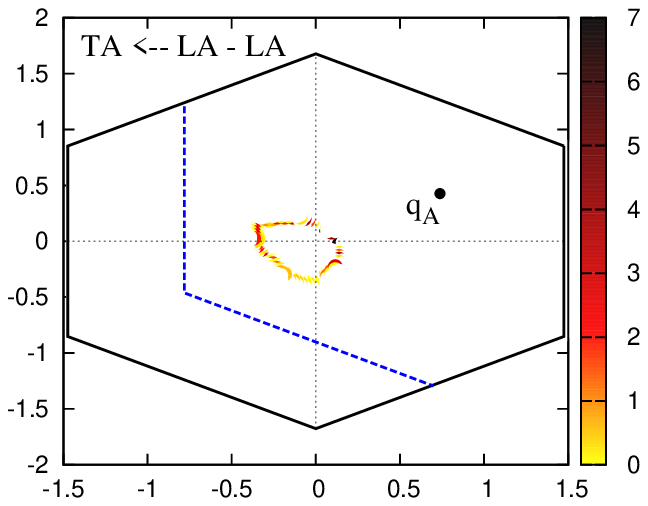}
\hspace{0.1cm}
\includegraphics[trim=0.5cm 0.1cm 0.40cm 0.1cm, clip, width=0.32\textwidth]{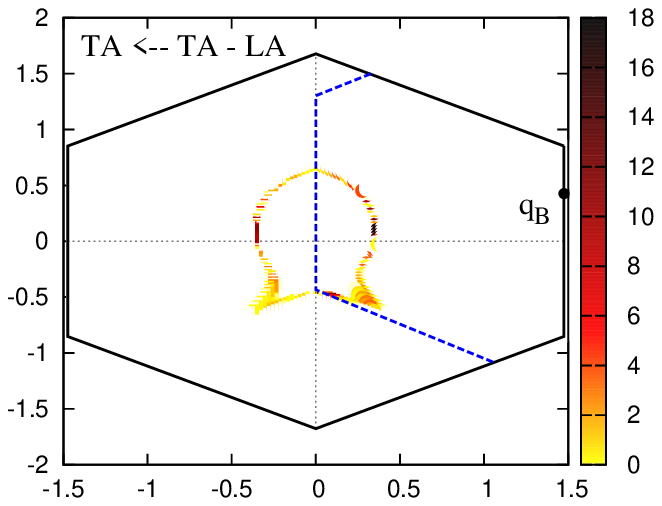}

\includegraphics[trim=0.5cm 0.1cm 0.40cm 0.1cm, clip, width=0.32\textwidth]{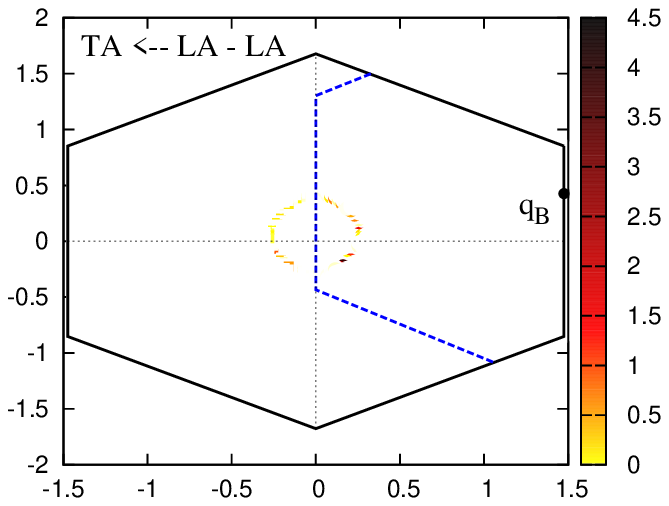}
%includegraphics[trim=0.5cm 0.1cm 0.40cm 0.1cm, clip, width=0.32\textwidth]{figs/qC-211-RI1.eps}
\hspace{0.1cm}
\includegraphics[trim=0.5cm 0.1cm 0.40cm 0.1cm, clip, width=0.32\textwidth]{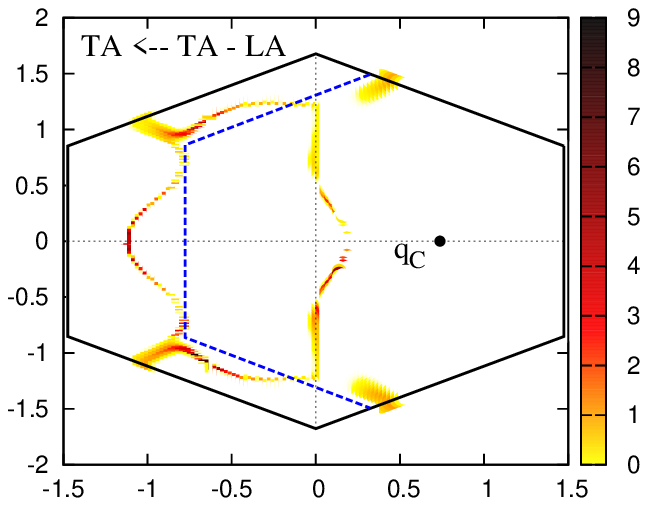}
\hspace{0.1cm}
\includegraphics[trim=0.5cm 0.1cm 0.40cm 0.1cm, clip, width=0.32\textwidth]{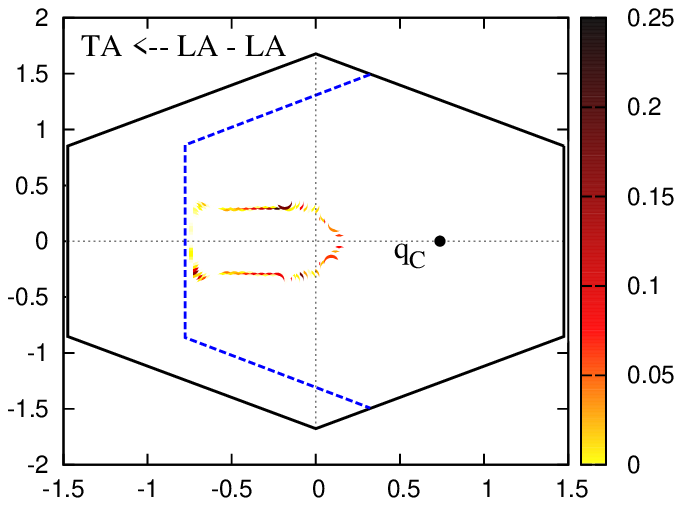}

%\includegraphics[width=0.45\textwidth]{figs/proc_all_sum_each_final.eps}
%\hspace{0.1cm}
\includegraphics[trim=0.5cm 0.1cm 0.40cm 0.1cm, clip, width=0.32\textwidth]{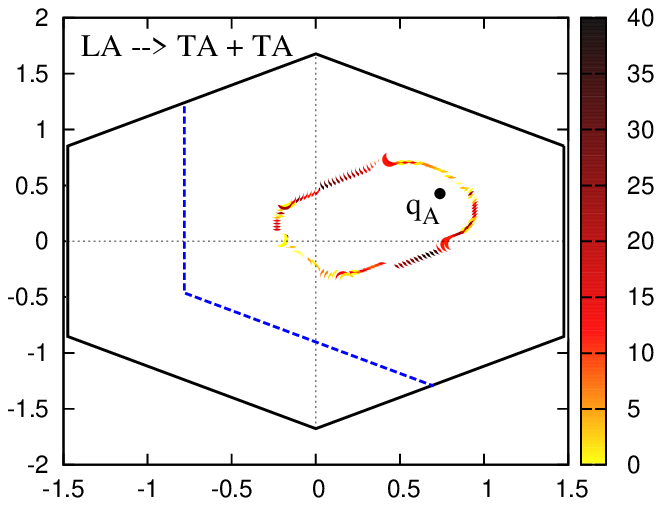}
\hspace{0.1cm}
\includegraphics[trim=0.5cm 0.1cm 0.40cm 0.1cm, clip, width=0.32\textwidth]{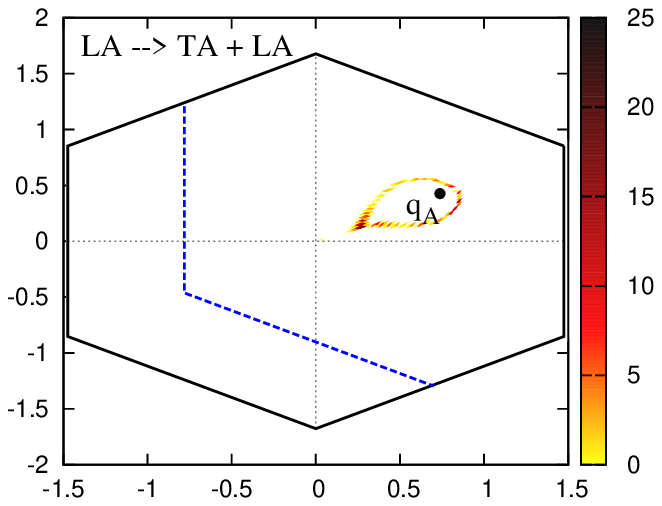}
\hspace{0.1cm}
\includegraphics[trim=0.5cm 0.1cm 0.40cm 0.1cm, clip, width=0.32\textwidth]{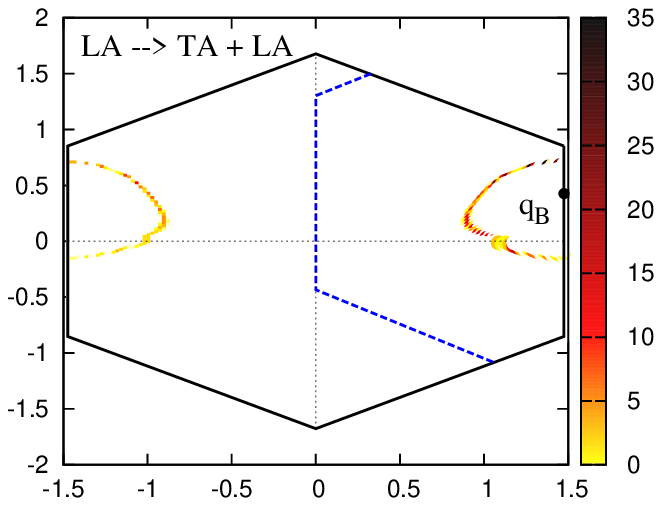}
\caption{(Color online)
Contour plots of the contribution of selected scattering channels along the BZ 
for the TA and LA acoustic phonon modes of graphene at 
$\vec{q}_A, \vec{q}_B$ and $\vec{q}_C$ (i. e. the middle points  of $\Gamma-$K$-$M$-\Gamma$). 
Units cm$^{-1}$ ($q_x$ (x-axis) and $q_y$ (y-axis) are given in units of $\AA^{-1}$). 
}\label{fig-sca}
\end{figure*}

Finally, the spatial distribution inside the BZ of the 
scattering channels (Fig.~\ref{fig-sca}) is also representative of 
the accurancy of the code (this figure is complementary to Fig.~\ref{damp_ZB-1}). 
Notice therein the perfect match of the blue dotted line 
between Normal and Umklapp contributions which, as we explained above,
require the inclusion of a reciprocal lattice vector $\vec{G}$.
%ofhich is determined 
%automatically respecting the symmetry properties. 

%%%%%%%%%%%%%%%%%%%%%%%%%%%%%%%%%%%%%%%%%%%%%%%%%%%%%%%%%%%

%%%%%%%%%%%%%%%%%%%%%%%%%%%%%%%%%%%%%%%%%%%%%%%%%%%%%%%%%%%


\begin{thebibliography}{99}

%1
\bibitem{book} J. M. Ziman, {\it Electrons and Phonons: The theory of 
Transport Phenomena in Solids}, Oxford University Press, Oxford (1960).

%2
\bibitem{maradudin}See e.g., A. A. Maradudin, {\it Dynamical Properties of Solids}, 
Vols. 1 and 2, G. K. Horton and A. A. Maradudin Eds., 
North-Holland Publ. Co, Amsterdam (1974).

%3
\bibitem{geim}K. S. Novoselov, A.K. Geim, 
S. V. Morosov, D. Jiang, Y. Zhang, S. V. Dubonos, I.V. Grigorieva,
and A. A. Frisov, Science {\bf 306}, 666 (2004). 

%4
\bibitem{geim2}K. S. Novoselov, D. Jiang, F. Schedin, T. J. Booth, 
V. V. Khotkevich, S. V. Morozov, 
and A. K. Geim, Science {\bf 102}, 10451 (2005).

%5
\bibitem{castro} A. H. Castro Neto, F. Guinea, N. M. R. Peres, K. S.
Novoselov, and A. K. Geim, Reviews of Modern Physics {\bf 81}, 109 (2009).

%6
\bibitem{meyer}J. C. Meyer, A. K. Geim, M. I. Katsnelson, K. S: Novoselov, 
D. Obergfell, S. Roth, C. Girit, and A. Zette, Nature (London) {\bf 446}, 60 (2007).

%7
\bibitem{yoon} D. Yoon, Y-W. Son, and H. Cheong, Nano Letters {\bf 11}, 3227 (2011).

%8
\bibitem{bao} W. Bao, F. Miao, Z. Chen, H. Zhang, W. Janoy, C. Dames, and C. N. Lau,
Nat. Nanotechnol. {\bf 4}, 562 (2009).

%9
\bibitem{Balandin-exp}A. A. Balandin, S. Ghosh, W. Bao, I. Calizo, 
D. Teweldebrhan, F. Miao, and C. N. Lau, Nano Lett. {\bf 8}, 902 (2008).

%10
\bibitem{science}H. Seol, I. Jo, A. L. Moore, L. Lindsay, Z. H. Aitken, M.
T. Pettes, X. Li, Z. Yao, R. Huang, D. Broido, N. Mingo,
R. S. Ruoff, and L. Shi, Science {\bf 328}, 213 (2010).

%11
\bibitem{wang}Z. Wang, R. Xie, C. T. Bui, D. Liu, X. Ni, B. Li,
and J. T. L. Long, Nano Lett. {\bf 11}, 112 (2011).

%12
\bibitem{bala}A. A. Balandin, Nature Materials {\bf 10}, 569 (2011). 

%13
\bibitem{Lifshitz}I. M. Lifshitz. Zh. Eksp. Teor. Fiz. {\bf 22}, 475 (1952).

%14
\bibitem{mounet} N. Mounet and N. Marzari, Phys. Rev. B {\bf 71}, 205214 (2005).

%15
\bibitem{faso3}K. V. Zakharchenko, M. I. Katsnelson, and A. Fasolino,
Phys. Rev. Lett. {\bf 102}, 046808 (2009). 

\bibitem{franco} A. L. C. da Silva, L. Candido, J. N. Teisceira Rabelo, G. Q. Hai, 
and F. M. Peeters, EPL {\bf 107}, 56004 (2014).

%16
\bibitem{fasolino}A. Fasolino, J. H. Los, and M. I. Katsnelson, 
Nat. Mater. {\bf 6}, 858 (2007).

%17
\bibitem{nano-marzari} N. Bonini, J. Garg, and N. Marzari, Nano Lett. {\bf 12}, 2673 (2012).

%18
\bibitem{lorenzo}L. Paulatto, F. Mauri, and M. Lazzeri,
Phys. Rev. B {\bf 87}, 214303 (2013). 

%19
\bibitem{amorin}B. Amorin, R. Roldan, E. Cappellutti, A. Fasolino, 
F. Guinea and M. I. Katnelson, Phys. Rev. B {\bf 89}, 224307 (2014).

%20
\bibitem{chaikin}P. M. Chaikin and T. C. Lubensky, 
{\it Principles of condensed matter physics}, Cambridge University Press, Cambridge (1995).

%21
\bibitem{membrane}D. Nelson, T. Piran and S. Weinberg, 
{\it Statistical Mechanics of Membranes and Surface}, 
Word Scientiﬁc, Singapore (1989).

%22
\bibitem{Landau}L. D. Landau and E. M. Lifshitz, {\it Elastizit\"atstheorie} 
(Akademie-Verlag, Berlin, 1968).

\bibitem{Nelson-Peliti}D. R. Nelson and L. Peliti, J. Physique {\bf 48}, 1085 (1987).

%23
\bibitem{Mohr}M. Mohr, J. Maultzsch, E. Dobardzie, S. Reich, I. Milosevic, M. Damnjanovic, 
A. Bosak, M. Krisch, and C. Thomsen, Phys. Rev. B {\bf 76}, 035439 (2007).

%24
\bibitem{nikas}D. L. Nika, A. S. Askerov, and A. A. Balandin, Nano Letters {\bf 12}, 3238 (2012).

%25
\bibitem{nika-1}D. L. Nika and A. A. Balandin, 
J. Phys.: Condens. Matter {\bf 24} 233203 (2012).

%26
\bibitem{hindawi} T. Feng and X. Ruan, Journal of Nanomaterials 
{\bf 2014}, 206370 (2014). 

\bibitem{broido1} L. Lindsay, D. A. Broido, and N. Mingo, 
Phys. Rev. B {\bf 82}, 115427 (2010). 

\bibitem{fisher1} D. Singh, J. Y. Murthy, and T. S. Fisher, J. Appl. Phys.
{\bf 110}, 094312 (2011).

%27
\bibitem{Born} M. Born and K. Huang, {\it Dynamical Theory of Crystal Lattices}, 
(Oxford University Press, New York, 1952).

%28
\bibitem{Saito}R. Saito, G. Dresselhaus, and M. S. Dresselhaus, 
{\it Physical Properties of Carbon Nanotubes}  
(Imperial College, London, 1998).

%29
\bibitem{Reich} S. Reich, C. Thomson, and J. Maultzsch, {\it Carbon Nanotubes}
(Wiley-VCH, Weinheim, 2004).

%30
\bibitem{peirels} R. Peierls, {\it Quantum theory of Solids}, (Oxford) 1955.

\bibitem{Karl}K. H. Michel and B. Verberck, Phys. Rev. B {\bf 78}, 085424 (2008).

\bibitem{lindsay1}L. Lindsay, D. A. Broido, and N. Mingo, 
Phys. Rev. B {\bf 80}, 125407 (2009).

\bibitem{Nickow}R. Nicklow, N. Wakabayashi, and H. G. Smith,
Phys. Rev. B {\bf 5}, 4951 (1971).


%%%%
\bibitem{Leibfried}G. Leibfried and W. Ludwig, {\it Theory of Anharmonic Effects in Crystals},  
Solid State Physics vol. {\bf 12}, edited by F. Seitz and D. Turnbull (Academic press, New York, 1661). 

%33
\bibitem{gotze}W. G\"otze and K. H. Michel, Phys. Rev. {\bf 157}, 738 (1967).

%34
\bibitem{Kokkedee}J. Kokkedee, Physica {\bf 28}, 374 (1962).

%35
\bibitem{Maradudin1}A. A. Maradudin and A. E. Fein, Phys. Rev. {\bf 128}, 2589 (1962).

%36

\bibitem{Mariani}E. Mariani and F. von Oppen, Phys. Rev Lett. {\bf 100}, 076801 (2010).

%\bibitem{costa1}S. Costamagna and A. Dobry, Phys. Rev. B {\bf 83}, 233401 (2011).

%\bibitem{pablo1}P. Scuracchio and A. Dobry, Phys. B {\bf 87}, 165411 (2013).
%Bending mode fluctuations and structural stability of graphene nanoribbons

\bibitem{pablo2}P. Scuracchio, S. Costamagna, F. M. Peeters, and A. Dobry, Phys. Rev. B {\bf 90}, 035429 (2014).
%Role of atomic vacancies and boundary conditions on ballistic thermal transport in graphene nanoribbons
%P. Scuracchio, S. Costamagna, F. M. Peeters, and A. Dobry
%Phys. Rev. B 90, 035429 – Published 21 July 2014

\bibitem{hanfland}M. Hanfland, H. Beister, and K. Syassen, Phys. Rev. B {\bf 39}, 12598 (1989).

\bibitem{ferrari}T. M. G. Mohiuddin, A. Lombardo, R. R. Nair, A. Bonetti, G. Savini, R. Golil, N. Bonini,
D. M. Basko, C. Galiotis, N. Marzari, K,. S. Novoselov, A. K. Geim and, A. C. Ferrari, 
Phys. Rev. B {\bf 79}, 205433 (2009).


\bibitem{Landau1}L. D. Landau and E. M. Lifshitz, {\it Statistische Physik},
Teil 1, Akademie-Verlag, Berlin (1987).

\bibitem{sevik}C. Sevik, Phys. Rev. B {\bf 89} 035422 (2014).

\bibitem{linas}
S. Linas, Y. Magnin, B. Poinsot, O. Boisron, G. D. Forster, Z. Han, D. Kalita, V. Bouchiat, V. Martinez, R. Fulcrand, F. Tournus, V. Dupuis, F. Rabilloud, L. Bardotti, F. Calvo,
preprint arXiv:1411.7840v1 (2014).
%%%%%%%% Numerical

\bibitem{santoro2} F. Franchini, G. Santoro, V. Bortolani, A. A. Maradudin,
and R. F. Willis, Phys. Rev. B {\bf 45}, 11982 (1992).

\bibitem{greco} A. Greco, S. Koval, and R. Migoni, J. Phys.: Condens. Matter {\bf 4}, 5291 (1992).

\bibitem{fultz1} X. Tang, C. L. Li, and B. Fultz, Phys. Rev. B {\bf 82}, 184301(2010).

\bibitem{fultz2} X. Tang and B. Fultz, Phys. Rev. B {\bf 84}, 054303 (2011).

\bibitem{balandin1} D. L. Nika, E. P. Pokatilov, A. S. Askerov, and A. A.
Balandin, Phys. Rev. B {\bf 79}, 155413 (2009).



\bibitem{lindsay2} L. Lindsay, D. A. Broido, and N. Mingo, 
Phys. Rev. B {\bf 82}, 115427 (2010).

\bibitem{haro} E. Haro, M. Balkanski, R. F. Wallis, and K. H. Wanser,
Phys. Rev. B {\bf 34}, 5358 (1986).

\bibitem{seol} E. Haro-Poniatowski, J. L. Escamilla-Reyes, and K. H.
Wanser, Phys. Rev. B {\bf 53}, 12121 (1996).

\bibitem{fisher2} D. Singh, J. Y. Murthy, and T. S. Fisher, J. Appl. Phys.
{\bf 110}, 044317 (2011).

\bibitem{lapack}http://www.netlib.org/lapack 

\bibitem{sort}
%http://blog.sciencenet.cn/blog-345795-422819.html?COLLCC=1302924382
L. F. Huang and Z. Zeng, 
%Lattice dynamics and disorder-induced lattice contraction in functionalized graphene, 
J. Appl. Phys. {\bf 113}, 083524 (2013).

\bibitem{santoro1} M. Zoli, G. Santoro, V. Bortolani, A. A. Maradudin, and
R. F. Willis, Phys. Rev. B {\bf 41}, 7507 (1990).


\end{thebibliography}
\end{document}